\documentclass{article}

\pdfoutput=1

\usepackage[nonatbib, final]{neurips_2020_mod}

\usepackage[utf8]{inputenc} 
\usepackage[T1]{fontenc}    
\usepackage[hypertexnames=false]{hyperref}
\usepackage{url}            
\usepackage{booktabs}       
\usepackage{amsfonts,amsmath,amsthm}       
\usepackage{nicefrac}       
\usepackage{microtype}      
\usepackage{graphicx}      

\usepackage{cancel}
\usepackage{algorithm}
\usepackage[noend]{algpseudocode}
\usepackage[usestackEOL]{stackengine}
\usepackage{xr-hyper}
\usepackage{hyperref}

\newtheorem{mydef}{Definition}

\newtheorem{mythmm}{Theorem}
\newtheorem*{myremark}{Remark}
\newtheorem{mylemma}{Lemma}

\newtheorem{mydefsm}{Definition}
\newtheorem{mypropsm}{Proposition}

\newtheorem{mycorsm}{Corollary}
\newtheorem{mylemmasm}{Lemma}

\newtheorem{innercustomgeneric}{\customgenericname}
\providecommand{\customgenericname}{}
\newcommand{\newcustomtheorem}[2]{%
  \newenvironment{#1}[1]
  {%
   \renewcommand\customgenericname{#2}%
   \renewcommand\theinnercustomgeneric{##1}%
   \innercustomgeneric
  }
  {\endinnercustomgeneric}
}

\newcustomtheorem{customlemma}{Lemma}
\newcustomtheorem{customtheorem}{Theorem}

\theoremstyle{definition}

\def\ci{\perp\!\!\!\perp}

\newcommand{\G}{\mathcal{G}}
\newcommand{\M}{\mathcal{M}(\mathcal{G})}

\newcommand{\MstatA}{\mathcal{M}(\mathcal{G})_{statA}}

\newcommand{\PG}{\mathcal{P}(\mathcal{G})}

\newcommand{\PGstatA}{\mathcal{P}(\mathcal{G})_{statA}}
\newcommand{\PGstatAO}{\mathcal{P}(\mathcal{G})_{statAO}}

\newcommand{\C}{\mathcal{C}(\mathcal{G})}
\newcommand{\Xobs}{\mathbf{X}}
\newcommand{\pa}{pa}
\newcommand{\an}{an}
\newcommand{\adj}{adj}
\newcommand{\spouse}{sp}

\newcommand{\omark}{$\circ$}
\newcommand{\DSEP}{\text{D-Sep}}
\newcommand{\PDSEP}{\text{Possible-D-Sep}}
\newcommand{\taumax}{\tau_{\max}}
\newcommand{\SSet}{\mathcal{S}}
\newcommand{\SSep}{\SSet}
\newcommand{\apds}{apds_t}
\newcommand{\napds}{napds_t}
\newcommand{\ppc}{p}
\newcommand{\porder}{p_{ts}}
\newcommand{\totalorder}{{$<$ }}
\newcommand{\tailhead}{{\rightarrow}}
\newcommand{\headtail}{{\leftarrow}}
\newcommand{\headhead}{{\leftrightarrow}}
\newcommand{\ohead}{{\circ\!{\rightarrow}}}
\newcommand{\oo}{{\circ\!{\--}\!\circ}}
\newcommand{\heado}{{{\leftarrow}\!\circ}}
\newcommand{\astast}{{\ast\!{\--}\!\ast}}
\newcommand{\asthead}{{\ast\!{\rightarrow}}}
\newcommand{\headast}{{{\leftarrow}\!\ast}}
\newcommand{\asto}{{\ast\!{\--}}\!\!\circ\!}
\newcommand{\oast}{{\circ\!{\--}}\!\!\!\ast}
\newcommand{\ostar}{{\circ\!{\--}\!\star}}
\newcommand{\staro}{{\star\!{\--}\!\circ}}
\newcommand{\starhead}{{\star\!{\rightarrow}}}
\newcommand{\headstar}{{{\leftarrow}\!\star}}

\newcommand{\LMM}[1]{{\mathbin{\ensurestackMath{\stackon[-1pt]{#1}{{\scriptscriptstyle L}}}}}}
\newcommand{\RMM}[1]{{\mathbin{\ensurestackMath{\stackon[-1pt]{#1}{{\scriptscriptstyle R}}}}}}
\newcommand{\QMM}[1]{{\mathbin{\ensurestackMath{\stackon[-2pt]{#1}{{\scriptscriptstyle ?}}}}}}
\newcommand{\EMM}[1]{{\mathbin{\ensurestackMath{\stackon[-2pt]{#1}{{\scriptscriptstyle !}}}}}}

\newcommand{\SMM}[1]{{\mathbin{\ensurestackMath{\stackon[-2pt]{#1}{{\scriptstyle \ast}}}}}}
\newcommand{\LMMnoast}[1]{{\mathbin{\ensurestackMath{\stackon[-0pt]{#1}{{\hspace*{-0.5mm}\scriptscriptstyle L}}}}}}
\newcommand{\RMMnoast}[1]{{\mathbin{\ensurestackMath{\stackon[-0pt]{#1}{{\hspace*{-0.5mm}\scriptscriptstyle R}}}}}}
\newcommand{\QMMnoast}[1]{{\mathbin{\ensurestackMath{\stackon[-0pt]{#1}{{\hspace*{-0.5mm}\scriptscriptstyle ?}}}}}}
\newcommand{\EMMnoast}[1]{{\mathbin{\ensurestackMath{\stackon[-0pt]{#1}{{\hspace*{-0.5mm}\scriptscriptstyle !}}}}}}

\newcommand{\SMMnoast}[1]{{\mathbin{\ensurestackMath{\stackon[-0pt]{#1}{{\scriptstyle \ast}}}}}}
\newcommand{\R}{\mathcal{R}}
\newcommand{\ER}[1]{\mathcal{R}#1^\prime}

\newcommand{\nprelim}{k}
\newcommand{\onlyright}{{\rightarrow}}
\newcommand{\leftright}{{\leftrightarrow}}
\newcommand{\oright}{{\circ\!{\rightarrow}}}

\title{High-recall causal discovery for autocorrelated time series with latent
confounders}
\author{%
Andreas Gerhardus\\
German Aerospace Center\\
Institute of Data Science\\
07745 Jena, Germany\\
\texttt{andreas.gerhardus@dlr.de}
\And
Jakob Runge\\
German Aerospace Center\\
Institute of Data Science\\
07745 Jena, Germany\\
\texttt{jakob.runge@dlr.de}
}

\begin{document}

\maketitle

\begin{abstract}
We present a new method for linear and nonlinear, lagged and contemporaneous constraint-based causal discovery from observational time series in the presence of latent confounders. We show that existing causal discovery methods such as FCI and variants suffer from low recall in the autocorrelated time series case and identify low effect size of conditional independence tests as the main reason. Information-theoretical arguments show that effect size can often be increased if causal parents are included in the conditioning sets. To identify parents early on, we suggest an iterative procedure that utilizes novel orientation rules to determine ancestral relationships already during the edge removal phase. We prove that the method is order-independent, and sound and complete in the oracle case. Extensive simulation studies for different numbers of variables, time lags, sample sizes, and further cases demonstrate that our method indeed achieves much higher recall than existing methods for the case of autocorrelated continuous variables while keeping false positives at the desired level. This performance gain grows with stronger autocorrelation. 
At \href{https://github.com/jakobrunge/tigramite}{\nolinkurl{github.com/jakobrunge/tigramite}} we provide Python code for all methods involved in the simulation studies.
\end{abstract}

\section{Introduction}
Observational causal discovery \cite{Spirtes2000,Peters2018} from time series is a challenge of high relevance to many fields of science and engineering if experimental interventions are infeasible, expensive, or unethical. Causal knowledge of direct and indirect effects, interaction pathways, and time lags can help to understand and model physical systems and to predict the effect of interventions \cite{Pearl2000}. Causal graphs can also guide interpretable variable selection for prediction and classification tasks.
Causal discovery from time series faces major challenges \cite{Runge2019a} such as unobserved confounders, high-dimensionality, and  nonlinear dependencies, to name a few. Few frameworks can deal with these challenges and we here focus on constraint-based methods pioneered in the seminal works of Spirtes, Glymour, and Zhang \cite{Spirtes2000,Zhang2008}. We demonstrate that existing latent causal discovery methods strongly suffer from low recall in the time series case where identifying lagged and contemporaneous causal links is the goal and autocorrelation is an added, ubiquitous challenge. Our main \textbf{theoretical contributions} lie in identifying low effect size as a major reason why current methods fail and in introducing a novel sound, complete, and order-independent causal discovery algorithm that yields strong gains in recall for autocorrelated continuous data. Our \textbf{practical contributions} lie in extensive numerical experiments that can serve as a future benchmark and in open-source Python implementations of our and major previous time series causal discovery algorithms. \textbf{The paper is structured as follows:} After briefly introducing the problem and existing methods in Sec.~\ref{sec:causal_discovery}, we describe our method and theoretical results in Sec.~\ref{sec:lpcmci}. Section~\ref{sec:numerics} provides numerical experiments followed by a discussion of strengths and weaknesses as well as an outlook in Sec.~\ref{sec:conclusion}. The paper is accompanied by Supplementary Material (SM).

\section{Time series causal discovery in the presence of latent confounders} \label{sec:causal_discovery}

\subsection{Preliminaries}
We consider multivariate time series $\mathbf{V}^j = (V^j_t, V^j_{t-1},\ldots)$ for $j=1,\ldots,\tilde{N}$ that follow a stationary discrete-time structural vector-autoregressive process described by the structural causal model (SCM)
\begin{align}\label{eqmain:SVARProcess}
V^j_t = f_j(\pa(V^j_t), \eta^j_t) \quad \text{with } j = 1, \ldots, \tilde{N} \ .
\end{align}
The measurable functions $f_j$ depend non-trivially on all their arguments, the noise variables $\eta^j_t$ are jointly independent, and the sets $\pa(V^j_t) \subseteq (\mathbf{V}_t, \mathbf{V}_{t-1}, \ldots, \mathbf{V}_{t-\porder})$ define the causal parents of $V^j_t$. Here, $\mathbf{V}_t = (V^1_t, V^2_t, \ldots)$ and $\porder$ is the order of the time series. Due to stationarity the causal relationship of the pair of variables $(V^i_{t-\tau}, V^j_t)$, where $\tau \geq 0$ is known as lag, is the same as that of all time shifted pairs $(V^i_{t^\prime-\tau}, V^j_{t^\prime})$. This is why below we always fix one variable at time $t$. We assume that there are no cyclic causal relationships, which as a result of time order restricts the contemporaneous ($\tau = 0)$ interactions only. We allow for unobserved variables, i.e., we allow for observing only a subset $\Xobs = \{\mathbf{X}^1, \ldots ,\mathbf{X}^N\} \subseteq \mathbf{V} = \{\mathbf{V}^1, \mathbf{V}^{2}, \ldots\} $ of time series with $N \leq \tilde{N}$. We further assume that there are no selection variables and assume the faithfulness \cite{Spirtes2000} condition, which states that conditional independence (CI) in the observed distribution $P(\bold{V})$ generated by the SCM implies d-separation in the associated time series graph $\G$ over variables $\bold{V}$.

We assume the reader is familiar with the Fast Causal Inference (FCI) algorithm \cite{Spirtes1995, Spirtes2000, Zhang2008} and related graphical terminology, see Secs.~\ref{sec:background} and \ref{sec:fci} of the SM for a brief overview. Importantly, the MAGs (maximal ancestral graphs) considered in this paper can contain directed ($\tailhead$) and bidirected ($\headhead$) edges (interchangeably also called links). The associated PAGs (partial ancestral graphs) may additionally have edges of the type $\ohead$ and $\oo$.

\subsection{Existing methods}
The \textbf{tsFCI} algorithm \cite{Entner2010} adapts the constraint-based \textbf{FCI} algorithm to time series. It uses time order and stationarity to restrict conditioning sets and to apply additional edge orientations. \textbf{SVAR-FCI} \cite{malinsky2018causal} uses stationarity to also infer additional edge removals. There are no assumptions on the functional relationships or on the structure of confounding. \textbf{Granger causality} \cite{Granger1969} is another common framework for inferring the causal structure of time series. It cannot deal with contemporaneous links (known as instantaneous effects in this context) and may draw wrong conclusions in the presence of latent confounders, see e.g. \cite{Peters2018} for an overview. The \textbf{ANLTSM} method \cite{Chu2008} restricts contemporaneous interactions to be linear, and latent confounders to be linear and contemporaneous. \textbf{TS-LiNGAM} \cite{TSLINGAM} is based on \textbf{LiNGAM} \cite{Shimizu2006} that is rooted in the structural causal model framework \cite{Peters2018, Spirtes2016}. It allows for contemporaneous effects, assumes linear interactions with additive non-Gaussian noise, and might fail in the presence of confounding. The \textbf{TiMINo} \cite{PetersJSLMZK2013} method restricts interactions to an identifiable function class or requires an acyclic summary graph. Yet another approach are \textbf{Bayesian score-based or hybrid methods} \cite{Chickering2002, Tsamardinos2006}. These often become computationally infeasible in the presence of unobserved variables, see \cite{Jabbari2017} for a discussion, or make restrictive assumptions about functional dependencies or variable types.

In this paper we follow the constraint-based approach that allows for general functional relationships (both for lagged and contemporaneous interactions), general types of variables (discrete and continuous, univariate and multivariate), and that makes no assumption on the structure of confounding. The price of this generality is that we will not be able to distinguish all members of a Markov equivalence class (although time order and stationarity allow to exclude some members of the equivalence class). Due to its additional use of stationarity we choose SVAR-FCI rather than tsFCI as a baseline and implement the method, restricted to no selection variables, in Python. As a second baseline we implement \textbf{SVAR-RFCI}, which is a time series adaption of RFCI along the lines of SVAR-FCI (also restricted to no selection variables). The \textbf{RFCI} algorithm \cite{Colombo2012} is a modification of FCI that does not execute FCI's potentially time consuming second edge removal phase.

\subsection{On maximum time lag, stationarity, soundness, and completeness}\label{sec:onstationarity}
In time series causal discovery the assumption of stationarity and the length of the chosen time lag window $t - \taumax \leq t^\prime \leq t$ play an important role. In the causally sufficient case ($\mathbf{X} = \mathbf{V})$ the causal graph stays the same for all $\taumax\geq \porder$. Not so in the latent case: Let $\M^{\taumax}$ be the MAG obtained by marginalizing over all unobserved variables and also all generally observed variables at times $t^\prime < t -\taumax$. Then, increasing the considered time lag window by increasing $\taumax$ may result in the removal of edges that are fully contained in the original window, even in the case of perfect statistical decisions. In other words, $\M^{\tau_{\max, 1}}$ with $\tau_{\max, 1} < \tau_{\max, 2}$ need not be a subgraph of $\M^{\tau_{\max, 2}}$. Hence, $\taumax$ may be regarded more as an analysis choice than as a tunable parameter. For the same reason stationarity also affects the definition of MAGs and PAGs that are being estimated. For example, SVAR-FCI uses stationarity to also remove edges whose separating set extends beyond the chosen time lag window. It does, therefore, in general not determine a PAG of $\M^{\taumax}$. To formalize this let $i)$ $\MstatA^{\taumax}$ be the MAG obtained from $\M^{\taumax}$ by enforcing repeating adjacencies, let $ii)$ $\PGstatA^{\taumax}$ be the maximally informative PAG for the Markov equivalence class of $\MstatA^{\taumax}$, which can be obtained from running the FCI orientation rules on $\MstatA^{\taumax}$, and let $iii)$ $\PGstatAO^{\taumax}$ be the PAG obtained when additionally enforcing time order and repeating orientations at each step of applying the orientation rules. Note that $\PGstatAO^{\taumax}$ may have fewer circle marks, i.e., may be more informative than $\PGstatA^{\taumax}$. Our aim is to estimate $\PGstatAO^{\taumax}$. We say an algorithm is \textit{sound} if it returns a PAG for $\MstatA^{\taumax}$, and \textit{complete} if it returns $\PGstatAO^{\taumax}$. Below we write $\M = \MstatA^{\taumax}$ and $\PG = \PGstatAO^{\taumax}$ for simplicity.

\subsection{Motivational example}
\begin{figure*}[t]  
\centering
\includegraphics[width=.33\linewidth]{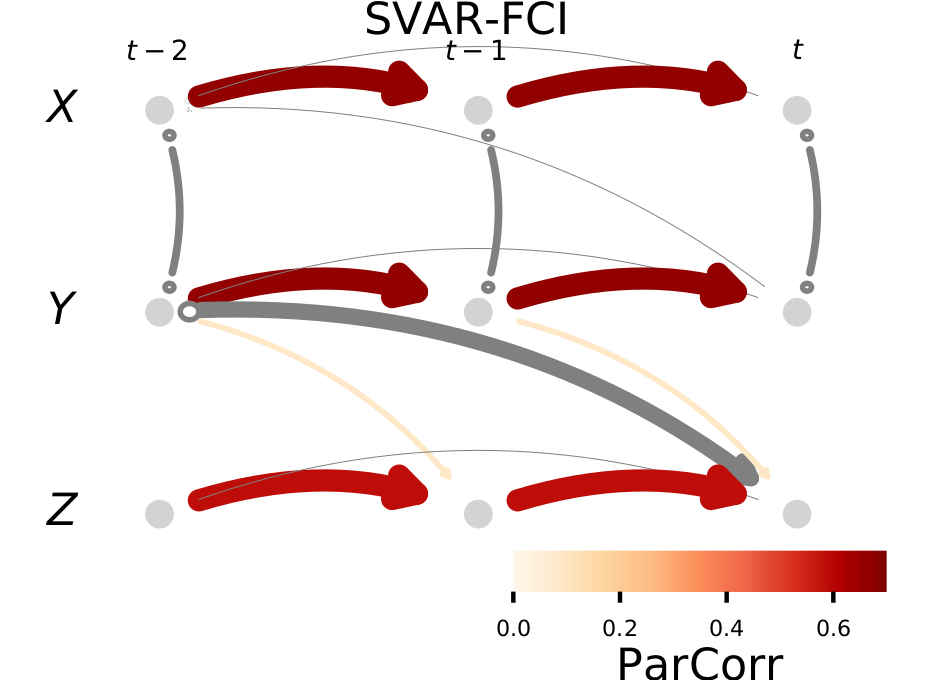}%
\includegraphics[width=.33\linewidth]{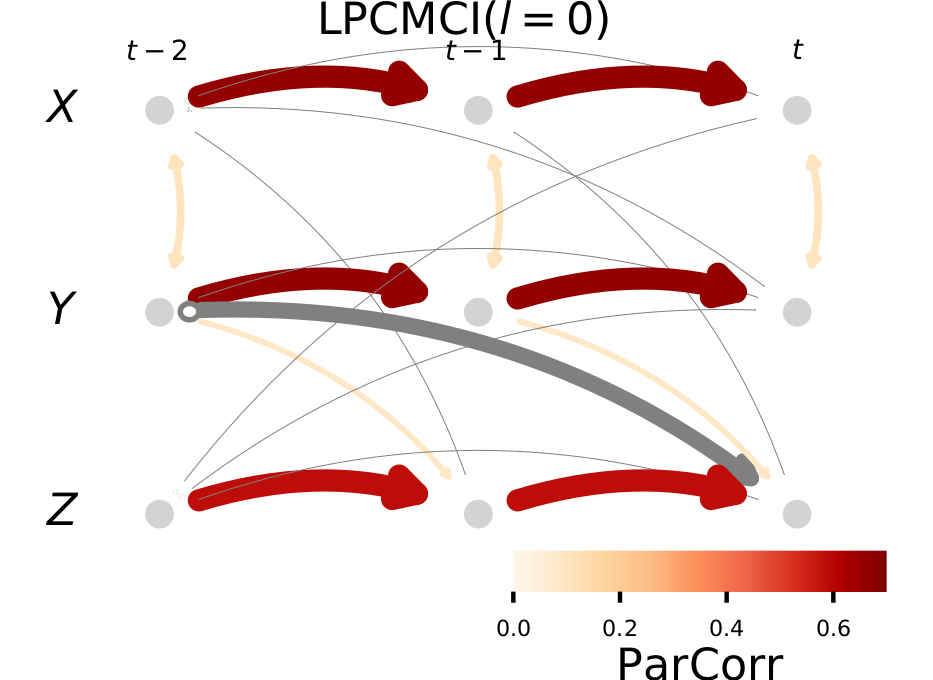}%
\includegraphics[width=.33\linewidth]{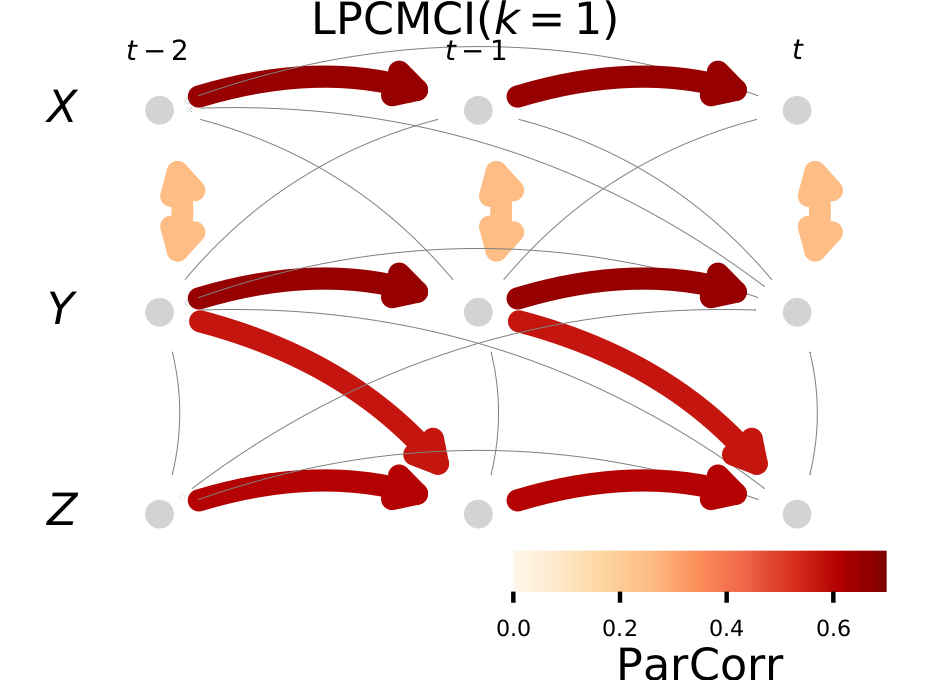}%
\caption{Latent confounder example of the model in eq.~\eqref{eqmain:numericalmodel} (Sec.~\ref{sec:numerics}) with linear ground truth links shown for the LPCMCI case (right panel). All auto-coefficients are $0.9$, all cross-coefficients are $0.6$ (colored links), false links or links with false orientations are grey. True and false adjacency detection rates shown as link width. Detection rates based on $500$ realizations run at $\alpha=0.01$ for $T=500$.}
\label{fig:example}
\end{figure*}

We illustrate the challenge posed by unobserved variables with the example of Fig.~\ref{fig:example}. SVAR-FCI with the partial correlation (ParCorr) CI test correctly identifies the auto-links but misses the true lagged link $Y_{t-1}\tailhead Z_t$ and returns a false link $Y_{t-2}\tailhead Z_t$ instead. In most realizations the algorithm fails to detect the contemporaneous adjacency $X_{t}\headhead Y_t$ and, if detected, fails to orient it as bidirected. The reason are wrong CI tests in its edge removal and orientation phases. When it iterates through conditioning sets of cardinality $\ppc = 0$ in the \textbf{edge removal phase}, the correlation $\rho(X_t; Y_t)$ is non-significant in many realizations since the high autocorrelation of both $X$ and $Y$ increases their variance and \emph{decreases} their signal-to-noise ratio (the common signal due to the latent confounder). Further, for $\ppc=1$ also the lagged correlation $\rho(Y_{t-1}; Z_t| Y_{t-2})$ often is non-significant and the true link $Y_{t-1} \tailhead Z_t$ gets removed. Here conditioning away the autocorrelation of $Y_{t-1}$ decreases the signal while the noise level in $Z_t$ is still high due to $Z$'s autocorrelation. This false negative has implications for further CI tests since $Y_{t-1}$ won't be used in subsequent conditioning sets: The path $Y_{t-2}\tailhead Y_{t-1} \tailhead Z_{t}$ can then not be blocked anymore and the false positive $Y_{t-2}\tailhead Z_{t}$ remains even after the next removal phase. In the \textbf{orientation phase} of SVAR-FCI rule $\R1$ yields tails for all auto-links. Even if the link $X_t \oo Y_t$ is detected, it is in most cases not oriented correctly. The reason again lies in wrong CI tests: In principle the collider rule $\R0$ should identify $X_t \headhead Y_t$ since the middle node of the triple $X_{t-1} \ohead X_t \oo Y_t$ does \emph{not} lie in the separating set of $X_{t-1}$ and $Y_t$ (and similarly for $X$ and $Y$ swapped). In practice $\R0$ is implemented with the majority rule \cite{Colombo2014} to avoid order-dependence, which involves further CI test given subsets of the adjacencies of $X_{t-1}$ and $Y_t$. SVAR-FCI here finds independence given $Y_{t-1}$ (correct) but also given $X_t$ (wrong, due to autocorrelation). Since the middle node $X_t$ is in exactly half of the separating sets, the triple is marked as ambiguous and left unoriented. The same applies when $X$ and $Y$ are swapped.

Autocorrelation is only one manifestation of a more general problem we observe here: Low signal-to-noise ratio due to an `unfortunate' choice of conditioning sets that leads to \textit{low effect size} (here partial correlation) and, hence, low statistical power of CI tests. Wrong CI tests then lead to missing links, and these in turn to false positives and wrong orientations. In the following we analyze effect size more theoretically and suggest a general idea to overcome this issue.

\section{Latent PCMCI} \label{sec:lpcmci}

\subsection{Effect size in causal discovery}\label{sec:MCI}
The detection power of a true link $X^i_{t - \tau} \asthead X^j_t$, where below we write $A = X^i_{t-\tau}$ and $B = X^j_t$ to emphasize that the discussion also applies to the non-time series case, quantifies the probability of the link not being erroneously removed due to a wrong CI test. It depends on $i)$ the sample size (usually fixed), $ii)$ the CI tests' significance level $\alpha$ (fixed by the researcher as the desired false positives level), $iii)$ the CI tests' estimation dimensions (kept at a minimum by SVAR-FCI's design to preferentially test small conditioning sets), and $iv)$ the effect size. We here define effect size as the minimum of the CI test statistic values $I(A;B|\SSet)$ taken over all conditioning sets $\SSet$ that are being tested (for fixed $A$ and $B$). As observed in the motivating example, this minimum can become very small and hence lead to low detection power. The central idea of \textbf{our proposed method Latent PCMCI (LPCMCI)} is to increase effect size by $a)$ restricting the conditioning sets $\SSet$ that need to be tested in order to remove all wrong links, and by $b)$ extending those sets $\SSet$ that do need to be tested with so called \emph{default conditions} $\SSet_{def}$ that increase the CI test statistic values and at the same time do not induce spurious dependencies. Regarding $a)$, Lemma~\ref{lemma:notconditiononfuture} proves that it is sufficient to only consider conditioning sets that consist of ancestors of $A$ or $B$ only. Regarding $b)$, and well-fitting with $a)$, Lemma~\ref{lemma:conditiononparents} proves that no spurious dependencies are introduced if $\SSet_{def}$ consist of ancestors of $A$ or $B$ only. Further, the following theorem shows that taking $\SSet_{def}$ as the union of the parents of $A$ and $B$ (without $A$ and $B$ themselves) improves the effect size of LPCMCI over that of SVAR-FCI. This generalizes the \emph{momentary conditional independence} (MCI) idea that underlies the PCMCI and PCMCI$^+$ algorithms \cite{Runge2018d,Runge2020a} to causal discovery with latent confounders. We state the theorem in an information theoretic framework, where $I$ denotes (conditional) mutual information and $\mathcal{I}(A;B;C|D)\equiv I(A;B|D)-I(A;B|C \cup D)$ the \emph{interaction information}.
\begin{mythmm}[LPCMCI effect size] \label{thm:effect_size}
Let $A \asthead B$ (with $A = X^i_{t-\tau}$ and $B = X^j_t$) be a link ($\tailhead$ or $\headhead$) in $\M$. Consider the default conditions $\SSet_{def} =\pa(\{A, B\}, \M) \setminus \{A, B\}$ and denote $\Xobs^*=\Xobs \setminus \SSet_{def}$. Let $\mathbf{S}=\arg\min_{\SSet\subseteq \Xobs^* \setminus \{A, B\}} I(A; B|\SSet \cup \SSet_{def})$ be the set of sets that define LPCMCI's effect size.
If $i)$ there is $\SSet^* \in \mathbf{S}$ with $\SSet^* \subseteq \adj(A, \M)\setminus \SSet_{def}$ or $\SSet^* \subseteq \adj(B, \M)\setminus \SSet_{def}$ and $ii)$ there is a proper subset $\mathcal{Q}\subset \SSet_{def}$ such that $\mathcal{I}(A; B; \SSet_{def}\setminus \mathcal{Q}|\SSet^* \cup \mathcal{Q}) < 0$, then
\begin{align} \label{eqmain:effect_size}
\min_{\SSet\subseteq \Xobs^* \setminus \{A, B\}} I(A; B|\SSet \cup \SSet_{def}) &> \min_{\tilde{\SSet}\subseteq \Xobs\setminus \{A, B\}} I(A; B|\tilde{\SSet})\,.
\end{align}
If the assumptions are not fulfilled, then (trivially) "$\geq$" holds in eq.~\eqref{eqmain:effect_size}.
\end{mythmm}
The second assumption only requires that \emph{any} subset $\SSet_{def} \setminus{Q}$ of the parents contains information that increases the information between $A$ and $B$. A sufficient condition for this is detailed in Corollary~\ref{thm:effect_size_corollary}.

These considerations lead to two design principles behind LPCMCI: First, when testing for conditional independence of $A$ and $B$, discard conditioning sets that contain known non-ancestors of $A$ and $B$. Second, use known parents of $A$ and $B$ as default conditions. Unless the higher effect size is overly counteracted by the increased estimation dimension (due to conditioning sets of higher cardinality), this leads to higher detection power and hence higher recall of true links. While we do not claim that our choice of default conditions as further detailed in Sec.~\ref{sec:pseudocode} is optimal, our numerical experiments in Sec.~\ref{sec:numerics} and the SM indicate strong increases in recall for the case of continuous variables with autocorrelation. In \cite{Runge2018d,Runge2020a} it is discussed that, in addition to higher effect size, conditioning on the parents of both $A$ and $B$ also leads to better calibrated tests which in turn avoids inflated false positives. Another benefit is that fewer conditioning sets need to be tested, which is also the motivation for a default conditioning on known parents in \cite{Sanghack2020}. 

The above design principles are only useful if some (non-)ancestorships are known before all CI test have been completed. LPCMCI achieves this by entangling the edge removal and edge orientation phases, i.e., by learning ancestral relations before having removed all wrong links. For this purpose we below develop novel orientation rules. These are not necessary in the causally sufficienct setting considered by PCMCI$^+$ \cite{Runge2020a} because there the default conditions need not be limited to ancestors of $A$ or $B$ (although PCMCI$^+$ tries to keep the number of default conditions low). While not considered here, background knowledge about (non-)ancestorships can easily be incorporated.

\subsection{Introducing middle marks and LPCMCI-PAGs}\label{sec:middlemarksmain}
To facilitate early orientation of edges we give an unambiguous causal interpretation to the graph at every step of the algorithm. This is achieved by augmenting edges with \textit{middle marks}. Using generic variable names $A$, $B$, and $C$ indicates that the discussion also applies to the non-time series case.

Middle marks are denoted above the link symbol and can be `?', `L', `R', `!', or `' (empty). The `L' (`R') on $A \LMM{\astast} B$ ($A \RMM{\astast} B$) asserts that if $A < B$ ($B < A$) then $B \notin \an(A, \G)$ or there is no $\SSet \subseteq \pa(A, \M)$ that m-separates $A$ and $B$ in $\M$. Here \totalorder is any total order on the set of variables. Its choice is arbitrary and does not influence the causal information content, the sole purpose being to disambiguate $A \LMM{\astast} B$ from $A \RMM{\astast} B$. Moreover, `$\ast$' is a wildcard that may stand for all three edge marks (tail, head, circle) that appear in PAGs. Further, the `!' on $A \EMM{\astast} B$ asserts that both $A \LMM{\astast} B$ and $A \RMM{\astast} B$ are true, and the empty middle mark on $A \astast B$ says that $A \in \adj(B, \M)$. Lastly, the `?' on $A \QMM{\astast} B$ doesn't promise anything. Non-circle edge marks (here potentially hidden by the `$\ast$' symbol) still convey their standard meaning of ancestorship and non-ancestorship, and the absence of an edge between $A$ and $B$ still asserts that $A \notin \adj(B, \M)$. We call a PAG $\C$ whose edges are extended with middle marks a LPCMCI-PAG for $\M$, see Sec.~\ref{sec:lpcmcipags} in the SM for a more formal definition. The `$\ast$' symbol is also used as a wildcard for the five middle marks.

Note that we are \textit{not} changing the quantity we are trying to estimate, this is still the PAG $\PG$ as explained in Sec.~\ref{sec:onstationarity}. The notion of LPCMCI-PAGs is used in intermediate steps of LPCMCI and has two advantages. First, $A \astast B$ is reserved for $A \in \adj(B, \M)$ and thus has an unambiguous meaning at every point of the algorithm, unlike for (SVAR-)FCI and (SVAR-)RFCI. In fact, even if LPCMCI is interrupted at any arbitrary point it still yields a graph with unambiguous and sound causal interpretation. Second, middle marks carry fine-grained causal information that allows to determine definite adjacencies early on:
\begin{mylemma}[Ancestor-parent-rule] \label{lemma:middlemarks}
In LPCMCI-PAG $\C$ one may replace $\textbf{1.)}$ $A \EMMnoast{\tailhead} B$ by $A \tailhead B$, $\textbf{2.)}$ $A \LMMnoast{\tailhead} B$ for $A > B$ by $A \tailhead B$, and $\textbf{3.)}$  $A \RMMnoast{\tailhead} B$ for $A < B$ by $A \tailhead B$.
\end{mylemma}
When LPCMCI has converged all middle marks are empty and hence $\C$ is a PAG. We choose a total order consistent with time order, namely $X^i_{t-\tau} < X^j_t$ iff $\tau > 0$ or $\tau = 0$ and $i < j$. Lagged links can then be initialized with edges $\LMM{\ohead}$ (contemporaneous links as $\QMM{\oo}$).

\subsection{Orientations rules for LPCMCI-PAGs}\label{sec:algoorient}
We now discuss rules for edge orientation in LPCMCI-PAGs. For this we need a definition:
\begin{mydef}[Weakly minimal separating sets] \label{def:weaklyminimal}
In MAG $\M$ let $A$ and $B$ be m-separated by $\SSet$. The set $\SSet$ is a weakly minimal separating set of $A$ and $B$ if $i)$ it decomposes as $\SSet = \SSet_1 \dot{\cup}\, \SSet_2$ with $\SSet_1 \subseteq \an(\{A, B\}, \M)$ such that $ii)$ if $\SSet^\prime = \SSet_1 \dot{\cup}\, \SSet_2^\prime$ with $\SSet_2^\prime \subseteq \SSet_2$ m-separates $A$ and $B$ then $\SSet_2^\prime = \SSet_2$. The pair $(\SSet_1, \SSet_2)$ is called a weakly minimal decomposition of $\SSet$.
\end{mydef}
This generalizes the notion of minimal separating sets, for which additionally $\SSet_1 = \emptyset$. Since LPCMCI is designed to extend conditioning sets by known ancestors, the separating sets it finds are in general not minimal. However, they are still weakly minimal. The following Lemma, a generalization of the unshielded triple rule \cite{Colombo2012}, is central to orientations in LPCMCI-PAGs:
\begin{mylemma}[Strong unshielded triple rule] \label{lemma:strongutr}
Let $A \SMM{\astast} B \SMM{\astast} C$ be an unshielded triple in LPCMCI-PAG $\C$ and $\SSet_{AC}$ the separating set of $A$ and $C$.
$\textbf{1.)}$
If $i)$ $B \in \SSet_{AC}$ and $ii)$ $\SSet_{AC}$ is weakly minimal, then $B \in \an(\{A, C\}, \G)$.
$\textbf{2.)}$
Let $\mathcal{T}_{AB} \subseteq \an(\{A, B\}, \M)$ and $\mathcal{T}_{CB} \subseteq \an(\{C, B\}, \M)$ be arbitrary. If $i)$ $B \notin \SSet_{AC}$, $ii)$ $A$ and $B$ are not m-separated by $\SSet_{AC} \cup \mathcal{T}_{AB} \setminus \{A, B\}$, $iii)$ $C$ and $B$ are not m-separated by $\SSet_{AC} \cup \mathcal{T}_{CB} \setminus \{C, B\}$, then $B \notin \an(\{A, C\}, \G)$. The conditioning sets in $ii)$ and $iii)$ may be intersected with the past and present of the later variable.
\end{mylemma}
Part 2.) of this Lemma generalizes the FCI collider rule $\R0$ to rule $\ER0$ (of which there are several variations when restricting to particular middle marks), and part 1.) generalizes $\R1$ to $\ER{1}$. Rules $\R2$ and $\R8$ generalize trivially to triangles in $\C$ with arbitrary middle marks, giving rise to $\ER2$ and $\ER8$. Rules $\R3$, $\R9$ and $\R10$ are generalized to $\ER3$, $\ER9$ and $\ER{10}$ by adding the requirement that the middle variables of certain unshielded colliders are in the separating set of the two outer variables, and that these separating sets are weakly minimal. Since there are no selection variables, rules $\R5$, $\R6$ and $\R7$ are not applicable. Rule $\ER4$ generalizes the discriminating path rule \cite{Colombo2012} of RFCI. These rules are complemented by the replacements specified in Lemma~\ref{lemma:middlemarks} and a rule for updating middle marks. Precise formulations of all rules are given in Sec.~\ref{sec:orientationrules} of the SM.

We stress that these rules are applicable at every point of the algorithm and that they may be executed in any order. This is different from the (SVAR-)FCI orientation phase which requires that prior to orientation a PAG has been found. Also (SVAR-)RFCI orients links only once an RFCI-PAG has been determined, and both (SVAR-)FCI and (SVAR-)RFCI require that all colliders are oriented before applying their other orientation rules.

The relevance of the novel orientation rules bears on them allowing to determine ancestorships and non-ancestorships already after only few CI tests have been performed. This is utilized in LPCMCI by entangling the edge removal and edge orientation phases, which then allows to implement the idea of the PCMCI and PCMCI$^+$ algorithms \cite{Runge2018d,Runge2020a} to increase the effect sizes of CI tests (and hence the recall of the algorithm, see Theorem~\ref{thm:effect_size} and the subsequent discussion) by conditioning on known parents also in the causally insufficient case considered here (where latent confounders are allowed). The aspect of determining ancestorships with only few CI tests is similar in spirit to an approach taken in the recent work \cite{mastakouri2020necessary_fix}, which considers the narrower but important task of causal feature selection in time series with latent confounders: The SyPI algorithm introduced there does not aim at finding the full PAG $\PG$ but rather at finding ancestors of given target variables. Under several assumptions on the connectivity pattern of the time series graph $\G$ the work \cite{mastakouri2020necessary_fix} presents conditions that are sufficient for a given variable (the potential cause) to be an ancestor of another given variable (the potential effect). For certain types of ancestors, namely for parents that belong to a time series which in the summary graph is not confounded with the time series of the target variable, these conditions are even necessary. These findings allow SyPI to determine (some) ancestorships with only two CI tests per pair of potential cause and potential effect. It would be interesting to investigate whether the problem of causal feature selection as framed in \cite{mastakouri2020necessary_fix} can benefit from some of the ideas presented here, for example from the novel orientation rules (which do not require restrictions on the connectivity pattern of $\G$) or the idea to increase the effect sizes of CI tests by conditioning on known parents. Similarly, it would be interesting to explore whether the ideas behind SyPI can be utilized to further improve the statistical performance of algorithms that approach the more general task of finding the full PAG.

\subsection{The LPCMCI algorithm}\label{sec:pseudocode}
\textbf{LPCMCI} is a constraint-based causal discovery algorithm that utilizes the findings of Sec.~\ref{sec:MCI} to increase the effect size of CI tests. High-level pseudocode is given in \textbf{Algorithm \ref{algo:lpcmci}}. After initializing $\C$ as a complete graph, the algorithm enters its \textit{preliminary phase} in lines 2 to 4. This involves calls to \textbf{Algorithm \ref{algo:ancestral}} (pseudocode in Sec.~\ref{sec:pseudocode2} of the SM), which removes many (but in general not all) false links and, while doing so, repeatedly applies the orientation rules introduced in the previous section. These rules identify a subset of the (non-)ancestorships in $\G$ and accordingly mark them by heads or tails on edges in $\C$. This information is then used as prescribed by the two design principles of LPCMCI that were explained in Sec.~\ref{sec:MCI}: The non-ancestorships further constrain the conditioning sets $\SSet$ of subsequent CI tests, the ancestorships are used to extend these sets to $\SSet \cup \SSet_{def}$ where $\SSet_{def} = \pa(\{X^i_{t-\tau}, X^j_t\}, \C)$ are the by then known parents of those variables whose independence is being tested. All parentships marked in $\C$ after line 3 are remembered and carried over to an elsewise re-initialized $\C$ before the next application of Alg.~\ref{algo:ancestral}. Conditioning sets can then be extended with known parents already from the beginning. The purpose of this iterative process is to determine an accurate subset of the parentships in $\G$. These are then passed on to the \textit{final phase} in lines 5 - 6, which starts with one final application of Alg.~\ref{algo:ancestral}. At this point there may still be false links because Alg.~\ref{algo:ancestral} may fail to remove a false link between variables $X^i_{t-\tau}$ and $X^j_t$ if neither of the two is an ancestor of the other. This is the purpose of \textbf{Algorithm \ref{algo:nonancestral}} (pseudocode in Sec.~\ref{sec:pseudocode2} of the SM) that is called in line 6, which thus plays a similar role as the second removal phase in (SVAR-)FCI. Algorithm \ref{algo:nonancestral} repeatedly applies orientation rules and uses identified (non-)ancestorships in the same way as Alg.~\ref{algo:ancestral}. As stated in the following theorems, LPCMCI will then have found the PAG $\PG$. Moreover, its output does not depend on the order of the $N$ time series variables $X^j$. The number $k$ of iterations in the preliminary phase is a hyperparameter and we write LPCMCI($k=k_0$) when specifying $k = k_0$. Stationarity is enforced at every step of the algorithm, i.e., whenever an edge is removed or oriented all equivalent time shifted edges (called `homologous' in \cite{Entner2010}) are removed too and oriented in the same way.
\begin{algorithm*}[h!]
\caption{LPCMCI}
\begin{algorithmic}[1]
\Require
Time series dataset $\Xobs = \{\mathbf{X}^1, \ldots ,\mathbf{X}^N\}$,
maximal considered time lag $\taumax$,
significance level $\alpha$,
CI test ${\rm CI}(X,\,Y,\,\SSet)$,
non-negative integer $\nprelim$
\State
Initialize $\C$ as complete graph with $X^i_{t-\tau} \LMM{\oright} X^j_t$ ($0 < \tau \leq \taumax$) and $X^i_{t-\tau} \QMM{\oo} X^j_t$ ($\tau = 0$)
\For{$0 \leq l \leq \nprelim - 1 $}
  \State
  Remove edges and apply orientations using Algorithm \ref{algo:ancestral}
  \State
  Repeat line 1, orient edges as $X^i_{t-\tau} \QMMnoast{\tailhead} X^j_t$ if $X^i_{t-\tau} \SMMnoast{\tailhead} X^j_t$ was in $\C$ after line 3
\EndFor
\State
Remove edges and apply orientations using Algorithm \ref{algo:ancestral}
\State
Remove edges and apply orientations using Algorithm \ref{algo:nonancestral}
\State
\Return PAG $\C = \PG = \PGstatAO^{\taumax}$
\end{algorithmic} \label{algo:lpcmci}
\end{algorithm*}

\begin{mythmm}[LPCMCI is sound and complete] \label{thm:sound}
Assume that there is a process as in eq.~\eqref{eqmain:SVARProcess} without causal cycles, which generates a distribution $P$ that is faithful to its time series graph $\G$. Further assume that there are no selection variables, and that we are given perfect statistical decisions about CI of observed variables in $P$. Then LPCMCI is sound and complete, i.e., it returns the PAG $\PG$.
\end{mythmm}

\begin{mythmm}[LPCMCI is order-independent]\label{thm:orderindependent}
The output of LPCMCI does not depend on the order of the $N$ time series variables $X^j$ (the $j$-indices may be permuted).
\end{mythmm}

\subsection{Back to the motivational example in Fig.~\ref{fig:example}}
The \textbf{first iteration} ($l=0$) of LPCMCI also misses the links $Y_{t-1} \tailhead Z_{t}$ and finds $X_{t} \astast Y_{t}$ in only few realizations (we here suppress middle marks for simpler notation), but orientations are already improved as compared to SVAR-FCI. Rule $\ER1$ applied after $\ppc = 1$ orients the auto-links $X_{t-1} \tailhead X_{t}$ and $Y_{t-1} \tailhead Y_{t}$. This leads to the parents sets $\pa(X_t, \C) = \{X_{t-1}\}$ and $\pa(Y_t, \C) = \{Y_{t-1}\}$, which are then used as default conditions in subsequent CI tests. This is relevant for orientation rule $\ER0$ that tests whether the middle node of the unshielded triple $X_{t-1} \ohead X_t \oo Y_t$ does \emph{not} lie in the separating set of $X_{t-1}$ and $Y_t$. Due to the extra conditions the relevant partial correlation $\rho(X_{t-1}; Y_t|X_t, X_{t-2}, Y_{t-1})$ now correctly turns out significant. This identifies $X_t$ as collider and (since the same applies with $X$ and $Y$ swapped) the bidirected edge $X_{t} \headhead Y_{t}$ is correctly found. The \textbf{next iteration} ($l = 1$) then uses the parents obtained in the $l=0$ iteration, here the autodependencies plus the (false) link $Y_{t-2}\tailhead Z_t$, as default conditions already from the beginning for $\ppc = 0$. While the correlation $\rho(X_t; Y_t)$ used by SVAR-FCI is often non-significant, the partial correlation $\rho(X_t; Y_t|X_{t-1},Y_{t-1})$ is significant since the autocorrelation noise was removed and effect size increased (indicated as link color in Fig.~\ref{fig:example}) in accord with Theorem \ref{thm:effect_size}. Also the lagged link is correctly detected because $\rho(Y_{t-1}; Z_t|Y_{t-2},Z_{t-1})$ is larger than $\rho(Y_{t-1}; Z_t|Y_{t-2})$. The false link $Y_{t-2} \tailhead Z_t$ is now removed since the separating node $Y_{t-1}$ was retained. This wrong parentship is then also not used for default conditioning anymore. Orientations of bidirected links are facilitated as before and $Y_{t-1} \tailhead Z_t$ is oriented by rule $\ER1$.

\section{Numerical experiments} \label{sec:numerics}
\begin{figure*}[t]  
\centering
\includegraphics[width=.49\linewidth]{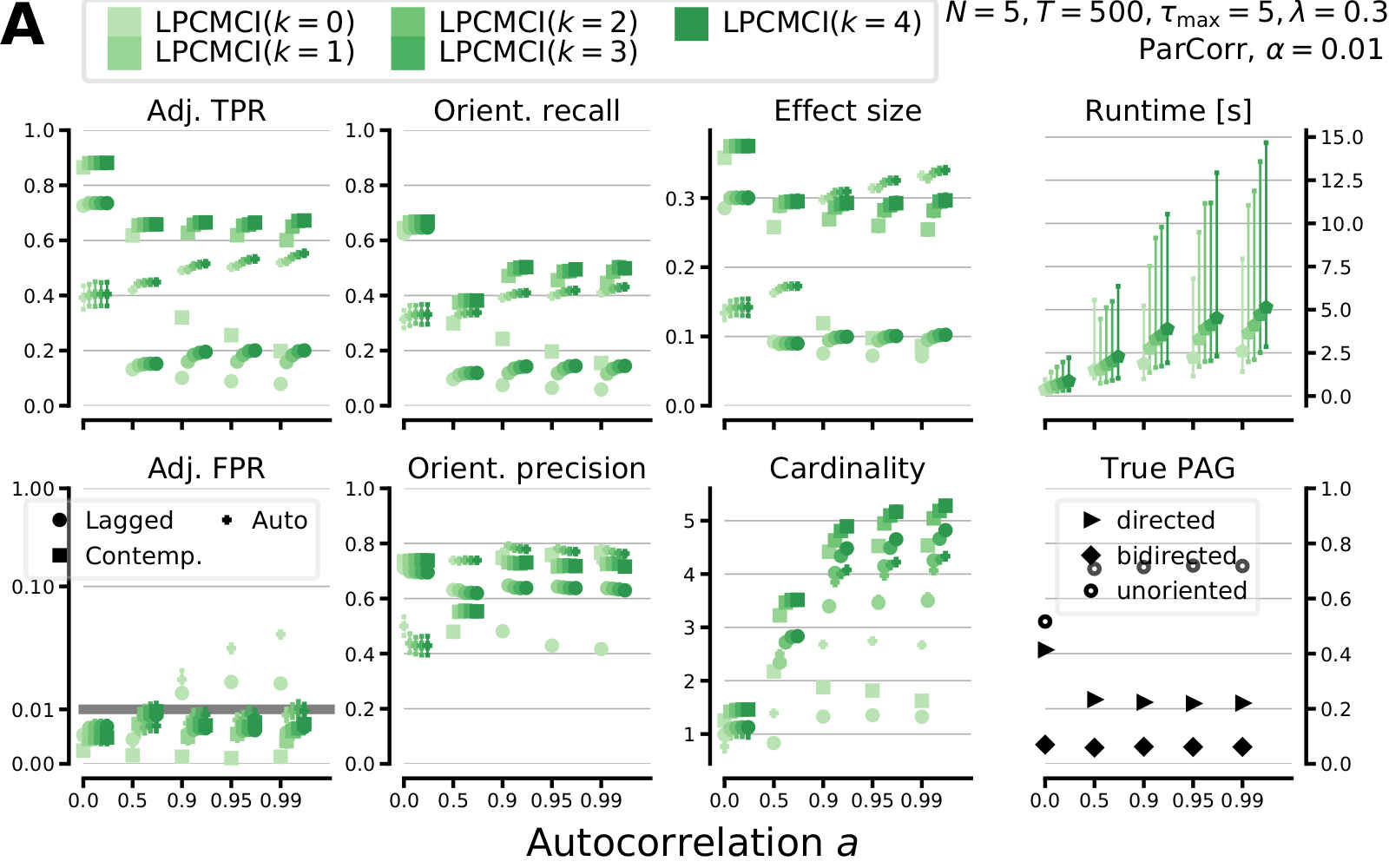}\hspace*{3pt}
\includegraphics[width=.49\linewidth]{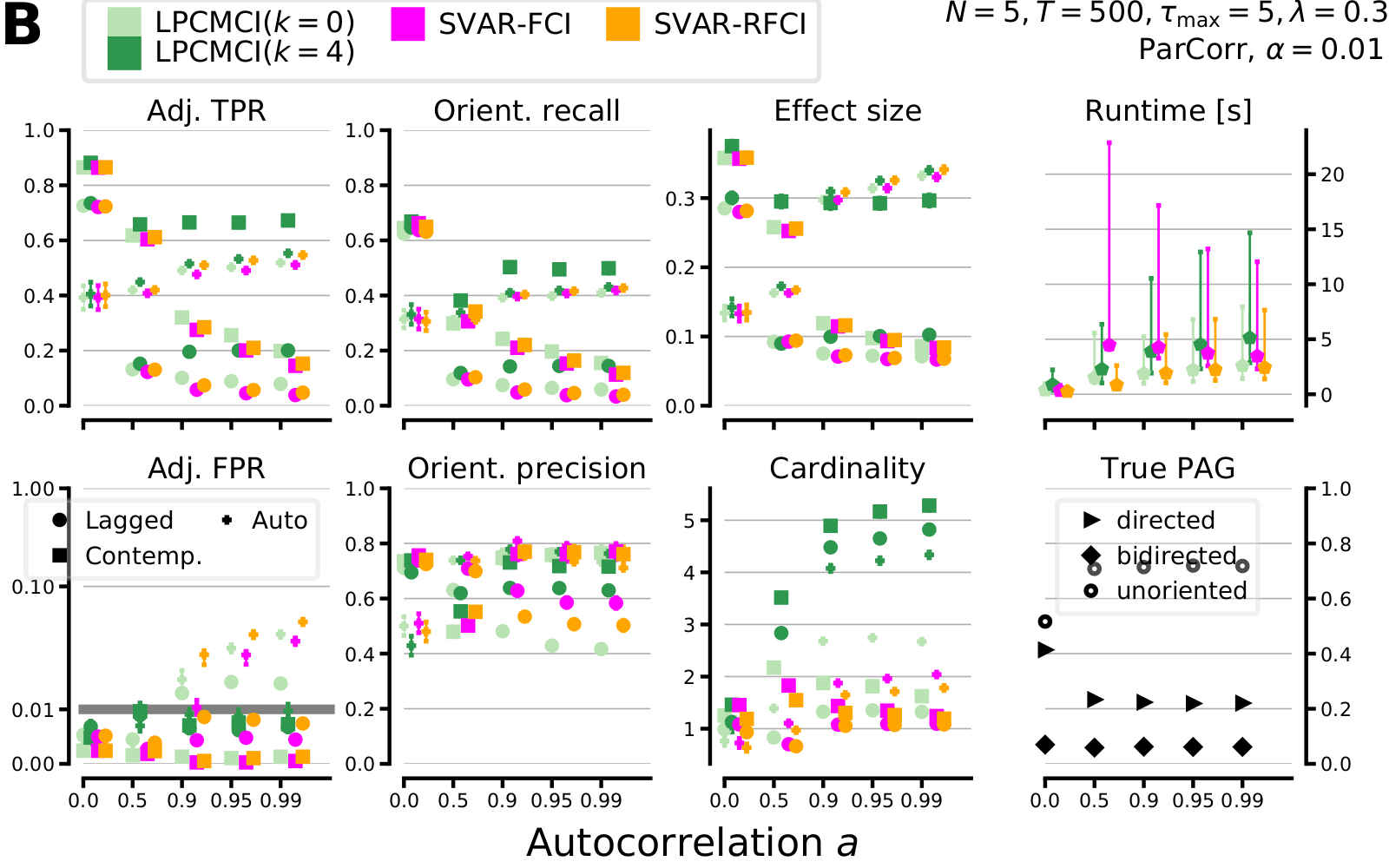}%

\includegraphics[width=.49\linewidth]{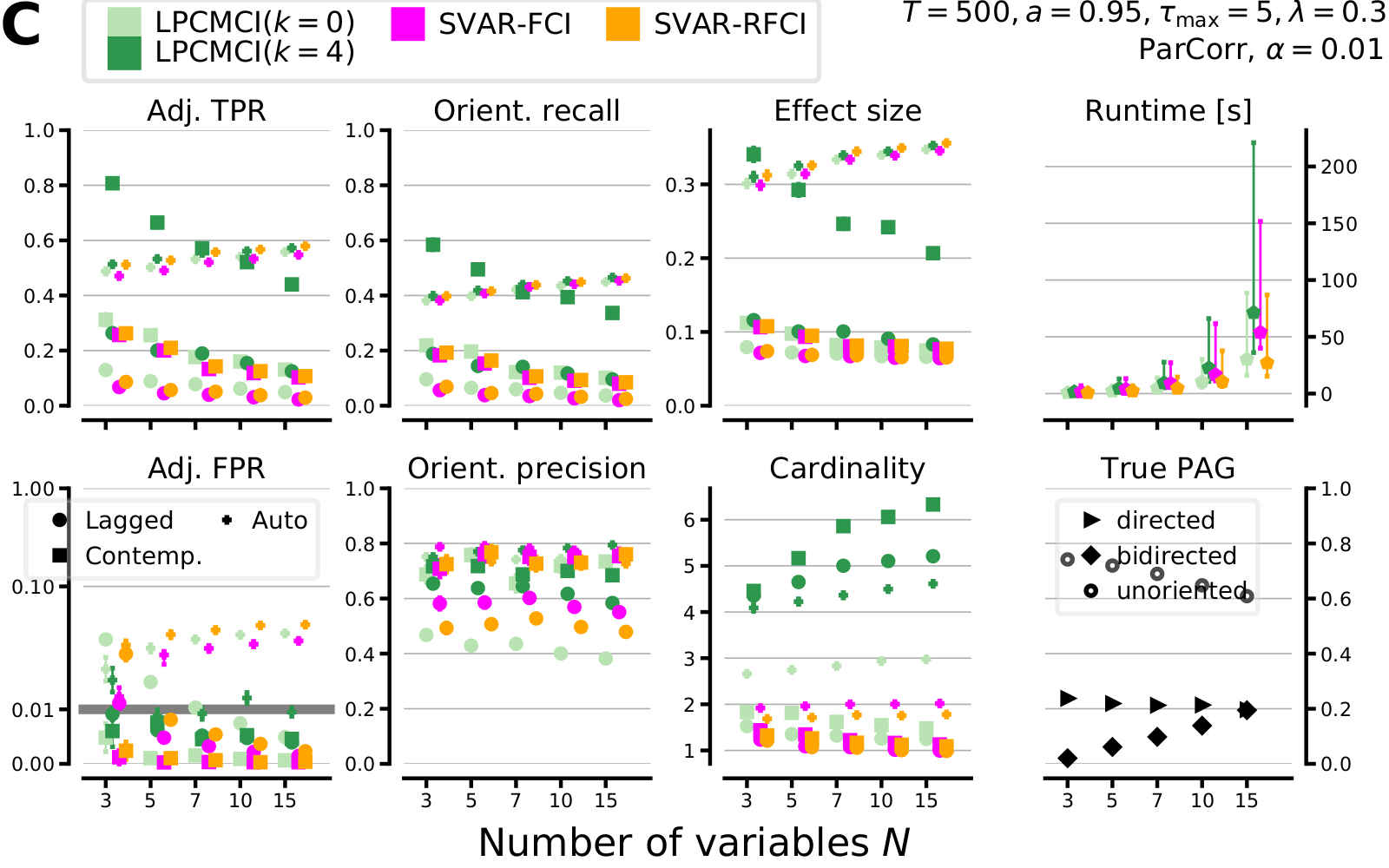}\hspace*{3pt}
\includegraphics[width=.49\linewidth]{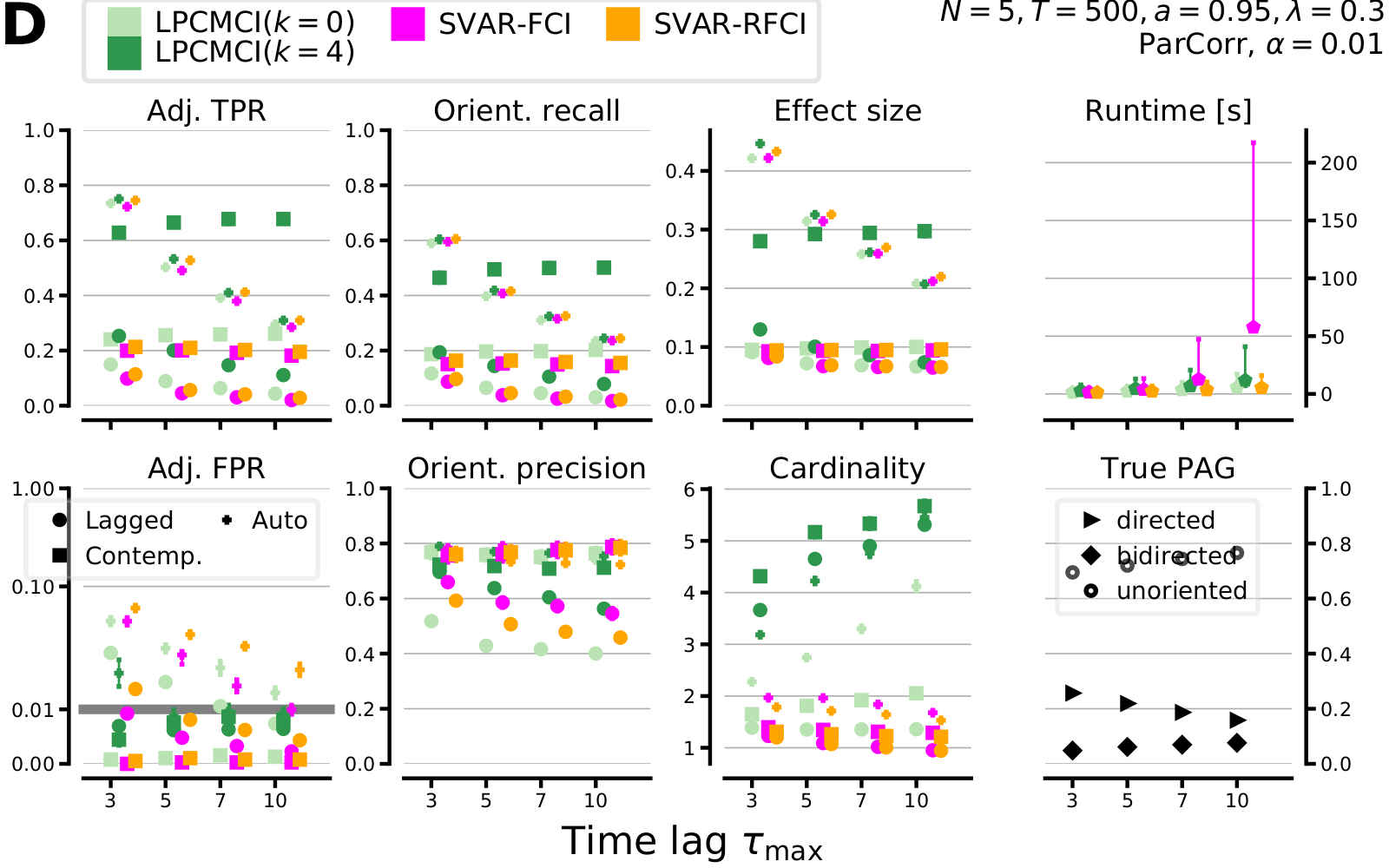}%
\caption{
Results of numerical experiments for (\textbf{A}) LPCMCI($k$) for different $k$, LPCMCI compared to SVAR-FCI and SVAR-RFCI for (\textbf{B}) varying autocorrelation, for (\textbf{C}) number of variables $N$, and for (\textbf{D}) maximum time lag $\tau_{\max}$ (other parameters indicated in upper right of each panel).
}
\label{fig:experiments}
\end{figure*}

We here compare LPCMCI to the SVAR-FCI and SVAR-RFCI baselines with CI tests based on linear partial correlation (ParCorr), for an overview of further experiments presented in the SM see the end of this section. To limit runtime we constrain the cardinality of conditioning sets to $3$ in the second removal phase of SVAR-FCI and in Alg.~\ref{algo:nonancestral} of LPCMCI (excluding the default conditions $\SSet_{def}$, i.e., $|\SSet| \leq 3$ but $|\SSet\cup\SSet_{def}| > 3$ is allowed). We generate datasets with this variant of the SCM in eq.~\eqref{eqmain:SVARProcess}:
\begin{align} \label{eqmain:numericalmodel}
V_t^j &= a_j V^j_{t-1} + \textstyle{\sum_i} c_i f_{i}(V^i_{t-\tau_i}) + \eta^j_t\quad \text{for}\quad j\in\{1,\ldots,\tilde{N}\}
\end{align}
Autocorrelations $a_j$ are drawn uniformly from $[\max(0, a-0.3), a]$ for some $a$ as indicated in Fig.~\ref{fig:experiments}. For each model we in addition randomly choose $L=\tilde{N}$ linear (i.e., $f_i = id$) cross-links with the corresponding non-zero coefficients $c_i$ drawn uniformly from $\pm[0.2, 0.8]$. 30\% of these links are contemporaneous (i.e., $\tau_i=0$), the remaining $\tau_i$ are drawn from $[1,\,p_{ts}=3]$. The noises $\eta^j\sim \mathcal{N}(0,\sigma_j^2)$ are \emph{iid} with $\sigma_j$ drawn from $[0.5, 2]$. We only consider stationary models. From the $\tilde{N}$ variables of each model we randomly choose $N=\lceil (1 - \lambda)\tilde{N}\rceil$ for $\lambda=0.3$ as observed. As discussed in Sec.~\ref{sec:onstationarity}, the true PAG $\PG$ of each model depends on $\tau_{\max}$. In Fig.~\ref{fig:experiments} we show the relative average numbers of directed, bidirected, and (partially) unoriented links. For performance evaluation true positive ($=$ recall) and false positive rates for adjacencies are distinguished between lagged cross-links ($i\neq j$), contemporaneous, and autodependency links. False positives instead of precision are shown to investigate whether methods can control these below the $\alpha$-level. Orientation performance is evaluated based on edgemark recall and precision. In Fig.~\ref{fig:experiments} we also show the average of minimum absolute ParCorr values as an estimate of effect size and the average maximum cardinality of all tested conditioning sets. All metrics are computed across all estimated graphs from $500$ realizations of the model in eq.~\eqref{eqmain:numericalmodel} at time series length $T$. The average and 90\% range of runtime estimates were evaluated on Intel Xeon Platinum 8260.

In \textbf{Fig.~\ref{fig:experiments}A} we show LPCMCI for $k=0, \ldots, 4$ against increasing autocorrelation $a$. Note that $a=0$ implies a different true PAG than $a>0$. The largest gain, both in recall and precision, comes already from $k=0$ to $k=1$. For higher $k$ LPCMCI maintains false positive control and orientation precision, and improves recall before converging at $k = 4$. The gain in recall is largely attributable to improved effect size. On the downside, larger $k$ increase cardinality (estimation dimension) and runtime. However, the runtime increase is only marginal because later $l$-steps converge faster and the implementation caches CI test results. 
\textbf{Fig.~\ref{fig:experiments}B} shows a comparison of LPCMCI with SVAR-FCI and SVAR-RFCI against autocorrelation, depicting LPCMCI for $k=0$ and $k = 4$. Already LPCMCI($k=0$) has higher adjacency and orientation recall than SVAR-FCI and SVAR-RFCI for increasing autocorrelation while they are on par for $a=0$. This comes at the price of precision, especially lagged orientation precision. LPCMCI($k=4$) has more than 0.4 higher contemporaneous orientation recall and still 0.1 higher lagged orientation recall than SVAR-FCI and SVAR-RFCI. Lagged precision is higher for high autocorrelation and contemporaneous precision is slightly lower. LPCMCI($k=4$) maintains high recall for increasing autocorrelation $a\geq0.5$ while SVAR-FCI and SVAR-RFCI's recall sharply drops. These results can be explained by improved effect size while the increased cardinality ($\approx 5$) of separating sets is still moderate compared to the sample size $T=500$. LPCMCI($k=0$) has similar low runtime as SVAR-RFCI, for LPCMCI($k=4$) it is comparable to that of SVAR-FCI.
In \textbf{Fig.~\ref{fig:experiments}C} we show results for different numbers of variables $N$. As expected, all methods have decreasing adjacency and orientation recall for higher $N$, but LPCMCI starts at a much higher level. For $N=3$ both SVAR-FCI and SVAR-RFCI cannot control false positives for lagged links while for larger $N$ false positives become controlled. The reason is the interplay of ill-calibrated CI tests for smaller $N$ due to autocorrelation (inflating false positives) with sequential testing for larger $N$ (reducing false positives), as has been discussed in \cite{Runge2018d,Runge2020a} for the similar PC algorithm \cite{Spirtes1991}. LPCMCI better controls false positives here, its decreasing recall can be explained by decreasing effect size and increasing cardinality. Runtime becomes slightly larger than that of SVAR-FCI for larger $N$.
\textbf{Fig.~\ref{fig:experiments}D} shows results for different maximum time lags $\tau_{\max}$. Note that these imply different true PAGs, especially since further lagged links appear for larger $\tau_{\max}$. All methods show a decrease in lagged recall and precision, whereas contemporaneous recall and precision stay almost constant. For SVAR-FCI there is an explosion of runtime for higher $\tau_{\max}$ due to excessive searches of separating sets in its second removal phase. In LPCMCI this is partially overcome since the sets that need to be searched through are more restricted.

In Sec.~\ref{sec:experiments} in the SM we present further numerical experiments. This includes more combinations of model parameters $N,\,a,\,\lambda,\,T$, \textbf{nonlinear} models together with the nonparametric GPDC CI test \cite{Runge2018d}, and a comparison to a \textbf{residualization} approach. In these cases the results are largely comparable to those above regarding relative performances. For \textbf{non-time series} models we find that, although all findings of Secs.~\ref{sec:MCI} through \ref{sec:pseudocode} still apply, LPCMCI($k$) is on par with the baselines for $k=0$ while it shows inflated false positives for $k = 4$. Similarly, for models of \textbf{discrete variables} together with a $G$-test of conditional independence LPCMCI($k$) performs comparable to the baselines for $k = 0$ and gets worse with increasing $k$. A more detailed analysis of LPCMCI's performance in these two cases, non-time series and discrete models, is subject to future research.

\section{Application to real data}\label{sec:real_data}
We here discuss an application of LPCMCI to average daily discharges of rivers in the upper Danube basin, measurements of which are made available by the Bavarian Environmental Agency at \href{https://www.gkd.bayern.de/en/}{https://www.gkd.bayern.de}. We consider measurements from the Iller at Kempten ($X$), the Danube at Dillingen ($Y$), and the Isar at Lenggries ($Z$). While the Iller discharges into the Danube upstream of Dillingen with the water from Kempten reaching Dillingen within about a day, the Isar reaches the Danube downstream of Dillingen. We thus expect a contemporaneous link $X_t \tailhead Y_t$ and no direct causal relationships between the pairs $X, Z$ and $Y, Z$. Since all variables may be confounded by rainfall or other weather conditions, this choice allows to test the ability of detecting and distinguishing directed and bidirected links. To keep the sample size comparable with those in the simulation studies we restrict to the records of the past three years (2017-2019). We set $\taumax = 2$ and apply LPCMCI($k$) for $k = 0, \ldots, 4$ and $\alpha = 0.01$ with ParCorr CI tests. Restricting the discussion to contemporaneous links, LPCMCI correctly finds $X_t \tailhead Y_t$ for $k = 1, \ldots, 4$ and for $k = 0$ wrongly finds $X_t \headhead Y_t$. For all $k$ it infers the bidirected link $X_t \headhead Z_t$, which is plausible due to confounding by weather. For $k = 3, 4$ LPCMCI wrongly finds the directed link $Z_t \tailhead Y_t$, which should either be absent or bidirected. The results are similar for $\alpha = 0.05$, with the difference that LPCMCI then always correctly finds $X_t \tailhead Y_t$ but wrongly infers $Z_t \tailhead Y_t$ also for $k = 1, 2$. In comparison, SVAR-FCI with ParCorr CI tests finds the contemporaneous adjacencies $Y_t \oo X_t \oo Z_t$ for $\alpha = 0.01, 0.03, 0.05, 0.08, 0.1, 0.3, 0.5$ and $Y_t \heado X_t \oo Z_t$ for $\alpha = 0.8$. The estimated PAGs are shown in Sec.~\ref{sec:real_data_supplement} of the SM.

We note that since the discharge values show extreme events caused by heavy rainfall, the assumption of stationarity is expected to be violated. For other analyses of the dataset of average daily discharges see \cite{Asadi2015, Engelke2020, Mhalla2020, Gnecco2020_fix}. More detailed applications to and analyses of LPCMCI on real data are subject to future research.

\section{Discussion and future work}\label{sec:conclusion}
Major \textbf{strengths} of LPCMCI lie in its significantly improved recall as compared to the SVAR-FCI and SVAR-RFCI baselines for autocorrelated continuous variables, which grows with autocorrelation and is particularly strong for contemporaneous links. At the same time LPCMCI (for $k > 0$) has better calibrated CI test leading to better false positive control than the baselines. We cannot prove false positive control, but are not aware of any such proof for other constraint-based algorithms in the challenging latent, nonlinear, autocorrelated setting considered here. 
A general \textbf{weakness}, which also applies to (SVAR-)FCI and (SVAR-)RFCI, is the faithfulness assumption. If violated in practice this may lead to wrong conclusions. We did not attempt to only assume the weaker form of adjacency-faithfulness \cite{Ramsey2006}, which to our knowledge is however generally an open problem in the causally insufficient case. Moreover, like all constraint-based methods, our method cannot distinguish all members of Markov equivalence classes like methods based on the SCM framework such as e.g. TS-LiNGAM \cite{TSLINGAM} and TiMINo \cite{PetersJSLMZK2013} do. These, however, restrict the type of dependencies.
\textbf{Concluding}, this paper shows how causal discovery in autocorrelated time series benefits from increasing the effect size of CI tests by including causal parents in conditioning sets. The LPCMCI algorithm introduced here implements this idea by entangling the removal and orientation of edges. As demonstrated in extensive simulation studies, LPCMCI achieves much higher recall than the SVAR-FCI and SVAR-RFCI baselines for autocorrelated continuous variables.
We further presented novel orientation rules and an extension of graphical terminology by the notions of middle marks and weakly minimal separating sets. Code for all studied methods is provided as part of the \textit{tigramite} Python package at \href{https://github.com/jakobrunge/tigramite}{https://github.com/jakobrunge/tigramite}.
\textbf{In future work} one may relax assumptions of LPCMCI to allow for selection bias and non-stationarity. Background knowledge about (non-)ancestorships may be included without any conceptual modification.
Since the presented orientation rules are applicable at any point and thus able to determine (non-)ancestorships already after having performed only few CI tests, the rules may also be useful for causal feature selection in the presence of hidden confounders, a task that for time series has recently been considered in \cite{mastakouri2020necessary_fix}. Lastly, it would be interesting to combine the ideas presented here with ideas from the structural causal model framework.

\clearpage
\section*{Broader Impact}
Observational causal discovery is especially important for the analysis of systems where experimental manipulation is impossible due to ethical reasons, e.g., in climate research or neuroscience. Our work focuses on the challenging time series case that is of particular relevance in these fields. Understanding causal climate mechanisms from large observational satellite datasets helps climate researchers in understanding and modeling climate change as a main challenge of humanity. Since all code will be published open-source, our methods can be used by anyone. Causal discovery is a rather fundamental topic and we deem the potential for misuse as low.

\begin{ack}
We thank the anonymous referees for considered and helpful comments that helped to improve the paper. Thanks also goes to Christoph Käding for proof-reading.

DKRZ (Deutsches Klimarechenzentrum) provided computational resources (grant no.~1083).
\end{ack}

\clearpage
\bibliography{library_new}
\bibliographystyle{apalike}


\newcommand{\sectionsupplement}[1]{%
  \section*{#1}
}

\clearpage
\sectionsupplement{\Large Supplementary material}

In this supplementary material we present a brief overview of the FCI algorithm and related graphical terminology as well as details, proofs, further simulation studies, and figures for illustrating the application to the real data example that have been omitted from the main text for reasons of space.

We always assume that there are no selection variables. When saying that $\SSet$ is a separating set of $A$ and $B$ the exclusions $A \notin \SSet$ and $B \notin \SSet$ are implicit. The term \textit{subset} without the attribute \textit{proper} refers to both proper subsets and the original set itself, although in formulas we make this explicit by using the symbol $\subseteq$ instead of $\subset$. We switch between using variable names such as $X^i_{t-\tau}$ and $X^j_t$ that make the time structure explicit, and generic names such as $A$ and $B$ that do not make this explicit (using generic names does, however, \textit{not imply} that there is no time structure). The precise configurations of numerical experiments are given in the respective panel label and figure caption.


\renewcommand\theequation{S\arabic{equation}}
\setcounter{equation}{0}
\renewcommand\thefigure{S\arabic{figure}}    
\setcounter{figure}{0}   
\renewcommand\thesection{S\arabic{section}}    
\setcounter{section}{0} 
\renewcommand\thealgorithm{S\arabic{algorithm}}   
\setcounter{table}{0} 
\renewcommand\thetable{S\arabic{table}}    
\setcounter{algorithm}{1} 

\section{Relevant graphical terminology and notation}\label{sec:background}
The structural causal model (SCM) in eq.~\eqref{eqmain:SVARProcess} can be graphically represented by its time series graph (also known as full time graph) $\G$ \cite{Spirtes2000, Pearl2000, Peters2018}. This graph contains a node for each variable in the SCM (we use the words \textit{node} and \textit{variable} interchangeably in this context) and an edge (link, words again used interchangeably) $X^i_{t-\tau} \onlyright X^j_t$ if and only if $X^i_{t-\tau} \in \pa(X^j_t)$. It can be understood as a directed acyclic graph (DAG) with infinite extension and repeating structure along the time axis. The parents $\pa(X^j_t, \G) =\pa(X^j_t)$ of $X^j_t$ are the set of nodes $X^i_{t-\tau}$ with $X^i_{t-\tau} \onlyright X^j_t$ in $\G$, the ancestors $\an(X^j_t, \G)$ are the set of nodes connected to $X^j_t$ by a directed path in $\G$ together with $X^j_t$ itself (so every node is an ancestor of itself), and the adjacencies $\adj(X^j_t, \G)$ the set of nodes connected to $X^j_t$ by any edge in $\G$. Parents are a special case of ancestors. We call $X^j_t$ a descendant of $X^i_{t-\tau}$ if $X^i_{t-\tau}$ is an ancestor of $X^j_t$ (this implies that every node is a descendant of itself). A link between $X^i_{t-\tau}$ and $X^j_t$ is \textit{lagged} if $\tau > 0$, \textit{contemporaneous} if $\tau = 0 $, for $i = j$ we speak of an \textit{autodependency link}, and for $i \neq j$ of a \textit{cross link}.

In the presence of unobserved variables so called maximal ancestral graphs (MAGs) \cite{richardson2002} provide an appropriate graphical language for representing causal relationships. Since in this paper we assume the absence of selection variables, the relevant MAGs $\mathcal{M}$ contain two types of edges: directed `$\onlyright$' and bidirected `$\leftright$'. These edges are interpreted as composite objects constituted by the symbols at their ends (edge marks), which can be an (arrow-)head (`>' or `<') or a tail (`-'). These edge marks carry a causal meaning: Tails convey ancestorships in $\G$, i.e., $X^i_{t-\tau} \onlyright X^j_t$ in $\mathcal{M}$ asserts that $X^i_{t-\tau} \in \an(X^j_t, \G)$; heads convey non-ancestorships in $\G$, i.e., $X^i_{t-\tau} \onlyright X^j_t$ and $X^i_{t-\tau} \leftright X^j_t$ in $\mathcal{M}$ say that $X^j_t \notin \an(X^i_{t-\tau}, \G)$. As an immediate consequence of time order there cannot be a link $X^i_{t-\tau} \headtail X^j_t$ for $\tau > 0$ (an effect cannot precede its cause). Parents, ancestors and adjacencies are defined in the same way as for DAGs, and the spouses $\spouse(X^j_t, \mathcal{M})$ of $X^j_t$ are the set of nodes $X^i_{t-\tau}$ with $X^i_{t-\tau} \leftright X^j_t$ in $\mathcal{M}$. Two variables are connected by an edge in $\mathcal{M}$ if and only if they cannot be d-separated by a subset of observed variables in $\G$, and d-separation in $\G$ restricted to observed variables is equivalent to m-separation in $\mathcal{M}$ \cite{Pearl1988, Verma1990, richardson2002}. The parents (ancestors, adjacencies, spouses) of a set of variables are defined as the union of parents (ancestors, adjacencies, spouses) of the individual variables. Example: $\pa(\{A, B\}, \cdot) = \pa(A, \cdot) \cup \pa(B, \cdot)$.

The Markov equivalence class of a MAG is the set of all MAGs that yield the exact same set of m-separations \cite{Zhang2008}. These are graphically represented by partial ancestral graphs (PAGs), in which the set of allowed edge marks is extended by the circle mark `\omark' \cite{Zhang2008}. Such a graph is said to be a PAG for MAG $\mathcal{M}$ if $i)$ it has the same nodes and adjacencies as $\mathcal{M}$ and if $ii)$ all its non-circle edge marks are shared by all members in the Markov equivalence class of $\mathcal{M}$. It is further said to be \textit{maximally informative} if for all its circle marks there is some member of the equivalence class in which there is a tail instead and some other member in which there is a head instead. The wildcard symbol `$\ast$' may stand for all three possible edge marks (head, tail, circle). This is a notational device only, there are no `$\ast$' marks in PAGs.

\section{Some background on FCI}\label{sec:fci}
The Fast Causal Inference (FCI) algorithm is an algorithm for constraint-based causal discovery in the presence of unobserved variables \cite{Spirtes1995, Spirtes2000, Zhang2008}. It allows for both latent confounders and selection variables, although in this paper we assume the absence of selection variables. Under the assumptions of faithfulness \cite{Spirtes2000}, acyclicity, and the existence of an underlying SCM the algorithm determines the maximally informative PAG from perfect statistical decisions of conditional independencies in the distribution $P$ generated by the SCM. The algorithm is based on the following fact:
\begin{mypropsm}[m-separation by subsets of {\DSEP} sets \cite{Spirtes2000}]\label{thm:dsep}
Let $A$ and $B$ be two nodes such that $A \notin \adj(B, \mathcal{M})$ and $B \notin \an(A, \mathcal{M})$, then they are m-separated by some subset of $\DSEP(B, A, \mathcal{M})$. Here:
\end{mypropsm}
\begin{mydefsm}[{\DSEP} sets \cite{Spirtes2000}]\label{def:dsep}
Node $V \in \mathcal{M}$ is in $\DSEP(B, A, \mathcal{M})$ if and only if $i)$ it is not $B$ and $ii)$ there is a path $p_V$ between $B$ and $V$ such that $iia)$ all nodes on $p_V$ are in $\an(\{A, B\}, \mathcal{M})$ and $iib)$ all non end-point nodes on $p_V$ are colliders on $p_V$.
\end{mydefsm}
A node $B$ is a collider on a path $p$ if the two edges on $p$ involving $B$ both have a head at $B$, as e.g. in $A \asthead B \headast C$, otherwise it is a non-collider. Together with acyclicity Proposition~\ref{thm:dsep} guarantees that non-adjacent variables $A$ and $B$ are m-separated by a subset of $\DSEP(B, A, \mathcal{M})$ or a subset of $\DSEP(A, B, \mathcal{M})$. However, $\mathcal{M}$ is initially unknown and the {\DSEP} sets cannot be determined without prior work. Therefore, starting from the complete graph over the set of variables, FCI first performs tests of CI given subset of $\pa(B, \mathcal{M}^\prime)$ and $\pa(A, \mathcal{M}^\prime)$ where $\mathcal{M}^\prime$ is the (changing) graph that the algorithm operates on. Whenever two variables are found to be conditionally independent given some subset of variables, the edge between them is removed and their separating set is remembered. This removes some, but in general not all false links. Second, the algorithm orients all resulting unshielded triples $A \astast B \astast C$ in $\mathcal{M}^\prime$ (these are triples $A \astast B \astast C$ such that $A$ and $C$ are not adjacent) as colliders $A \asthead B \headast C$ if $B$ is not in the separating set of $A$ and $C$ (rule $\R0$). We note that at this point head marks are not guaranteed to convey non-ancestorships, but those unshielded triples in $\mathcal{M}^\prime$ that are part of $\mathcal{M}$ are oriented correctly. This is enough to determine the {\PDSEP} sets, see \cite{Spirtes2000}, which are supersets of the {\DSEP} sets define above. Third, FCI performs tests of CI given subsets of $\PDSEP(B, A, \mathcal{M}^\prime)$ and $\PDSEP(A, B, \mathcal{M}^\prime)$. This removes all false links. Fourth, all previous orientations are undone, $\R0$ is applied once more and then followed by exhaustive application of the ten rules $\R1$ through $\R10$. Tests of CI are preferentially made given smaller conditioning sets $\SSet$, i.e., FCI first tests sets with $|\SSet| = \ppc = 0$, then those with $|\SSet| = \ppc = 1$ and so on.

\section{LPCMCI-PAGs}\label{sec:lpcmcipags}
Section \ref{sec:middlemarksmain} introduced middle marks and LPCMCI-PAGs. We here give a more formal definition of these notions. Recall that we assume the absence of selection variables.
\begin{mydefsm}[LPCMCI-PAGs] \label{def:LPCMCI_PAG}
Consider a simple graph $\C$ over the same set of variables as $\M$ with edges of the type $\SMMnoast{\tailhead}$, $\SMMnoast{\headhead}$, $\SMM{\ohead}$, and $\SMM{\oo}$ where the wildcard `$\ast$' can stand for the five possible middle marks `?', `L', `R', `!', or `' (empty). Such $\C$ is a LPCMCI-PAG for $\G$ with respect to total order \totalorder if for any probability distribution $P$ that is Markov relative and faithful to $\G$ the following seven conditions hold:
\begin{enumerate}
  \item If $A \SMM{\asthead} B$ in $\C$, then $B \notin \an(A, \G)$.
  \item If $A \SMMnoast{\tailhead} B$ in $\C$, then $A \in \an(B, \G)$.
  \item If $A \notin \adj(B, \C)$, then $A \notin \adj(B, \M)$.
  \item If $A \LMM{\astast} B$ in $\C$ for $A < B$, then $B \notin \an(A, \G)$ or there is no $\SSet \subseteq \pa(A, \M)$ that m-separates $A$ and $B$ in $\M$.
  \item If $A \RMM{\astast} B$ in $\C$ for $A < B$, then $A \notin \an(B, \G)$ or there is no $\SSet \subseteq \pa(B, \M)$ that m-separates $A$ and $B$ in $\M$.
  \item If $A \EMM{\astast} B$ in $\C$, then both $A \LMM{\astast} B$ and $A \RMM{\astast} B$ would be correct.
  \item If $A \astast B$ in $\C$, then $B \in \adj(A, \M)$.
\end{enumerate}
\end{mydefsm}
The first two points give the same causal meaning to head and tail edge marks as they have in MAGs and PAGs. We repeat that while this definition involves a fixed total order \totalorder, its choice is arbitrary and without influence on the conveyed causal information. Moreover, the definition does not depend on time order. Also note that if all middle marks in $\C$ are empty, then $\C$ is a PAG for $\M$ (guaranteed by the first, second, third, and seventh point). Parents, ancestors, descendants, spouses, and adjacencies in $\C$ are defined (and denoted) in the same way as for MAGs and PAGs, i.e., without being influenced by middle marks.

\section{Orientation rules for LPCMCI-PAGs}\label{sec:orientationrules}
The following is a list of rules for orienting edges in LPCMCI-PAGs. These are extensions of the standard FCI rules \cite{Zhang2008} as well as the unshielded triple rule and discriminating path rule of RFCI \cite{Colombo2012}. If a rule proposes to orient the same edge mark as both tail and head, this is resolved by putting a conflict mark `x' instead. The edge mark wildcard `$\ast$' is redefined to stand for the circle, head, tail or conflict mark; the second wildcard symbol `$\star$' excludes the conflict mark. For two reasons we explicitly present and prove also those rules that generalize without much modification: To demonstrate their validity for LPCMCI-PAGs, and to show in which cases the rules also apply to structures with conflict marks.

If $X \SMM{\astast} Y \SMM{\astast} Z$ is an unshielded triple we write $\SSet_{XZ}$ for the separating set of $X$ and $Z$. Many rules require that $\SSet_{XZ}$ be weakly minimal and $Y \in \SSet_{XZ}$. In all these case the requirement of weak minimality can be dropped if $X \astast Y \astast Z$, i.e., if both middle marks on $X \SMM{\astast} Y \SMM{\astast} Z$ are empty. For this reason the standard FCI orientation rules are implied as special cases.

\underline{$\mathbf{\ER{0}a}$}:
For all unshielded triples $A \SMM{\astast} B \SMM{\astast} C$: If $ia)$ $A \astast B$ or $ib)$ $A$ and $B$ are conditionally dependent given $[\SSet_{AC} \cup \pa(\{A,B\}, \C)] \setminus \{A, B, \text{nodes in the future of both $A$ and $B$}\}$, $iia)$ $C \astast B$ or $iib)$ $C$ and $B$ are conditionally dependent given $[\SSet_{AC} \cup \pa(\{C,B\}, \C)] \setminus \{C, B, \text{nodes in the future of both $C$ and $B$}\}$, $iii)$ none of the edge mark`$\ast$'s at $B$ on $A \SMM{\astast} B \SMM{\astast} C$ is `-' or `x', and $iv$) $B \notin \SSet_{AC}$, then mark the unshielded triple for orientation as collider $A \SMM{\asthead} B \SMM{\headast} C$. Condition $ib)$ need only be checked if not $ia)$, $iib)$ need only be checked if not $iia)$, and $iv)$ need only be checked if all previous conditions are true. If $ib)$ or $iib)$ find a conditional independence, mark the corresponding edge(s) for removal.

\underline{$\mathbf{\ER{0}b}$}:
For all unshielded triples $A \SMM{\asthead} B \EMM{\ostar} C$ and for all unshielded triples $A \SMM{\asthead} B \RMM{\ostar} C$ with $B < C$ and for all unshielded triples $A \SMM{\asthead} B \LMM{\ostar} C$ with $B > C$: If $ia)$ $A \asthead B$ or $ib)$ $A$ and $B$ are conditionally dependent given $[\SSet_{AC} \cup \pa(\{A,B\}, \C)] \setminus \{A, B, \text{nodes in the future of both $A$ and $B$}\}$, and $ii)$ $B \notin \SSet_{AC}$, then mark the edge between $B$ and $C$ for orientation as $B \SMM{\headstar} C$ (the middle mark remains as it was before). Condition $ib)$ need only be checked if not $ia)$. If $ib)$ finds a conditional independence, mark the corresponding edge for removal.

\underline{$\mathbf{\ER{0}c}$}:
For all unshielded triples $A \astast B \EMM{\ostar} C$ and for all unshielded triples $A \astast B \RMM{\ostar} C$ with $B < C$ and for all unshielded triples $A \astast B \LMM{\ostar} C$ with $B > C$: If $B \notin \SSet_{AC}$, then mark the edge between $B$ and $C$ for orientation as $B \SMM{\headstar} C$ (the middle mark remains as it was before).

\underline{$\mathbf{\ER{0}d}$}:
For all unshielded triples $A \asto B \oast C$ and for all unshielded triples $A \asthead B \oast C$: If $B \notin \SSet_{AC}$, then mark the unshielded triple for orientation as collider $A \asthead B \headast C$.

\underline{$\mathbf{\ER1}$}:
For all unshielded triples $A \SMM{\asthead} B \SMM{\ostar} C$: If $\SSet_{AC}$ is weakly minimal and $B \in \SSet_{AC}$, then mark the edge between $B$ and $C$ for orientation as $B \SMMnoast{\tailhead} C$.

\underline{$\mathbf{\ER2}$}:
For all $A \SMMnoast{\tailhead} B \SMM{\asthead} C$ with $A \SMM{\staro} C$ and for all $A \SMM{\asthead} B \SMMnoast{\tailhead} C$ with $A \SMM{\staro} C$: Mark the edge between $A$ and $C$ for orientation as $A \SMM{\starhead} C$.

\underline{$\mathbf{\ER3}$}:
For all unshielded triples $A \SMM{\asthead} B \SMM{\headast} C$ with $A \SMM{\staro} D \SMM{\ostar} C$ and $D \SMM{\staro} B$: If $\SSet_{AC}$ is weakly minimal and $D \in \SSet_{AC}$, then mark the edge between $D$ and $B$ for orientation as $D \SMM{\starhead} B$.

\underline{$\mathbf{\ER4}$}:
Use the discriminating path rule of \cite{Colombo2012} with the following modification: When the rule instructs to test whether any pair $(A, B)$ of variables is conditionally independent given any set $\SSet$, then $i)$ if $A$ and $B$ are connected by an edge with empty middle mark do not make this test, and $ii)$ else replace $\SSet$ with $[\SSet \cup \pa(\{A,B\}, \C)] \setminus \{A, B, \text{nodes in the future of both $A$ and $B$}\}$.

\underline{$\mathbf{\ER8}$}:
For all $A \SMMnoast{\tailhead} B \SMMnoast{\tailhead} C$ with $A \SMM{\ostar} C$: Mark the edge between $A$ and $C$ for orientation as $A \SMMnoast{\tailhead} C$.

\underline{$\mathbf{\ER9}$}:
For all $A_1 \SMM{\ohead} A_{n}$ for which $a)$ there is an uncovered potentially directed path from $A_1$ to $A_{n}$ through $A_2, \ldots, A_{n-1}$ (in this order) such that $b)$ $A_2$ is not adjacent to $A_{n}$: If for all $k = 1, \ldots, n-1$ $ia)$ $A_{k} \SMMnoast{\tailhead} A_{k+1}$ or $ib)$ $\SSet_{A_{k+1}A_{k-1}}$ is weakly minimal and $A_{k} \in \SSet_{A_{k+1}A_{k-1}}$ (with the convention $A_0 = A_{n}$), then mark the edge between $A_1$ and $A_{n}$ for orientation as $A_1 \SMMnoast{\tailhead} A_{n}$.

\underline{$\mathbf{\ER{10}}$}:
For all $A \SMM{\ohead} D$ for which $a)$ there is $B_n \SMMnoast{\tailhead} D \SMMnoast{\headtail} C_{m}$, $b)$ an uncovered potentially directed path $p_B$ from $A \equiv B_0$ to $B_{n}$ through $B_1, \ldots, B_{n-1}$ (in this order), $c)$ an uncovered potentially directed path $p_C$ from $A \equiv C_0$ to $C_{m}$ through $C_1, \ldots, C_{m-1}$ (in this order) such that $d)$ $B_1$ and $C_{1}$ are not adjacent: If $i)$ $\SSet_{B_1C_1}$ is weakly minimal and $A \in \SSet_{B_1C_1}$, $ii)$ for all $k = 0, \ldots, n-2$ $iia)$ $B_{k+1} \SMMnoast{\tailhead} B_{k+2}$ or $iib)$ $\SSet_{B_{k+2}B_k}$ is weakly minimal and $B_{k+1} \in \SSet_{B_{k+2}B_k}$, and $iii)$ for all $k = 0, \ldots, m-2$ $iiia)$ $C_{k+1} \SMMnoast{\tailhead} C_{k+2}$ or $iiib)$ $\SSet_{C_{k+2}C_k}$ is weakly minimal and $C_{k+1} \in \SSet_{C_{k+2}C_k}$, then mark the edge between $A$ and $D$ for orientation as $A \SMMnoast{\tailhead} D$.

These rules orient edge marks. They are complemented by the following two rules for updating middle marks:

\underline{\textbf{APR}:} (ancestor-parent-rule, see Lemma~\ref{lemma:middlemarks})
Replace all edges $A \EMMnoast{\tailhead} B$ by $A \tailhead B$, all edges $A \LMMnoast{\tailhead} B$ with $A > B$ by $A \tailhead B$, and all edges $A \RMMnoast{\tailhead} B$ with $A < B$ by $A \tailhead B$.

\underline{\textbf{MMR}:} (middle-mark-rule)
Replace all edges $A \QMM{\asthead} B$ with $A < B$ by $A \LMM{\asthead} B$, all edges $A \QMM{\asthead} B$ with $A > B$ by $A \RMM{\asthead} B$, all edges $A \RMM{\asthead} B$ with $A < B$ by $A \EMM{\asthead} B$, and all edges $A \LMM{\asthead} B$ with $A > B$ by $A \EMM{\asthead} B$.

\section{Pseudocode for Algorithms \ref{algo:ancestral} and \ref{algo:nonancestral}}\label{sec:pseudocode2}
In Sec.~\ref{sec:pseudocode} of the main text we give pseudocode for LPCMCI in Algorithm \ref{algo:lpcmci}. This involves calls to Algorithms \ref{algo:ancestral} and \ref{algo:nonancestral}, for which we here provide pseudocode and further explanations.

\begin{algorithm*}[ht]
\caption{Ancestral removal phase}
\begin{algorithmic}[1]
\Require
LPCMCI-PAG $\C$,
memory of minimal test statistic values $I^{\min}(\cdot, \cdot)$,
memory of separating sets SepSet$(\cdot, \cdot)$,
time series dataset $\Xobs = \{\mathbf{X}^1, \ldots ,\mathbf{X}^N\}$,
maximal considered time lag $\taumax$,
significance level $\alpha$,
CI test ${\rm CI}(X,\,Y,\,\SSet)$
\Repeat{ starting with $\ppc = 0$}
  \For{$-1\leq m \leq \taumax$} 
    \ForAll{ordered pairs of variables $(X^i_{t-\tau}, X^j_t)$ adjacent in $\C$ with $X^i_{t-\tau} < X^j_t$}
      \If{$(m = - 1$ and $i \neq j)$ or $(m\geq 0$ and $\tau \neq m$ or $i = j)$}{ continue with next pair}
      \EndIf
      \State
      $\SSet_{def} = \pa(\{X^i_{t-\tau}, X^j_t\}, \C)$
      \If{the middle mark is `?' or `L'}
        \State
        $\SSet_{search} = \apds(X^j_t, X^i_{t-\tau}, \C) \setminus \SSet_{def}$, ordered according to $I^{\min}(X^j_t, \cdot)$
        \If{$|\SSet_{search}| < \ppc$}{ update middle mark with `R' according to Lemma \ref{lemma:middlemarkupdate}}
        \EndIf
        \ForAll{subsets $\SSet \subseteq \SSet_{search}$ with $|\SSet| = \ppc$}
          \State
          $(\text{$p$-value},\,I) \gets$ \Call{CI}{$X^i_{t-\tau},\,X^j_{t},\,\SSet \cup \SSet_{def}}$
                \State
                $I^{\min}(X^i_{t-\tau},X^j_t)= I^{\min}(X^j_t, X^i_{t-\tau}) =\min(|I|,I^{\min}(X^i_{t-\tau},X^j_t))$
                \If{$p$-value $> \alpha$}
                  \State
                  mark edge for removal, add $\SSet \cup \SSet_{def}$ to $\text{SepSet}(X^i_{t-\tau}, X^j_t)$
                  \State
                  break innermost for-loop
                \EndIf
        \EndFor
      \EndIf
      \State
      repeat lines 6 - 14 with $X^i_{t-\tau}$ and $X^j_t$ as well as `R' and `L' swapped
    \EndFor
    \State
    remove all edges that are marked for removal from $\C$
  \EndFor
  \If{any edge has been removed in line 16}
    \State
    run Alg.~\ref{algo:orientation} using $[\text{APR}, \text{MMR}, \ER8, \ER2, \ER1, \ER9, \ER{10}]$, orient lagged links only
    \State
    let $\ppc = 0$
  \Else { increase $\ppc$ to $\ppc + 1$}
  \EndIf
\Until{there are no other middle marks than `!' or `' (empty)}
\State
run Alg.~\ref{algo:orientation} using $[\text{APR}, \text{MMR}, \ER8, \ER2, \ER1, \ER0d, \ER0c, \ER3, \R4, \ER9, \ER{10}, \ER0b, \ER0a]$
\State
\Return
$\C$, $I^{\min}(\cdot, \cdot)$, SepSet$(\cdot, \cdot)$
\end{algorithmic} \label{algo:ancestral}
\end{algorithm*}
\textbf{Algorithm \ref{algo:ancestral}} removes the edges between all pairs $(X^i_{t-\tau}, X^j_t)$ of variables that are not adjacent in $\M$ and for which one of them is an ancestor of the other (it may also removed edges between some pairs of non-adjacent variables for which neither one of them is ancestor of the other, but this is not guaranteed). To this end the algorithm tests for CI given $\SSet \cup \SSet_{def}$, where the cardinality $|\SSet| = \ppc$ of $\SSet \subseteq \SSet_{search} = \apds(X^j_t, X^i_{t-\tau}, \C) \setminus \SSet_{def}$ is successively increased. The $\apds$ sets are defined in Sec.~\ref{sec:apdsnapds} below, they exclude all variables that have already been identified as non-ancestors of $X^j_t$. This reflects the first design principle behind LPCMCI, see Sect.~\ref{sec:MCI}. The default conditioning set $\SSet_{def} = \pa(\{X^i_{t-\tau}, X^j_t\}, \C)$ consists of all variables that have been marked as parents of $X^i_{t-\tau}$ or $X^j_t$ in $\C$, which implies that they are ancestors of $X^i_{t-\tau}$ or $X^j_t$ in $\G$. The extension of $\SSet$ to $\SSet \cup \SSet_{def}$ reflects the second design principle behind LPCMCI, see Sect.~\ref{sec:MCI}, and according to Lemma \ref{lemma:conditiononparents} cannot destroy m-separations. The parentships used to define $\SSet_{def}$ are found by the application of orientation rules in line 18 (with Alg.~\ref{algo:orientation}, see further below in this section) that are made if at least one edge was removed in the current step of the repeat-loop (or have been passed on from an earlier iteration in the preliminary phase of LPCMCI). It is then necessary to restart with $\ppc = 0$, otherwise future separating sets might not be weakly minimal. The rules may also find non-ancestorships, these then further restrict the $\apds$ sets. Another novelty is that some edges are tested and removed (if found insignificant) before other edges are tested, see lines 2, 4 and the indentation of line 16. To be precise: All autodependency links are tested first, followed by cross links starting with lag $\tau = 0$ and moving to lag $\tau = \taumax$ in steps of one. This ordering does not depend on the ordering of the $N$ time series variables $X^j$ and does therefore not introduce order-dependence in the sense studied in \cite{Colombo2014}. The algorithm converges once all middle marks in $\C$ are `!' or empty. By means of the APR rule (see Lemma \ref{lemma:middlemarks} or Sect.~\ref{sec:orientationrules}) all edges with a tail mark will then have an empty middle mark, i.e., they cannot be m-separated and do not need further testing. Line 11 updates a memory for keeping track of the minimum test statistic value across all previous CI tests for a given pair of variables (the memory is initialized to plus infinity when line 1 of Algorithm \ref{algo:lpcmci} is executed). These values are used to sort $\SSet_{search}$ in line 7 such that $X^{l}_{t-\tau_l}$ appears before $X^{k}_{t-\tau_k}$ in $\SSet_{search}$ if $I^{\min}(X^j_t, X^{l}_{t-\tau_l}) > I^{\min}(X^j_t, X^{k}_{t-\tau_k})$. Note that in line 18 only a select subset of rules is applied and that these are only used to orient lagged links. Moreover, in line 22 we choose to apply the standard rule $\R4$ rather than the modified rule $\ER4$. The reason for this is that, as observed in \cite{Colombo2014}, the discriminating path rule (on which $\ER4$ is based) becomes computationally intensive when applied in an order-independent way involving conflict resolution. We found these choices to work well in practice but do not claim their optimality.

\begin{algorithm*}[h!]
\caption{Non-ancestral removal phase}
\begin{algorithmic}[1]
\Require
LPCMCI-PAG $\C$,
memory of minimal test statistic values $I^{\min}(\cdot, \cdot)$,
memory of separating sets SepSet$(\cdot, \cdot)$,
time series dataset $\Xobs = \{\mathbf{X}^1, \ldots ,\mathbf{X}^N\}$,
maximal considered time lag $\taumax$,
significance level $\alpha$,
CI test ${\rm CI}(X,\,Y,\,\SSet)$
\Repeat{ starting with $\ppc = 0$}
  \For{$-1\leq m \leq \taumax$}
    \ForAll{ordered pairs of variables $(X^i_{t-\tau}, X^j_t)$ adjacent in $\C$ with $X^i_{t-\tau} < X^j_t$}
      \If{the middle mark is empty}{ continue with next pair}
      \EndIf
      \If{$(m = - 1$ and $i \neq j)$ or $(m\geq 0$ and $\tau \neq m$ or $i = j)$}{ continue with next pair}
      \EndIf
      \State
      $\SSet^1_{def} = \pa(\{X^i_{t-\tau}, X^j_t\}, \C)$
      \State
      $\SSet^2_{def} = \text{nodes that have ever been in $\pa(\{X^i_{t-\tau}, X^j_t\}, \C)$ since re-initialization}$
      \State
      $\SSet^1_{search} = \napds(X^j_t, X^i_{t-\tau}) \setminus (\SSet^1_{def} \cup \SSet^2_{def})$, ordered according to $I^{\min}(X^j_t, \cdot)$
      \State
      $\SSet^2_{search} = \napds(X^i_{t-\tau}, X^j_t) \setminus (\SSet^1_{def} \cup \SSet^2_{def})$, ordered according to $I^{\min}(X^i_{t-\tau}, \cdot)$
      \If{$|\SSet^1_{search}| < \ppc $ or $\tau = 0$ and $|\SSet^2_{search}| < \ppc $}
        \State
        Update middle mark with `' according to Lemma \ref{lemma:middlemarkupdate}, continue with next pair
      \EndIf
      \ForAll{subsets $\SSet \subseteq \SSet^1_{search}$ with $|\SSet| = \ppc$}
        \State
        $\SSet_{def} = \SSet^1_{def} \cup [\SSet^2_{def} \cap \napds(X^j_t, X^i_{t-\tau}, \C)]$
        \State
        $(\text{$p$-value},\,I) \gets$ \Call{CI}{$X^i_{t-\tau},\,X^j_{t},\,\SSet \cup \SSet_{def}}$
            \State
            $I^{\min}(X^i_{t-\tau},X^j_t)= I^{\min}(X^j_t, X^i_{t-\tau}) =\min(|I|,I^{\min}(X^i_{t-\tau},X^j_t))$
            \If{$p$-value $> \alpha$}
              \State
              mark edge for removal, add $\SSet \cup \SSet_{def}$ to $\text{SepSet}(X^i_{t-\tau}, X^j_t)$
              \State
              break innermost for-loop
            \EndIf
      \EndFor
      \If{$\tau = 0$}
        \State
        run lines 12 - 18 with $\SSep^2_{search}$ replacing $\SSep^1_{search}$, and $X^i_{t-\tau}$ and $X^j_t$ swapped
      \EndIf
    \EndFor
    \State
    remove all edges that are marked for removal from $\C$
  \EndFor
  \If{any edge has been removed in line 21}
    \State
    run Alg.~\ref{algo:orientation} using the same rules as in line 22 of Alg.~\ref{algo:ancestral}
    \State
    let $\ppc = 0$
  \Else{ increase $\ppc$ to $\ppc + 1$}
  \EndIf
\Until{all middle marks in $\C$ are empty}
\State
run Alg.~\ref{algo:orientation} using the same rules as in line 22 of Alg.~\ref{algo:ancestral}
\State
\Return
$\C$, $I^{\min}(\cdot, \cdot)$, SepSet$(\cdot, \cdot)$
\end{algorithmic} \label{algo:nonancestral}
\end{algorithm*}
\textbf{Algorithm \ref{algo:nonancestral}} is structurally similar to Algorithm \ref{algo:ancestral}. Once called in line 6 of Algorithm \ref{algo:lpcmci}, all middle marks in $\C$ are `!' or empty. Whereas edges with empty middle mark are in $\M$ for sure, some edges with middle mark `!' might not be in $\M$. Those latter type of edges are between pairs of variables in which neither one of them is ancestor of the other. According to Lemma \ref{lemma:apdsnapds} below in combination with Proposition \ref{thm:dsep} such pairs are m-separated by some subset of $\napds(X^j_t, X^i_{t-\tau}, \C)$ as well as by some subset of $\napds(X^i_{t-\tau}, X^j_t, \C)$. These sets are defined in Sec.~\ref{sec:apdsnapds} below, they are the more restricted LPCMCI equivalent of the {\PDSEP} sets in FCI and the $pds_t$ sets in SVAR-FCI. For computational reasons the algorithm nevertheless only searches for separating sets in $\napds(X^j_t, X^i_{t-\tau}, \C)$, unless for $\tau = 0$ where order-independence dictates otherwise. This is the reason for the logical or-connection in line 10. As compared to Algorithm \ref{algo:ancestral}, the default conditioning is extended: According to Definition \ref{def:LPCMCI_PAG} a tail on an edge in $\C$ signifies ancestorship in $\G$. Since $\C$ is an LPCMCI-PAG at every point of LPCMCI, $X^i_{t-\tau}$ is an ancestor of $X^j_t$ if there ever was the link $X^i_{t-\tau} \SMMnoast{\tailhead} X^j_t$. This gives rise to the set $\SSet^2_{def}$ in line 6. In addition to the parents in $\C$, the algorithm also conditions per default on all nodes in $\SSet^2_{def}$ that are in the current $\napds$ set. This decreases the number of sets $\SSet$ that need to be searched through in the for-loop in line 12 at the price of a higher-dimensional conditioning set. Also this extended default conditioning cannot destroy m-separations. Non-ancestorships are used to constrain the $\napds$ sets in the first place, and prior to determining $\napds$ sets the collider rule $\ER0a$ must have been applied to all unshielded triples in $\C$. The algorithm converges once all middle marks are empty, followed by a final exhaustive rule application to guarantee completeness. 

\begin{algorithm*}[h!]
\caption{Orientation phase}
\begin{algorithmic}[1]
\Require
LPCMCI-PAG $\C$,
ordered list of rules $\mathfrak{r}$,
memory of minimal test statistic values $I^{\min}(\cdot, \cdot)$,
memory of separating sets SepSet$(\cdot, \cdot)$,
time series dataset $\Xobs = \{\mathbf{X}^1, \ldots ,\mathbf{X}^N\}$,
maximal considered time lag $\taumax$,
significance level $\alpha$,
CI test ${\rm CI}(X,\,Y,\,\SSet)$
\State
$i = 0$
\Repeat
  \State
  apply the $i$-th rule in $\mathfrak{r}$ to $\C$, do not modify $\C$ yet
  \If{the rule proposes any modification}
    \ForAll{edges marked for orientation}
      \State
      resolve conflicts among the proposed orientations
      \State
      apply the conflict resolved orientations $\C$
    \EndFor
    \ForAll{edges marked for removal}
      \State
      remove the edge from $\C$
      \State
      make the corresponding separating set weakly minimal
    \EndFor
    \State
    let $i = 0$
  \Else{ increase $i$ to $i + 1$}
  \EndIf
\Until{$ i \geq \text{len}(\mathfrak{r})$}
\State
\Return
$\C$, $I^{\min}(\cdot, \cdot)$, SepSet$(\cdot, \cdot)$
\end{algorithmic} \label{algo:orientation}
\end{algorithm*}
\textbf{Algorithm \ref{algo:orientation}} exhaustively applies a given set of orientation rules specified by an ordered list $\mathfrak{r}$. The rules are executed in this order and, once any rule has modified $\C$, the loop jumps back to the first rule. This can be used for a preferential execution of simpler and less time consuming rules. Rules $\ER0a$ and $\ER0b$ involve CI tests and may therefore remove some edges. The corresponding separating sets are not guaranteed to be weakly minimal, see Example 1 in the supplement paper to \cite{Colombo2012} for a counterexample. (There this example is used to show that the separating sets may not be minimal, however it is also a counterexample for weak minimality.) Since many other rules require weak minimality of separating sets, line 10 instructs to make them weakly minimal. This is implemented in the following way: A separating set of $X^i_{t-\tau}$ and $X^j_t$ that is not necessarily weakly minimal is made weakly minimal by successively removing single elements that are not known ancestors of $X^i_{t-\tau}$ and $X^j_t$ until the resulting set is no separating set anymore. In particular, there is no need to search through all subsets of the original separating set. The validity of this procedure owes to the equivalence of weak minimality and weak minimality of the second type, see Definition \ref{def:weaklyminimal2} and part 3.) of Lemma \ref{lemma:weaklyminimal} below. The algorithm also tests for potential conflicts among the proposed orientations and, if present, resolves them by putting the conflict mark `x'. Most rules require to know whether certain nodes are or are not in certain separating sets. Queries of the second type (\textit{Is node $B$ not in the separating set of nodes $A$ and $C$?}) are answered by a modified version of the majority rule proposed in \cite{Colombo2014}. Our modification consists of $i)$ searching for separating sets not in the adjacencies of $A$ and $C$ but rather in the relevant $\apds$ sets and $ii)$ including also those separating sets that were found by Algs.~\ref{algo:ancestral} and \ref{algo:nonancestral} in the majority vote. The second part of this modification is necessary to guarantee completeness (FCI with the unmodified majority rule is not complete, see Sec.~\ref{sec:noncompleteness} for an example). The modification does not introduce order-independence since $i)$ the sets $\SSet_{search}$, $\SSet^1_{search}$ and $\SSet^2_{search}$ are ordered by means of $I^{min}(\cdot, \cdot)$ and since $ii)$ line 13 of Alg.~\ref{algo:ancestral} and line 17 of Alg.~\ref{algo:nonancestral} instruct to \textit{add} $\SSet \cup \SSet_{def}$ to $\text{SepSet}(X^i_{t-\tau}, X^j_t)$ rather than saying \textit{write to}. Point $ii)$ is relevant for contemporaneous links: if in the same iteration of Alg.~\ref{algo:ancestral} (Alg.~\ref{algo:nonancestral}) a pair of variables is found to be conditionally independent given subsets of both $\apds(X^i_t, X^j_t, \C)$ and $\apds(X^i_t, X^j_t, \C)$ (both $\napds(X^i_t, X^j_t, \C)$ and $\napds(X^i_t, X^j_t, \C)$), \textit{both} separating sets are remembered. The search for separating sets involves the same default conditioning as in Alg.~\ref{algo:ancestral}. For queries of the first type (\textit{Is node $B$ in the separating set of nodes $A$ and $C$?}) we distinguish two cases. If $B$ is adjacent to both $A$ and $C$ and the middle mark of both edges is empty, then the query is answered in the same way as queries of the first type. Otherwise, the query is answered solely based on the separating sets found by Algs.~\ref{algo:ancestral} and \ref{algo:nonancestral}. Alternatively, one might also in this second case perform a majority-type search of additional separating sets, albeit restricted to separating sets of minimal cardinality due to the requirement of weak minimality (whereas this restriction is not necessary when $A$ and $B$ as well as $C$ and $B$ are connected by edges with empty middle marks). We do not claim optimality of these choices.

\section{A variant of LPCMCI without Alg.~\ref{algo:nonancestral}}
A variant of LPCMCI can be obtained by skipping the execution of Alg.~\ref{algo:nonancestral} in line 6. According to Lemma~\ref{lemma:withoutalg3} the estimated graph $\C^\prime$ is then still a LPCMCI-PAG. This implies that all causal information as conveyed by the absence of edges, by the presence of edges with their respective middle marks, as well as by heads and tails as detailed in Sec.~\ref{sec:lpcmcipags} remains correct. Lemma~\ref{lemma:withoutalg3} further says that $i)$ all edges in $\C^\prime$ that are of the form $\SMMnoast{\tailhead}$ are also in $\PG$ and that $ii)$ if $X^i_{t-\tau}$ and $X^j_t$ are adjacent in $\C^\prime$ but not in $\PG$ then neither of these variables is an ancestor of the other. This is analogous to RFCI-PAGs, see Theorem 3.4 in \cite{Colombo2012}. This variant of LPCMCI therefore compares to standard LPCMCI as (SVAR-)RFCI compares to (SVAR-)FCI.

As a side remark, even if LPCMCI is interrupted at any arbitrary point it still yields a graph with unambiguous and sound causal interpretation. This is implied by the fact that the graph $\C$ remains and LPCMCI-PAG at every step of the algorithm, which is proven in Lemmas~\ref{lemma:algo2} and \ref{lemma:algo3}.

\section{Definition and relevance of $\apds$ and $\napds$ sets}\label{sec:apdsnapds}
As explained in Sec.~\ref{sec:pseudocode2}, Algorithms \ref{algo:ancestral} and \ref{algo:nonancestral} respectively perform tests of CI given subsets of $\apds$ sets and $\napds$ sets. These are defined and motivated here.

In words $\apds(X^j_t, X^i_{t-\tau}, \C)$ is the set of all non-future adjacencies of $X^j_t$ other than $X^i_{t-\tau}$ that have not already been identified as non-ancestors of $X^j_t$, formally:
\begin{mydefsm}[$\apds$ sets]
The set $\apds(X^j_t, X^i_{t-\tau}, \C)$ is the set of all $X^k_{t-\tau^\prime}$ other than $X^i_{t-\tau}$ with $\tau^\prime \geq 0$ that are connected to $X^j_t$ by an edge without head at $X^k_{t-\tau^\prime}$.
\end{mydefsm}
All statements in this and the following definition are with respect to the graph $\C$. The definition of $\napds$ sets is more involved. It uses already identified (non-)ancestorships, time order and some general properties of {\DSEP} sets to provide a tighter approximate of the latter than the {\PDSEP} sets of FCI and $pds_t$ sets of SVAR-FCI do. Formally:
\begin{mydefsm}[$\napds$ sets]\label{def:napds}
$\textbf{1.)}$ The set $\napds(X^j_t, X^i_{t-\tau}, \C)$ is the union of $\napds^1(X^j_t, X^i_{t-\tau}, \C)$ and $\napds^2(X^j_t, X^i_{t-\tau}, \C)$. $\textbf{2.)}$ The set $\napds^1(X^j_t, X^i_{t-\tau}, \C)$ is $\apds(X^j_t, X^i_{t-\tau}, \C)$ without all variables $X^k_{t-\tau^\prime}$ that are connected to $X^i_{t-\tau}$ by an edge with tail at $X^i_{t-\tau}$. $\textbf{3.)}$ The set $\napds^2(X^j_t, X^i_{t-\tau}, \C)$ is the set of all variables $X^k_{t-\tau^\prime}$ that are connected to $X^j_t$ by a path $p$ with the following properties: $i)$ on $p$ there is no tail at any node other than $X^k_{t-\tau^\prime}$, $ii)$ the middle node of every unshielded triple on $p$ is a collider on $p$, $iii)$ $p$ does not contain $X^i_{t-\tau}$, $iv)$ the node $X^l_{t-\tilde{\tau}}$ adjacent to $X^j_t$ is not connected to $X^i_{t-\tau}$ by an edge with head at $X^l_{t-\tilde{\tau}}$, and is not after $X^i_{t-\tau}$, $v)$ all nodes on $p$ other than $X^j_t$ and $X^l_{t-\tilde{\tau}}$ are not connected to $X^j_t$ or $X^i_{t-\tau}$ by an edge with tail at $X^j_t$ or $X^i_{t-\tau}$, are not at the same time connected to both $X^j_t$ and $X^i_{t-\tau}$ by edges with a head at themselves, and are not after both $X^j_t$ and $X^i_{t-\tau}$.
\end{mydefsm}
The use of the $\apds$ and $\napds$ sets in Algorithms \ref{algo:ancestral} and \ref{algo:nonancestral} is due to the following result:
\begin{mylemmasm}[Relevance of $\apds$ and $\napds$ sets]\label{lemma:apdsnapds}
Let $A$ and $B$ be such that $A \notin \adj(B, \M)$. $\textbf{1.)}$ If $A \in \an(B, \G)$ then $\apds(B, A, \C) \supseteq \DSEP(B, A, \M)$. $\textbf{2.)}$ If $B \notin \an(A, \G)$, $A \notin \an(B, \G)$ and rule $\ER0a$ has been exhaustively applied to $\C$ then $\napds(B, A, \C) \supseteq \DSEP(B, A, \M)$.
\end{mylemmasm}
This remains true when Definition \ref{def:napds} is strengthened in the following way: Whenever the definition demands that there be no edge between $X^i_{t-\tau}$ (or $X^j_{t}$) and some node $X^m_{t-\tau_m}$ with head at $X^m_{t-\tau_m}$, add the requirement that there be a potentially directed path from $X^m_{t-\tau_m}$ to $X^i_{t-\tau}$ (or $X^j_{t}$). 

\section{Proofs}\label{sec:proofs}

\begin{customtheorem}{1}[LPCMCI effect size]
Let $A \asthead B$ (with $A = X^i_{t-\tau}$ and $B = X^j_t$) be a link ($\tailhead$ or $\headhead$) in $\M$. Consider the default conditions $\SSet_{def} =\pa(\{A, B\}, \M) \setminus \{A, B\}$ and denote $\Xobs^*=\Xobs \setminus \SSet_{def}$. Let $\mathbf{S}=\arg\min_{\SSet\subseteq \Xobs^* \setminus \{A, B\}} I(A; B|\SSet \cup \SSet_{def})$ be the set of sets that define LPCMCI's effect size.
If $i)$ there is $\SSet^* \in \mathbf{S}$ with $\SSet^* \subseteq \adj(A, \M)\setminus \SSet_{def}$ or $\SSet^* \subseteq \adj(B, \M)\setminus \SSet_{def}$ and $ii)$ there is a proper subset $\mathcal{Q}\subset \SSet_{def}$ such that $\mathcal{I}(A; B; \SSet_{def}\setminus \mathcal{Q}|\SSet^* \cup \mathcal{Q}) < 0$, then
\begin{align} \label{eq:effect_size}
\min_{\SSet\subseteq \Xobs^* \setminus \{A, B\}} I(A; B|\SSet \cup \SSet_{def}) &> \min_{\tilde{\SSet}\subseteq \Xobs\setminus \{A, B\}} I(A; B|\tilde{\SSet})\,.
\end{align}
If the assumptions are not fulfilled, then (trivially) "$\geq$" holds in eq.~\eqref{eq:effect_size}.
\end{customtheorem}

\begin{myremark}[]
Assuming the link between $X^i_{t-\tau}$ and $X^j_t$ to be of the form $X^i_{t-\tau} \asthead X^j_t$ is no restriction. If $X^i_{t-\tau} \headtail X^j_t$ then $\tau = 0$ by time order and we can swap the roles of $X^i_{t - \tau} = X^i_t$ and $X^j_t$.
\end{myremark}

\textbf{Proof of Theorem~\ref{thm:effect_size}.} \label{sec:effect_size_proof}
We start the proof of eq.~\eqref{eq:effect_size} by splitting up the set $\Xobs$ that occurs on its right hand side as follows:
\begin{align}
\min_{\SSet\subseteq \Xobs^*\setminus \{A, B\}} I(A; B|\SSet \cup \SSet_{def}) &> \min_{\tilde{\SSet}\subseteq \mathbf{X}\setminus\{A, B\}} I(A; B|\tilde{\SSet})\\
\Leftrightarrow \min_{\SSet\subseteq \Xobs^*\setminus \{A, B\}} I(A; B| \SSet \cup \SSet_{def}) &> \min_{\tilde{\SSet}\subseteq \Xobs^*\setminus \{A, B\}} \min_{\mathcal{Q}\subseteq \SSet_{def}} I(A; B|\tilde{\SSet} \cup \mathcal{Q})  \label{eq:effect_size_tmp1}
\end{align}
Note that for $\mathcal{Q} = \SSet_{def}$ the right hand side equals the left hand side. Therefore, eq.~\eqref{eq:effect_size_tmp1} becomes trivially true when ``$>$'' is replaced by ``$\geq$'', but as it stands with ``$>$'' it is equivalent to
\begin{align}
\min_{\SSet\subseteq \Xobs^*\setminus \{A, B\}} I(A; B|\SSet \cup \SSet_{def}) &> \min_{\tilde{\SSet}\subseteq \Xobs^*\setminus \{A, B\}} \min_{\mathcal{Q}\subset \SSet_{def},\, \mathcal{Q}\neq \SSet_{def}} I(A; B|\tilde{\SSet} \cup \mathcal{Q}) \ , \label{eq:effect_size_tmp2}
\end{align}
where $\mathcal{Q}$ is now restricted to be a \textit{proper} subset of $\SSet$. Let, as stated in the theorem, $\mathbf{S}$ be the set of sets that make the left hand side minimal. A \emph{sufficient} condition for eq.~\eqref{eq:effect_size_tmp2} is then the existence of $\SSet^* \in \mathbf{S}$ such that
\begin{align}
I(A; B|\SSet^* \cup \SSet_{def}) &> \min_{\mathcal{Q}\subset \SSet_{def}, \, \mathcal{Q}\neq \SSet_{def}} I(A; B|\SSet^* \cup \mathcal{Q})\,. \label{eq:effect_size_tmp3}
\end{align}
This implies eq.~\eqref{eq:effect_size_tmp2} because the left hand side of eq.~\eqref{eq:effect_size_tmp3} equals the left hand side of eq.~\eqref{eq:effect_size_tmp2} by definition of $\mathbf{S}$ and the right hand side of eq.~\eqref{eq:effect_size_tmp3} is greater or equal than the right hand side of eq.~\eqref{eq:effect_size_tmp2} because of the additional minimum operation in eq.~\eqref{eq:effect_size_tmp2}. By subtracting the left hand side of this inequality we get
\begin{align}
 \min_{\mathcal{Q} \subset \SSet_{def}, \, \mathcal{Q}\neq \SSet_{def}} \left[ I(A; B|\SSet^* \cup \mathcal{Q}) - I(A; B|\SSet^* \cup \mathcal{Q} \cup (\SSet_{def}\setminus \mathcal{Q}))\right] &< 0\,. \label{eq:effect_size_tmp4}
\end{align}
A difference of conditional mutual informations as in this equation defines a trivariate (conditional) interaction information $\mathcal{I}$ \cite{Abramson1963,Runge2015}, such that we can rewrite eq.~\eqref{eq:effect_size_tmp4} as
\begin{align}
 \min_{\mathcal{Q}\subset \SSet_{def}, \, \mathcal{Q}\neq \SSet_{def}} \mathcal{I}(A; B; \SSet_{def}\setminus \mathcal{Q}|\SSet^* \cup \mathcal{Q}) &< 0\,. \label{eq:effect_size_tmp5}
\end{align}
Contrary to conditional mutual information, the (conditional) interaction information can also attain negative values. This happens when an additional condition, here $\SSet_{def}\setminus \mathcal{Q}$, increases the conditional mutual information between $A$ and $B$. The second assumption of the theorem states that there is a proper subset $\mathcal{Q} \subset \SSet_{def}$ for which $\mathcal{I}(A; B; \SSet_{def}\setminus \mathcal{Q}|\SSet^* \cup \mathcal{Q}) < 0$. This implies eq.~\eqref{eq:effect_size_tmp5} and hence the main equation \eqref{eq:effect_size}.
\hfill $\square$

\begin{figure*}[t]
\centering
\includegraphics[width=.8\linewidth]{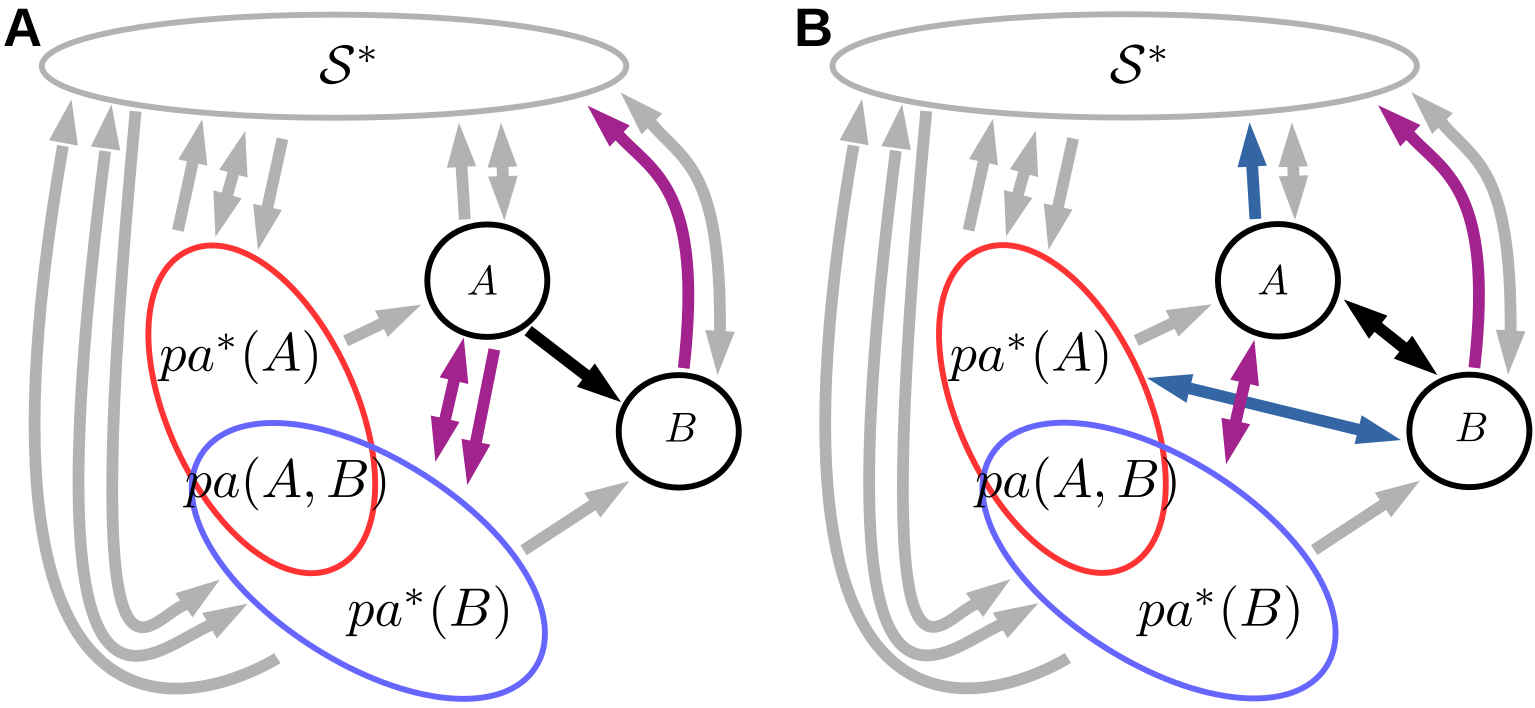}
\caption{Graph illustrating the two general cases of dependencies between $A = X^i_{t - \tau}$ and $B= X^j_t$ for proving Corollary~\ref{thm:effect_size_corollary}, namely (\textbf{A}) $A \to B$ and (\textbf{B}) $A \leftrightarrow B$. The multiple connections are to be understood between subsets of the respective sets such that the whole graph is still a MAG, i.e., that no (almost) directed cycles occur and that maximality is not violated. We omit the links within each subset. $\SSet^* \subseteq \Xobs \setminus \{\pa(\{A, B\}, \M), A, B\} $ denotes the conditions that make the LPCMCI effect size minimal. In panel (\textbf{A}) magenta connections are excluded by the assumptions of Corollary~\ref{thm:effect_size_corollary}, in panel (\textbf{B}) at least all magenta or all blue connections are excluded (they may both be excluded). These exclusions are, however, not sufficient to guarantee the assumptions of Corollary~\ref{thm:effect_size_corollary}.
}
\label{fig:proof}
\end{figure*}

We now state a Corollary of Theorem~\ref{thm:effect_size}, which details graphical assumptions that lead to an increase in effect size as required by eq.~\eqref{eq:effect_size_tmp5}. Fig.~\ref{fig:proof} illustrates these graphical criteria.
\begin{mycorsm}[LPCMCI effect size] \label{thm:effect_size_corollary}
Let $A \asthead B$ (with $A = X^i_{t-\tau}$ and $B = X^j_t$) be a link ($\tailhead$ or $\headhead$) in $\M$. Consider the default conditions $\SSet_{def} =\pa(\{A, B\}, \M) \setminus \{A, B\}$ and denote $\Xobs^*=\Xobs \setminus \SSet_{def}$. Let $\mathbf{S}=\arg\min_{\SSet\subseteq \Xobs^* \setminus \{A, B\}} I(A; B|\SSet \cup \SSet_{def})$ be the set of sets that define LPCMCI's effect size.

\textbf{1.)} Assume the link is of the form $A \tailhead B$. If $i)$ $\pa^*(B, \M) = \SSet_{def} \setminus \pa(A, \M)$ is non-empty (in words: $B$ has parents other than $A$ that are not at the same time also parents of $A$), and $ii)$ there is $\SSet^* \in \mathbf{S}$ with $\SSet^* \subseteq \adj(A, \M)\setminus \SSet_{def}$ or $\SSet^* \subseteq \adj(B, \M)\setminus \SSet_{def}$, and $iii)$ $B \notin \an(\SSet^*, \M)$, and $iv)$ there is no path between $A$ and $\pa^*(B, \M)$ that is active given $\pa(A, \M) \cup \SSet^*$, and $v)$ faithfulness holds, then
\begin{align} \label{eq:effect_size_corollary}
\min_{\SSet\subseteq \Xobs^* \setminus \{A, B\}} I(A; B|\SSet \cup \SSet_{def}) &> \min_{\tilde{\SSet}\subseteq \Xobs\setminus \{A, B\}} I(A; B|\tilde{\SSet})\,.
\end{align}

\textbf{2.)} Assume the links is of the form $A \headhead B$. The same inequality~\eqref{eq:effect_size_corollary} holds if the same assumptions $i)-v)$ as stated in 1.) hold or if these assumptions hold with the roles of $B$ and $A$ exchanged.

\textbf{3.)} If neither the assumptions of 1.) nor of 2.) are fulfilled, then (trivially) "$\geq$" holds in~\eqref{eq:effect_size_corollary}.
\end{mycorsm}

%
\textbf{Proof of Corollary ~\ref{thm:effect_size_corollary}.} \label{sec:effect_size_proof_cor}
Note that eq.~\eqref{eq:effect_size_corollary} and eq.~\eqref{eq:effect_size} are the same. All manipulations that have identified eq.~\eqref{eq:effect_size_tmp5} as a sufficient condition for eq.~\eqref{eq:effect_size} under the assumptions of Theorem \ref{thm:effect_size} are still valid under the assumptions of the corollary. Therefore, eq.~\eqref{eq:effect_size_tmp5} is what remains to be shown.

Since the interaction information is symmetric in its arguments before the ``$|$'', eq.~\eqref{eq:effect_size_tmp5} can be cast into the equivalent conditions:
\begin{align}
\min_{\mathcal{Q}\subset \SSet_{def}, \, \mathcal{Q}\neq \SSet_{def}} \mathcal{I}(A; B; \SSet_{def} \setminus \mathcal{Q}|\SSet^* \cup \mathcal{Q}) &< 0\\
\Leftrightarrow \min_{\mathcal{Q}\subset \SSet_{def}, \, \mathcal{Q}\neq \SSet_{def}} \left[ I(A; B|\SSet^* \cup \mathcal{Q}) - I(A; B|\SSet^*\cup \mathcal{Q}\cup (\SSet_{def} \setminus \mathcal{Q}))\right] &< 0 \\
\Leftrightarrow  \min_{\mathcal{Q}\subset \SSet_{def}, \, \mathcal{Q}\neq \SSet_{def}} \left[ I(A; \SSet_{def} \setminus \mathcal{Q}|\SSet^*\cup\mathcal{Q}) - I(A; \SSet_{def} \setminus \mathcal{Q}|\SSet^*\cup \mathcal{Q}\cup \{B\})\right] &< 0 \label{eq:effect_size_tmp6} \\
\Leftrightarrow  \min_{\mathcal{Q}\subset \SSet_{def}, \, \mathcal{Q}\neq \SSet_{def}} \left[ I(\SSet_{def} \setminus \mathcal{Q}; B|\SSet^* \cup \mathcal{Q}) - I(\SSet_{def} \setminus \mathcal{Q}; B|\SSet^* \cup \mathcal{Q} \cup \{A\})\right] &< 0 \label{eq:effect_size_tmp7}\,.
\end{align}
First consider the case $A\tailhead B$ in conjunction with eq.~\eqref{eq:effect_size_tmp6}. Independent of which $\mathcal{Q}$ minimizes the left hand side of this equation, a sufficient condition for its validity is the existence of a proper subset $\mathcal{Q}\subset \SSet_{def}$ for which the following two conditions hold:
\begin{align}
I(A; \SSet_{def} \setminus \mathcal{Q}|\SSet^* \cup \mathcal{Q}) &= 0 ~~\Longleftrightarrow ~~ A \ci \SSet_{def} \setminus \mathcal{Q}|\SSet^* \cup \mathcal{Q} \label{eq:effect_size_tmp8}\ ,\\ 
I(A; \SSet_{def} \setminus \mathcal{Q}|\SSet^*\cup\mathcal{Q}\cup\{B\}) &> 0 ~~\Longleftrightarrow~~ A ~\cancel{\ci}~ \SSet_{def} \setminus \mathcal{Q}|\SSet^*\cup\mathcal{Q}\cup\{B\}\label{eq:effect_size_tmp9} \,.
\end{align}
We choose $\mathcal{Q}=pa(A, \M)$ and hence get $\SSet_{def}\setminus \mathcal{Q}=pa^*(B, \M)$. Since by assumption $i)$ $pa^*(B, \M) = \SSet_{def}\setminus \mathcal{Q}$ is not empty, $\mathcal{Q}=pa(A, \M)$ is indeed a proper subset of $\SSet_{def}$. Further, eq.~\eqref{eq:effect_size_tmp8} is true by assumption $iv)$ and eq.~\eqref{eq:effect_size_tmp9} is true by the assumption of faithfulness together with the fact that the path $A \tailhead B \headtail pa^*(B, \M)$ is active given $\SSet^*\cup pa(A, \M) \cup \{B\}$. Since both conditions in eq.~\eqref{eq:effect_size_tmp8} and eq.~\eqref{eq:effect_size_tmp9} hold for this valid choice of $\mathcal{Q}$, part 1.) of the corollary is proven.

We note that assumption $iii)$ is needed: Otherwise conditioning on $\SSet^*$ opens the path $A \tailhead B \headtail pa^*(B, \M)$ since $B$ is an ancestor of a conditioned node, thus assumption $iv)$ could not be true. Assumption $iii)$ would be violated by the magenta connections shown in Fig.~\ref{fig:proof}.

In the case $A \headhead B$ we can either utilize eq.~\eqref{eq:effect_size_tmp6} or eq.~\eqref{eq:effect_size_tmp7}, depending on whether $B$ or $A$ (or both) contain non-empty non-shared parents for which eq.~\eqref{eq:effect_size_tmp8} and eq.~\eqref{eq:effect_size_tmp9} or the equivalent assumptions with $B$ and $A$ exchanged hold. Lastly, the case $A \headtail B$ is covered by part 1.) of this corollary with $B$ and $A$ exchanged. This proves part 2.) of the corollary.

Part 3.) follows because the minimum on the right hand side of eq.~\eqref{eq:effect_size_corollary} is taken over a superset of the set that the minimum on the left hand side is taken over.
\hfill $\square$

\begin{mylemmasm}[Inclusion of ancestors in separating sets]\label{lemma:conditiononparents}
Let $A$ and $B$ be m-separated given $\SSet$, and let $\SSet_{def} \subseteq \an(\{A,B\}, \M)\setminus \{A, B\}$ be arbitrary. Then, $A$ and $B$ are also m-separated given $\SSet^\prime = \SSet \cup \SSet_{def}$.
\end{mylemmasm}

\textbf{Proof of Lemma \ref{lemma:conditiononparents}.}
Assume without loss of generality that $\SSet_{def}$ is non-empty, else the statement is trivial. First, consider the case $\SSet_{def} \subseteq \an(B, \M)$ and assume $\SSet^\prime$ did not m-separate $A$ and $B$. This requires the existence of a path $p$ between $A$ and $B$ for which $a1)$ at least one non-collider on $p$ is in $\SSet$ or $a2)$ there is a collider on $p$ that is not an ancestor of $\SSet$, $b)$ none of the non-colliders on $p$ is in $\SSet^\prime$, and $c)$ all colliders on $p$ are ancestors of $\SSet^\prime$. Since $\SSet$ is a proper subset of $\SSet^\prime$, $a1)$ conflicts with $b)$. This means $a2)$ must be true, i.e, there is at least one collider on $p$ that is an ancestor of $\SSet^\prime \setminus \SSet = \SSet_{def} \setminus \SSet \subseteq \an(B, \M)$ and hence of $B$. Among all those colliders, let $C$ be the one closest to $A$ on $p$. According to $b)$ the sub-path $p_{AC}$ of $p$ from $A$ to $C$ is then active given $\SSet$ by construction. Since $C$ is an ancestor of $B$ there is at least one directed path $p_{CB}$ from $C$ to $B$. By definition of $C$ the path $p_{CB}$ does not cross any node in $\SSet$. Thus, $p_{CB}$ is active given $\SSet$.

We now construct a path from $A$ to $B$ that is active given $\SSet$, thereby reaching a contradiction. To this end, let $D$ be the node closest to $A$ on $p_{AC}$ that is also on $p_{CB}$ (such $D$ always exists, because $C$ is on both paths). Consider then the subpath $p_{AD}$ of $p_{AC}$ from $A$ to $D$, and the subpath $p_{DB}$ on $p_{CB}$ from $D$ to $B$. Since $p_{AC}$ and $p_{CB}$ are active given $\SSet$, also $p_{AD}$ and $p_{DB}$ are active given $\SSet$. By definition of $D$ the concatenation of $p_{AD}$ and $p_{DB}$ at their common end $D$ gives a path $p_{AB}$ from $A$ to $B$. Since $D$ is a non-collider on $p_{AB}$ (because $p_{DB}$ is out of $D$) and $D$ is not in $\SSet$ (because else $C$ would be an ancestor of $\SSet$), $p_{AB}$ is active given $\SSet$. Contradiction.

Second, since the Lemma does not make any distinction between $A$ and $B$, it is also true in case $\SSet_{def} \subseteq \an(A, \M)$. Third, write $\SSet = \SSet_{A} \dot{\cup}\, \SSet_{B}$ with $\SSet_{A} = \SSet \cap \an(A, \M)$ and $\SSet_B = \SSet \setminus \SSet_A \subseteq \an(B, \M)$. The statement then follows from applying the already proven special cases twice.
\hfill $\square$

\begin{mylemmasm}[Exclusion of non-ancestors and future from separating sets]\label{lemma:notconditiononfuture}
Let $A$ and $B$ be m-separated given $\SSet$, and let $\mathcal{U}$ be such that $\mathcal{U} \cap \an(\{A, B, \SSet \setminus \mathcal{U}\}, \M) = \emptyset$. Then, $A$ and $B$ are also m-separated given $\SSet^\prime = \SSet \setminus \mathcal{U}$. Two important special cases are: \textbf{Special case 1.)} $\mathcal{U} = \SSep \setminus \an(\{A, B\}, \M)$, which allows to restrict separating sets to ancestors. \textbf{Special case 2.)} $\mathcal{U} = \{\text{all nodes that are in the future of both $A$ and $B$}\}$, which allows to restrict separating sets to the present and past of the later variable.
\end{mylemmasm}

\textbf{Proof of Lemma \ref{lemma:notconditiononfuture}.}
Assume without loss of generality that $\mathcal{U}$ is non-empty, else the statement is trivial. Assume $\SSet^\prime$ did not m-separated $A$ and $B$. This requires the existence of a path $p$ between $A$ and $B$ for which $a1)$ at least one non-collider on $p$ is in $\SSet$ or $a2)$ there is a collider on $p$ that is not an ancestor of $\SSet$, $b)$ none of the non-colliders on $p$ is in $\SSet^\prime$, and $c)$ all colliders on $p$ are ancestors of $\SSet^\prime$. Since $\SSet^\prime$ is a proper subset of $\SSet$, $a2)$ conflicts with $c)$. This means $a1)$ must be true, i.e., there is a non-collider $D$ on $p$ in $\SSet \setminus \SSet^\prime = \SSet \cap \mathcal{U}$. In particular, $D$ is in $\mathcal{U}$. All nodes on $p$ are ancestors of $A$ or $B$ or of a collider on $p$. If $D$ is an ancestor of a collider on $p$, then by $c)$ it is also an ancestor of $\SSet^\prime = \SSet \setminus \mathcal{U}$. This shows that $D$ is also in $\an(\{A, B, \mathcal{S} \setminus \mathcal{U}\}, \M)$. Since $\mathcal{U} \cap \an(\{A, B, \mathcal{S} \setminus \mathcal{U}\}, \M) = \emptyset$, this is a contradiction.

\textbf{Special case 1.)} For $\mathcal{U} = \SSep \setminus \an(\{A, B\}, \M)$ we have $\SSet^\prime = \SSep \cap \an(\{A, B\}, \M)$ and $\an(\{A, B, \mathcal{S} \setminus \mathcal{U}\}, \M) = \an(\{A, B\}, \M)$. Hence, the condition is fulfilled. \textbf{Special case 2.)} For $\mathcal{U} = \{\text{all nodes that are in the future of both $A$ and $B$}\}$ we have $\an(\{A, B, \mathcal{S} \setminus \mathcal{U}\}, \M)  \subseteq \{\text{all nodes that are not after both $A$ and $B$}\}$. Hence, the condition is fulfilled.

Note that if $\mathcal{U}$ fulfills the above condition, a proper subset $\mathcal{U}^\prime$ of $\mathcal{U}$ does not necessarily fulfill the condition as well. Consider the example $A \tailhead C \headtail D \headtail B$. Here $\SSet = \{C, D\}$ m-separates $A$ and $B$, and $\mathcal{U} = \SSet$ fulfills the condition. However, $\mathcal{U}^\prime = \{C\}$ does not. This is why we need to require $\mathcal{U} \cap \an(\{A, B, \SSet \setminus \mathcal{U}\}, \M) = \emptyset$ and not just $\mathcal{U} \cap \an(\{A, B\}, \M) = \emptyset$.
\hfill $\square$

\begin{mylemmasm}[Some properties of {\DSEP} sets]\label{lemma:dsepsets}
Consider two distinct nodes $A, B \in \M$. Let $V \in \DSEP(B, A, \M)$ and path $p_V$ be as in Definition \ref{def:dsep}, and denote with $C \neq B$ the node on $p_V$ that is closest to $B$.
\textbf{1.)} If $A \notin \adj(B, \M)$, then $p_V$ does not contain $A$.
\textbf{2.)} If $B \notin \an(A, \G)$ and $p_V$ contains two nodes only, then $C$ is a parent or spouse of $B$.
\textbf{3.)} If $B \notin \an(A, \G)$ and $p_V$ contains more than two nodes, then $C$ is a spouse of $B$ and ancestor of $A$
\textbf{4.)} If $A \in \an(B, \G)$, then $\DSEP(B, A, \M) = \pa(B, \M)$.
\end{mylemmasm}

\textbf{Proof of Lemma \ref{lemma:dsepsets}.}
\textbf{1.)} Assume $p_V$ did contain $A$. The subpath of $p_V$ from $B$ to $A$ is then an inducing path between $B$ and $A$. Since $A$ and $B$ are not adjacent, this violates maximality of the MAG $M$.
\textbf{2.)} By construction $V = C$ is adjacent to $B$. Assume $C$ was a child of $B$. Since $C$ must be an ancestor of $A$ or $B$, $C$ must be an ancestor of $A$. Then $B$ is an ancestor of $A$, contrary to the assumption.
\textbf{3.)} According to the second part $C$ is a parent or spouse of $B$. If $C$ was a parent of $B$, $C$ would be a non-collider on $p_V$. This contradicts the definition of $p_V$, hence $C$ is a spouse of $B$. Moreover, since $C$ is an ancestor of $A$ or $B$, $C$ is an ancestor of $A$.
\textbf{4.)} The inclusion $\DSEP(B, A, \M) \supseteq \pa(B, \M)$ follows since if $V$ is a parent of $B$ then $V \tailhead B$ is a path $p_V$ as required by the definition. We now show the opposite inclusion $\DSEP(B, A, \M) \subseteq \pa(B, \M)$ by showing that $V \in \pa(B, \M)$. Case 1: $p_V$ has two nodes only. By the second part of this Lemma $V$ is a parent or spouse of $B$. Assume it was a spouse. Then $V$ must be an ancestor of $A$, which with $A \in \an(B, \G)$ gives $C \in \an(B, \G)$. But then $C$ cannot be a spouse of $B$. Case 2: $p_V$ has more than three nodes. By the third part of this Lemma we then get that $C$ is an ancestor of $A$, which agains leads to the contradiction $C \in \an(B, \G)$.
\hfill $\square$

\textbf{Proof of Lemma \ref{lemma:apdsnapds}.}
\textbf{1.)}
$A \in \an(B, \G)$ gives $\DSEP(B, A , \M) = \pa(B, \M)$ by part 4.) of Lemma \ref{lemma:dsepsets}. Consider $C \in pa(B, \M)$. Then, $C$ is adjacent to $B$ in $\C$ with a link that does not have a head at $C$. Moreover, $C$ cannot be after $B$. Since $A$ and $B$ are not adjacent, $C$ cannot be $A$. Hence $C$ in $\apds(B, A, \C)$.
\textbf{2.)}
Consider $V \in \DSEP(B, A, \M)$ and let the path $p_V$ be as in Definition \ref{def:dsep}. Case 1: $V$ is a parent of $B$ in $\C$. Then, as the proof of the first part of this Lemma shows, $V \in \apds(B, A, \C)$. Now assume $V$ was a child of $A$ in $\C$. Then $A \in \an(B, \G)$, contradicting the assumption. Hence $V \in \napds^1(B, A, \C)$. Case 2: $V$ is not a parent of $B$ in $\C$. We now show that $p_V$ is a path $p$ as required in $3.)$ of Definition \ref{def:napds} and hence $V \in \napds^2(B, A, \C)$. Let $C$ be the node on $p_V$ that is closest to $B$, which by 2.) and 3.) of Lemma \ref{lemma:dsepsets} is a spouse of $B$. $i)$ is true since all non end-point nodes on $p_V$ are colliders on $p_V$ together with the fact that $C$ is a spouse of $B$. $ii)$ is true for the same reason as $i)$ together with the fact that rule $\ER0a$ has been exhaustively applied, which guarantees that if an unshielded triple is a collider then it will be oriented as a collider. $iii)$ is true by 1.) of Lemma \ref{lemma:dsepsets}. $iv)$ is true since $C$ is an ancestor of $A$ by 3.) of Lemma \ref{lemma:dsepsets}. The second and third part of $v)$ are true since all nodes on $p_V$ are ancestors of $A$ or $B$. For the first part of $v)$ observe that if $V$ is a descendant of $A$ (or $B$) in $\C$, then since $V$ is an ancestor of $A$ or $B$ we would get $A \in \an(B, \G)$ (or $B \in \an(A, \G)$), a contradiction.
\hfill $\square$

\begin{mydefsm}[Weakly minimal separating sets of the second type] \label{def:weaklyminimal2}
In MAG $\M$ let $A$ and $B$ be m-separated by $\SSet$. The set $\SSet$ is a weakly minimal separating set of $A$ and $B$ of the second type if $i)$ there is a decomposition $\SSet = \SSet_1 \dot{\cup}\, \SSet_2$ with $\SSet_1 \subseteq \an(\{A, B\}, \M)$ such that $ii)$ if there is $S \in \SSet_2$ such that $\SSet^\prime = \SSet \setminus S$ is a separating set of $A$ and $B$ then $S \in \an(\{A, B\}, \M)$. The pair $(\SSet_1, \SSet_2)$ is called a weakly minimal decomposition of $\SSet$ of the second type.
\end{mydefsm}

\begin{mylemmasm}[Selected properties of weakly minimal separating sets] \label{lemma:weaklyminimal}
$\textbf{1.)}$
$\SSep$ is a weakly minimal separating set of the second type if and only if its canonical decomposition $(\mathcal{T}_1, \mathcal{T}_2)$ defined by $\mathcal{T}_1 = \SSet \cap \an(\{A, B\}, \M)$ and $\mathcal{T}_2 = \SSet \setminus{\mathcal{T}_1}$ is a weakly minimal decomposition of $\SSet$ of the second type.
$\textbf{2.)}$
If $\SSet$ is a weakly minimal separating set of $A$ and $B$ of the second type then $\SSet \subseteq \an(\{A, B\}, \M) \subseteq \an(\{A, B\}, \G)$.
$\textbf{3.)}$
$\SSet$ is a weakly minimal separating set of the second type if and only if it is a weakly minimal separating set.
$\textbf{4.)}$
$\SSet$ is a weakly minimal separating set of $A$ and $B$ if and only if it is a separating set of $A$ and $B$ and $\SSet \subseteq \an(\{A, B\}, \M) \subseteq \an(\{A, B\}, \G)$.
$\textbf{5.)}$
If $\SSet$ is a non-weakly minimal separating set of $A$ and $B$ then there is a proper subset $\SSet^\prime$ of $\SSet$ that is a weakly minimal separating set of $A$ and $B$.
\end{mylemmasm}

\textbf{Proof of Lemma \ref{lemma:weaklyminimal}.}
$\textbf{1.) if}$:
The existence of a weakly minimal decomposition of the second type implies weak minimality of the second type.
$\textbf{1.) only if:}$
By assumption there is some weakly minimal decomposition $(\SSet_1, \SSet_2)$ of the second type. By definition of the canonical decomposition and by condition $i)$ in Definition \ref{def:weaklyminimal2} the inclusions $\SSet_1 \subseteq \mathcal{T}_1$ and hence $\SSet_2 \supseteq \mathcal{T}_2$ hold. Assume the canonical decomposition were not a weakly minimal decomposition of $\SSet$ of the second type. Then there is some $S \in \mathcal{T}_2$ such that $\SSet^\prime = \SSet \setminus S$ is a separating set. Since $\SSet_2 \supseteq \mathcal{T}_2$ then also $S \in \SSet_2$, contradicting the assumption that $(\SSet_1, \SSet_2)$ is a weakly minimal decomposition of the second type.
$\textbf{2.)}$
Since $\SSep$ is weakly minimal of the second type, its canonical decomposition $(\mathcal{T}_1, \mathcal{T}_2)$ is a weakly minimal decomposition of $\SSep$ of the second type. We now show that $\mathcal{T}_2$ must be empty. Assume it was not and let $C_1, \ldots, C_n$ be its elements. Since by construction $C_1 \notin \an(\{A, B\}, \M)$ and since $(\mathcal{T}_1, \mathcal{T}_2)$ is weakly minimal decomposition of the second type, $A$ and $B$ are not m-separated by $\SSet^\prime = \SSet \setminus C_1$. This means there is a path $p$ that is active given $\SSet^\prime$ and blocked given $\SSet$. Hence, $C_1$ must be a non-collider on $p$. Together with $C_1 \notin \an(\{A, B\}, \M)$ this shows that $C_1$ is ancestor of some collider $D_1$ on $p$, which itself is an ancestor of $\SSet^\prime$ (else $p$ would not be active given $\SSet^\prime$). Hence, $C_1$ is an ancestor $\SSet^\prime$. Since $C_1 \notin \an(\{A, B\}, \M)$ and $\mathcal{T}_1 \subseteq \an(\{A, B\}, \M)$ , $C_1$ is an ancestor of  $\{C_2, ..., C_n\}$. If $n = 1$, this is a contradiction already. If $n > 1$ we may without loss of generality assume that $C_1$ is an ancestor of $C_2$. Hence, $C_2$ is not an ancestor of $C_1$. By applying the same argument to $C_2$, we conclude that $C_2$ is an ancestor of $\{C_3, ..., C_n\}$. Repeat this until reaching a contradiction. This shows $\mathcal{T}_2 = \emptyset$ and hence $\SSet \subseteq \an(\{A, B\}, \M) \subseteq \an(\{A, B\}, \G)$.
$\textbf{3.) if}$:
Condition $ii)$ in Definition \ref{def:weaklyminimal} is clearly stronger than $ii)$ in Definition \ref{def:weaklyminimal2}.
$\textbf{3.) only if}$:
Let $\SSet$ be a weakly minimal separating set of the second type, for which by part 2.) of this Lemma $\SSet \subseteq \an(\{A, B\}, \M)$. Therefore, $(\SSep, \emptyset)$ is a weakly minimal decomposition of $\SSet$, showing that $\SSet$ is weakly minimal.
$\textbf{4.) if}$:
$(\SSep, \emptyset)$ is a weakly minimal decomposition
$\textbf{4.) only if}$:
This follows from parts 2.) and 3.) of this Lemma.
$\textbf{5.)}$
According to (the first special case of) Lemma \ref{lemma:notconditiononfuture} $\SSet^\prime = \SSet \cap \an(\{A,B\}, \M)$ is a separating set. This $\SSet^\prime$ is weakly minimal according to part 4.) of this Lemma.
\hfill $\square$

\begin{customlemma}{1}[Ancestor-parent-rule]
In LPCMCI-PAG $\C$ one may replace $\textbf{1.)}$ $A \EMMnoast{\tailhead} B$ by $A \tailhead B$, $\textbf{2.)}$ $A \LMMnoast{\tailhead} B$ for $A > B$ by $A \tailhead B$, and $\textbf{3.)}$  $A \RMMnoast{\tailhead} B$ for $A < B$ by $A \tailhead B$.
\end{customlemma}

\textbf{Proof of Lemma \ref{lemma:middlemarks}.}
\textbf{2.)} By the fourth point in Definition \ref{def:LPCMCI_PAG}, $A \notin \an(B, \G)$ or there is no $\SSet \subseteq \pa(B, \M)$ that m-separates $A$ and $B$ in $\M$. The first option contradicts $A \LMMnoast{\tailhead} B$, so the second option must be true. Since $A \in \an(B, \G)$ gives $\DSEP(B, A, \M) = \pa(B, \M)$ according to part 4.) of Lemma \ref{lemma:dsepsets}, Proposition \ref{thm:dsep} then implies that $A$ and $B$ are not m-separated by any set. \textbf{3.)} Equivalent proof. \textbf{1.)} Recall that if $A \EMM{\astast} B$ in $\C$, then both $A \LMM{\astast} B$ and $A \RMM{\astast} B$ would be correct. The statement then follows since either 2.) or 3.) of this Lemma applies.
\hfill $\square$

\begin{customlemma}{2}[Strong unshielded triple rule]
Let $A \SMM{\astast} B \SMM{\astast} C$ be an unshielded triple in LPCMCI-PAG $\C$ and $\SSet_{AC}$ the separating set of $A$ and $C$.
$\textbf{1.)}$
If $i)$ $B \in \SSet_{AC}$ and $ii)$ $\SSet_{AC}$ is weakly minimal, then $B \in \an(\{A, C\}, \G)$.
$\textbf{2.)}$
Let $\mathcal{T}_{AB} \subseteq \an(\{A, B\}, \M)$ and $\mathcal{T}_{CB} \subseteq \an(\{C, B\}, \M)$ be arbitrary. If $i)$ $B \notin \SSet_{AC}$, $ii)$ $A$ and $B$ are not m-separated by $\SSet_{AC} \cup \mathcal{T}_{AB} \setminus \{A, B\}$, $iii)$ $C$ and $B$ are not m-separated by $\SSet_{AC} \cup \mathcal{T}_{CB} \setminus \{C, B\}$, then $B \notin \an(\{A, C\}, \G)$. The conditioning sets in $ii)$ and $iii)$ may be intersected with the past and present of the later variable.
\end{customlemma}

\textbf{Proof of Lemma \ref{lemma:strongutr}.}
\textbf{1.)} This follows immediately from part 4.) of Lemma \ref{lemma:weaklyminimal}. \textbf{2.)} By the contraposition of Lemma \ref{lemma:conditiononparents} condition $ii)$ implies that $A$ and $B$ are not m-separated by $\SSet_{AC}$, and similarly $iii)$ implies the same for $C$ and $B$. The additional claims made in the last sentence of the Lemma follow by the contraposition of Lemma \ref{lemma:notconditiononfuture}. The statement then follows from Lemma 3.1 in \cite{Colombo2012}. Although there minimality of $\SSep_{AC}$ is stated as an additional assumption, the proof given in the supplement to \cite{Colombo2012} does not use this assumption.
\hfill $\square$

\textbf{Proof of the orientation rules given in subsection \ref{sec:orientationrules}:}
Whenever neither a rule consequent nor the hypothetical manipulations involved in its proof require that a certain edge mark be oriented as head or tail, the rule also applies when that edge mark is the conflict mark `x'. This explains the use of `$\ast$' vs. `$\star$' marks in the rule antecedents. We repeat that if $X \SMM{\astast} Y \SMM{\astast} Z$ is an unshielded triple and a rule requires $Y \in \SSet_{XZ}$ with $\SSet_{XZ}$ weakly minimal, the requirement of weak minimality may be dropped if $X \astast Y \astast Z$. This is true since when $X \astast Y \astast Z$ we can conclude $Y \in \an(\{X, Z\}, \G)$ from $Y \in \SSet_{XZ}$ even if $\SSet_{XZ}$ is not weakly minimal.

\underline{$\mathbf{\ER{0}a}$}:
This follows from the second part Lemma \ref{lemma:strongutr}. Requirement $iii)$ is irrelevant in the case of perfect statistical decisions, it will then never be true given that $ia)$ or $ib)$ and $iia)$ or $iib)$ are true. 

\underline{$\mathbf{\ER{0}b}$}:
Assume $B \in \an(C, \G)$ were true. By Lemma \ref{lemma:middlemarks} then $B \tailhead C$ in $\C$ and hence in $\M$. Since one of $ia)$ or $iia)$ is true by assumption, there is a path $p_{AB}$ from $A$ to $B$ that is active given $[\SSet_{AC} \cup \pa(\{A,B\}, \C)] \setminus \{A, B, \text{nodes in the future of both $A$ and $B$}\}$. Due to Lemmas \ref{lemma:conditiononparents} and \ref{lemma:notconditiononfuture} and since $B \notin \SSet_{AC}$, $p_{AB}$ is also active given $\SSet_{AC}$. Since then every subpath of $p_{AB}$ is active given $\SSet_{AC}$ and since $\SSet_{AC}$ is a separating set of $A$ and $C$, $C$ cannot be on $p_{AB}$. When appending the edge $B \tailhead C$ to $p_{AB}$ we hence obtain a path $p_{AC}$. Since $B$ is a non-collider on $p_{AC}$ and $B \notin \SSet_{AC}$, $p_{AC}$ is active given $\SSet_{AC}$. Contradiction. Hence $B \notin \an(C, \G)$.

\underline{$\mathbf{\ER{0}c}$}:
Assume $B \in \an(C, \G)$ were true. By Lemma \ref{lemma:middlemarks} then $B \tailhead C$ in $\C$ and hence in $\M$. Moreover $A \headtail B$ or $A \headhead B$ or $A \tailhead B$ in $\M$ by assumption. In either case $A$, $B$ and $C$ form an unshielded triple in $\M$ with its middle node $B$ not being a collider. But then $B \in \SSet_{AC}$. Contradiction. Hence $B \notin \an(C, \G)$.

\underline{$\mathbf{\ER{0}d}$}:
Since all involved middle marks are empty this is just the standard FCI rule $\R0$.

\underline{$\mathbf{\ER{1}}$}:
From the first part of Lemma \ref{lemma:strongutr} we get $B \in \an(\{A, C\}, \G)$. Due to the head at $B$ on its edge with $A$ we know $B \notin \an(A, \G)$. Hence $B \in \an(C, \G)$.

\underline{$\mathbf{\ER{2}}$}:
Assume $C \in \an(A, \G)$ were true. Case 1: $A \SMMnoast{\tailhead} B \SMM{\asthead} C$. Due to transitivity of ancestorship then also $C \in \an(B, \G)$. This contradicts the head at $C$ on its edge with $B$. Case 2: $A \SMM{\asthead} B \SMMnoast{\tailhead} C$. Then $B \in \an(A , \G)$, contradicting the head at $B$ on its link with $A$. Hence $C \notin \an(A, \G)$.

\underline{$\mathbf{\ER{3}}$}:
Assume $B \in \an(D, \G)$ were true. By applying the first part of Lemma \ref{lemma:strongutr} to the unshielded triple $A \SMM{\staro} D \SMM{\ostar} C$ we deduce that $D \in \an(\{A, C\}, \G)$. Thus $B \in \an(\{A, C\}, \G)$, contradicting at least one of the heads at $B$ in the triple $A \SMM{\asthead} B \SMM{\headast} C$. Hence $B \notin \an(D, \G)$.

\underline{$\mathbf{\ER{4}}$}:
This follows from Lemma 3.2 in \cite{Colombo2012} together with $i)$ the contrapositions of Lemmas \ref{lemma:conditiononparents} and \ref{lemma:notconditiononfuture}, and $ii)$ that a pair of variables which in $\C$ is connected by an edge with empty middle mark then this pair of variables is also adjacent in $\M$.

\underline{$\mathbf{\ER{8}}$}:
Transitivity of ancestorship gives $A \in \an(C, \G)$, hence also $C \notin \an(A, \G)$.

\underline{$\mathbf{\ER{9}}$}:
Assume $A_{1} \notin \an(A_n, \G)$ were true, such that $A_1 \SMMnoast{\headhead} A_{n}$. From $ia)$ or from $ib)$ for $k = 1$ together with the first part of Lemma \ref{lemma:strongutr} applied to the unshielded triple $A_{n} \equiv A_0 \SMMnoast{\headhead} A_1 \SMM{\astast} A_2$ we then conclude $A_{1} \SMMnoast{\tailhead} A_2$. By successive application of Lemma \ref{lemma:strongutr} to the unshielded triple $A_{k-1} \SMMnoast{\tailhead} A_{k} \SMM{\astast} A_{k+1}$ together with $ia)$ or from $ib)$ for $k = 2, \ldots, n-1$ (in this order) we further conclude $A_{k} \SMMnoast{\tailhead} A_{k+1}$. This gives $A_1 \in \an(A_{n}, \G)$, a contradiction to the assumption. Hence $A_{1} \in \an(A_n, \G)$.

\underline{$\mathbf{\ER{10}}$}:
Application of the first part of Lemma \ref{lemma:strongutr} to the unshielded triple $B_1 \SMM{\astast} A \SMM{\astast} C_1$ gives $A \in \an(\{B_1, C_1\}, \G)$. Say, without loss of generality, $A \in \an(B_1, \G)$. By successive application of Lemma \ref{lemma:strongutr} to the unshielded triple $B_{k} \SMMnoast{\tailhead} B_{k+1} \SMM{\astast} B_{k+2}$ together with $ia)$ or from $ib)$ for $k = 0, \ldots, n-2$ (in this order) we further conclude $B_{k+1} \SMMnoast{\tailhead} B_{k+2}$. This shows that $A \in \an(D, \G)$.

\underline{\textbf{APR}:}
These are the replacements specified in Lemma \ref{lemma:middlemarks}, which was already proven above.

\underline{\textbf{MMR}:}
This follows immediately from the causal meaning of middle marks `L', `R', and `!' given in Definition \ref{def:LPCMCI_PAG}.
\hfill $\square$

\begin{mylemmasm}[Symbolic middle mark update] \label{lemma:middlemarkupdate}
Middle marks can be updated by the symbolic rules $\text{`?'} + \text{`$\ast$'} = \text{`$\ast$'}$, $\text{`$\ast$'} + \text{`'} = \text{`'}$ and $\text{`L'} + \text{`R'} = \text{`!'}$.
\end{mylemmasm}

\textbf{Proof of Lemma \ref{lemma:middlemarkupdate}.}
The first rule follows since the middle mark `?' does not make any statement, hence it is consistent with all other middle marks. The second rule follows since the statement made by the empty middle mark `' implies the statements made by all other middle marks. The third rule follows from the definition of the middle mark `!'.
\hfill $\square$

\begin{mylemmasm}[Algorithm \ref{algo:ancestral}]\label{lemma:algo2}
Assume Algorithm \ref{algo:ancestral} is being passed a LPCMCI-PAG $\C$ as well as the assumptions stated in Theorem \ref{thm:sound}. \textbf{1.)} $\C$ remains a LPCMCI-PAG at any point of the algorithm. \textbf{2.)} The algorithm converges. 
\end{mylemmasm}

\textbf{Proof of Lemma \ref{lemma:algo2}.}
Write $A = X^i_{t-\tau}$ and $B = X^j_t$.
\textbf{1.)}
Given faithfulness and perfect statistical decisions, edges are removed if and only if the corresponding nodes are m-separated by some subset of variables. The for-loop in line 3 considers ordered pairs $(A, B)$ only if $A < B$ with respect to the adopted total order \totalorder. According to Lemma \ref{lemma:conditiononparents} the default conditioning on parents as described by lines 5 and 10 does not destroy any m-separations. The algorithm therefore updates the edge between $A$ and $B$ with middle mark `R' only if $A$ and $B$ are not m-separated by any subset of $\apds(B, A, \C)$. Since $\pa(B, \M) \subseteq \apds(B, A, \C)$ holds, $A$ and $B$ are then not m-separated by any subset of $\pa(B, \M)$ and the update is correct. Similarly the update with middle mark `L' is correct. Note that the algorithm resets $\ppc = 0$ once any edge marks have been updated, i.e., once some default conditioning sets may potentially change. Therefore, all separating sets found by the algorithm are weakly minimal. More formally: The default conditioning set $\SSet_{def}$ corresponds to $\SSet_1$ in Definition \ref{def:weaklyminimal}, and $\SSet$ corresponds to $\SSet_2$. Whenever $\SSet_1$ changes, the algorithm restarts with $|\SSet_2| = \ppc = 0$ and keeps increasing $p$ by steps of one. If the algorithm finds that some pair of variables is conditionally independent given $\SSet_{def} \cup \SSet$, this pair of variables is not conditionally independent given $\SSet_{def} \cup \SSet^\prime$ for a proper subset $\SSet^\prime$ of $\SSet$. This is because CI given $\SSet_{def} \cup \SSet^\prime$ was tested before and rejected, if it would not have been rejected the edge would have been removed already. The statement then follows from correctness of the orientation rules, which is already proven.
\textbf{2.)}
If $A$ and $B$ are connected by a link with middle mark `?' or `L', the algorithm keeps testing for CI given subsets of $\apds(B, A, \C)$ until the link has been removed or updated with middle mark `R'. Similarly, if $A$ and $B$ are connected by a link with middle mark `?' or `R', the algorithm keeps testing for CI given subsets of $\apds(A, B, \C)$ until the link has been removed or update with middle mark `L'. There is no orientation rule that turns a middle mark `!' back into `?', `L', or `R', and there is no orientation rule that modifies an empty middle mark. With the update rules given in Lemma \ref{lemma:middlemarkupdate} this shows that all remaining edges will eventually have middle marks `!' or `' (empty). Then, the algorithm converges.
\hfill $\square$

\begin{mylemmasm}[An implication of middle mark `!']\label{lemma:!}
Assume $A \EMM{\astast} B$ in LPCMCI-PAG $\C$ but $A \notin \adj(B, \M)$. Then: \textbf{1.)} $A \notin \an(B, \G)$ and $B \notin \an(A, \G)$. \textbf{2.)} Assume further that $\ER0a$ has been exhaustively applied to $\C$. Then, $A$ and $B$ are m-separated by a subset of $\napds(B, A, \C)$ and by a subset of $\napds(A, B, \C)$.
\end{mylemmasm}

\textbf{Proof of Lemma \ref{lemma:!}.}
Without loss of generality we can assume that $A < B$.
\textbf{1.)} Assume $A \in \an(B, \G)$ were true. Then $A$ and $B$ would be m-separated by some subset of $\DSEP(B, A, \M)$ for which $\DSEP(B, A, \M) = \pa(B, \M)$ by 4.) of Lemma \ref{lemma:dsepsets}. This contradicts $A \RMM{\astast} B$ and hence $A \EMM{\astast} B$. Similarly $B \in \an(A, \G)$ contradicts $A \LMM{\astast} B$ and hence $A \EMM{\astast} B$.
\textbf{2.)}
This follows from the first part together with Lemma \ref{lemma:apdsnapds}.
\hfill $\square$

\begin{mylemmasm}[LPCMCI without Algorithm \ref{algo:nonancestral}]\label{lemma:withoutalg3}
Consider modifying the LPCMCI algorithm \ref{algo:lpcmci} by skipping line 6 of its pseudocode, i.e., by not executing Algorithm \ref{algo:nonancestral}. Denote the graph returned by this modified algorithm as $\C^\prime$. Then:
\textbf{1.)} $\C^\prime$ is a LPCMCI-PAG.
\textbf{2.)} If $A \SMMnoast{\tailhead} B$ in $\C^\prime$ then $A \in \adj(B, \M)$.
\textbf{3.)} If $A \SMM{\astast} B$ in $\C^\prime$ and $A \notin \adj(B, \M)$ then $A \notin \an(B, \G)$ and $B \notin \an(A, \G)$.
\end{mylemmasm}

\textbf{Proof of Lemma \ref{lemma:withoutalg3}.}
\textbf{1.)}
According to the MMR orientation rule the initialization of $\C$ in line 1 of Algorithm \ref{algo:lpcmci} produces an LPCMCI-PAG $\C$. Since Lemma \ref{lemma:algo2} proves that $\C$ is still an LPCMCI-PAG after line 3, this remains true when some parentships are carried over after the re-initialization in line 4. The statement thus follows from Lemma \ref{lemma:algo2}.
\textbf{2.)}
The proof of the second part of Lemma \ref{lemma:apdsnapds} implies that all edges in $\C^\prime$ have middle marks `!' or `' (empty). If $A \tailhead B$ in $\C^\prime$, then $A \in \adj(B, \M)$ by the seventh point in Definition \ref{def:LPCMCI_PAG}. If $ A \EMMnoast{\tailhead} B$ in $\C^\prime$, then $A \in \adj(B, \M)$ by the APR orientation rule, see Lemma \ref{lemma:middlemarks}.
\textbf{3.)}
According to the proof of the previous point $A \astast B$ or $A \EMM{\astast} B$, and since $A \notin \adj(B, \M)$ the seventh point in Definition \ref{def:LPCMCI_PAG} further restricts to $A \EMM{\astast} B$. The statement then follows from the first part of Lemma \ref{lemma:!}.
\hfill $\square$

\begin{mylemmasm}[Algorithm \ref{algo:nonancestral}]\label{lemma:algo3}
Assume Algorithm \ref{algo:nonancestral} is being passed a LPCMCI-PAG $\C$. \textbf{1.)} $\C$ remains a LPCMCI-PAG at any point of the algorithm. \textbf{2.)} The algorithm converges. 
\end{mylemmasm}

\textbf{Proof of Lemma \ref{lemma:algo3}.}
Write $A = X^i_{t-\tau}$ and $B = X^j_t$.
\textbf{1.)}
An edge between $A$ and $B$ is updated with the empty middle mark only if $A$ and $B$ are not m-separated by a subset of $\napds(B, A, \C)$ or $\tau = 0$ and $A$ and $B$ are not m-separated by a subset of $\napds(A, B, \C)$. Note that $\ER0a$ is exhaustively applied in line 22 of Alg.~\ref{algo:ancestral} as well as in line 23 of Alg.~\ref{algo:nonancestral}. According to Lemma \ref{lemma:!} the update is then correct. Apart from this the proof parallels the proof of 1.) of Lemma \ref{lemma:algo2}.
\textbf{1.)}
If $A$ and $B$ are connected by a link with middle mark `!', the algorithm keeps testing for CI given subsets of $\napds(B, A, \C)$ and if $\tau = 0$ also given subsets of $\napds(A, B, \C)$ until the link has been removed or updated with the empty middle mark. There is no orientation rule that turns a middle mark `!' back into `?', `L', or `R', and there is no orientation rule that modifies an empty middle mark. With the update rules given in Lemma \ref{lemma:middlemarkupdate} this shows that all remaining edges will eventually have empty middle marks. Then, the algorithm converges.
\hfill $\square$

\begin{customtheorem}{2}[LPCMCI is sound and complete]
Assume that there is a process as in eq.~\eqref{eqmain:SVARProcess} without causal cycles, which generates a distribution $P$ that is faithful to its time series graph $\G$. Further assume that there are no selection variables, and that we are given perfect statistical decisions about CI of observed variables in $P$. Then LPCMCI is sound and complete, i.e., it returns the PAG $\PG$.
\end{customtheorem}

\textbf{Proof of Theorem \ref{thm:sound}.}
\textbf{Soundness:} According to the MMR orientation rule the initialization of $\C$ in line 1 of Algorithm \ref{algo:lpcmci} produces an LPCMCI-PAG $\C$. Since Lemma \ref{lemma:algo2} proves that $\C$ is still an LPCMCI-PAG after line 3, this remains true when some parentships are carried over after the re-initialization in line 4. Stationarity both with respect to orientations and adjacencies is always enforced by construction. The statement then follows from Lemmas \ref{lemma:algo2} and \ref{lemma:algo3} together with the first, second, third, and seventh point in Definition \ref{def:LPCMCI_PAG}. \textbf{Completeness:} Note that after convergence of the while loop in Alg.~\ref{algo:nonancestral} all middle marks in $\C$ are empty. Since according to Lemma \ref{lemma:algo3} this $\C$ is a LPCMCI-PAG, the skeleton of $\C$ agrees with that of $\M$. Note again that stationarity both with respect to orientations and adjacencies is always enforced by construction, and that rules $R5$ through $R7$ do not apply due to the assumption of no selection variables. Completeness then follows since the orientation applied in line 27 of Algorithm \ref{algo:nonancestral} contain the FCI orientation rules $R0$ through $R4$ and $R8$ through $R10$ as special cases.
\hfill $\square$

\begin{customtheorem}{3}[LPCMCI is order-independent]
The output of LPCMCI does not depend on the order of the $N$ time series variables $X^j$ (the $j$-indices may be permuted).
\end{customtheorem}

\textbf{Proof of Theorem \ref{thm:orderindependent}.}
Both Algorithms \ref{algo:ancestral} and \ref{algo:nonancestral} remove edges only after the for-loop over ordered pairs has been completed. The ordering of ordered pairs imposed by the outer for-loop is order-independent. Note that the sets $\SSet_{search}$, $\SSet^1_{search}$ and $\SSet^2_{search}$ are ordered by means of $I^{min}$. Since this is an order-independent ordering, the break commands do not introduce order-dependence. The application of orientation rules is order-independent by construction of Algorithm \ref{algo:orientation}: Orientations and removals are only applied once the rule has been exhaustively applied, and conflicts are removed by means of the conflict mark `x'. Lastly, as discussed at the end of Sec.~\ref{sec:pseudocode2}, also the decision of whether a node is in a separating sets are made in an order-independent way.
\hfill $\square$

\section{Further numerical experiments}\label{sec:experiments}
On the following pages we present various further numerical experiments for evaluating and comparing the performances of LPCMCI and the SVAR-FCI and SVAR-RFCI baselines. For each setup we show results for significance levels $\alpha=0.01,\,0.05$ and, depending on the setup, for different autocorrelation values $a$, numbers of observed variables $N$, maximum time lag $\tau_{\max}$, fraction of unobserved variables $\lambda$, and sample sizes $T$. We focus the discussion on orientation recall and precision, runtimes, and control of false positives.

\begin{figure*}[t]  
\centering
\includegraphics[width=1\linewidth]{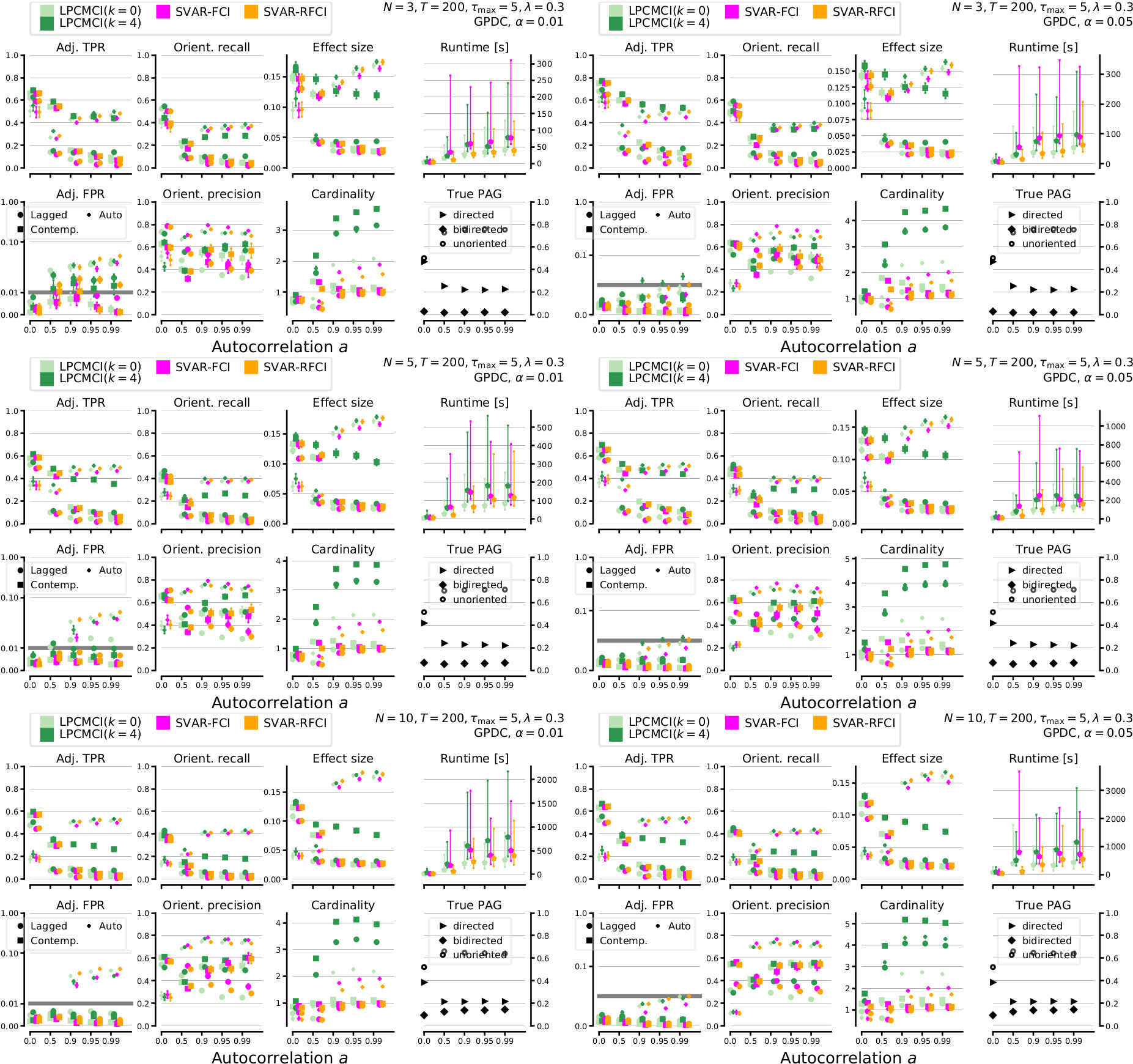}
\caption{
Results of numerical experiments for 
LPCMCI compared to SVAR-FCI and SVAR-RFCI (all with GPDC CI test \cite{Runge2018d}) for varying
 autocorrelation $a$
 for $T=200$
. The left (right) column shows results for significance level $\alpha=0.01$ ($\alpha=0.05$). 
The rows depict results for $N=3,\,5,\, 10$ (top and bottom).
All parameters are indicated in the upper right of each panel.
}
\label{fig:experiments_SM_autocorr_gpdc1}
\end{figure*}

\begin{figure*}[t]  
\centering
\includegraphics[width=1\linewidth]{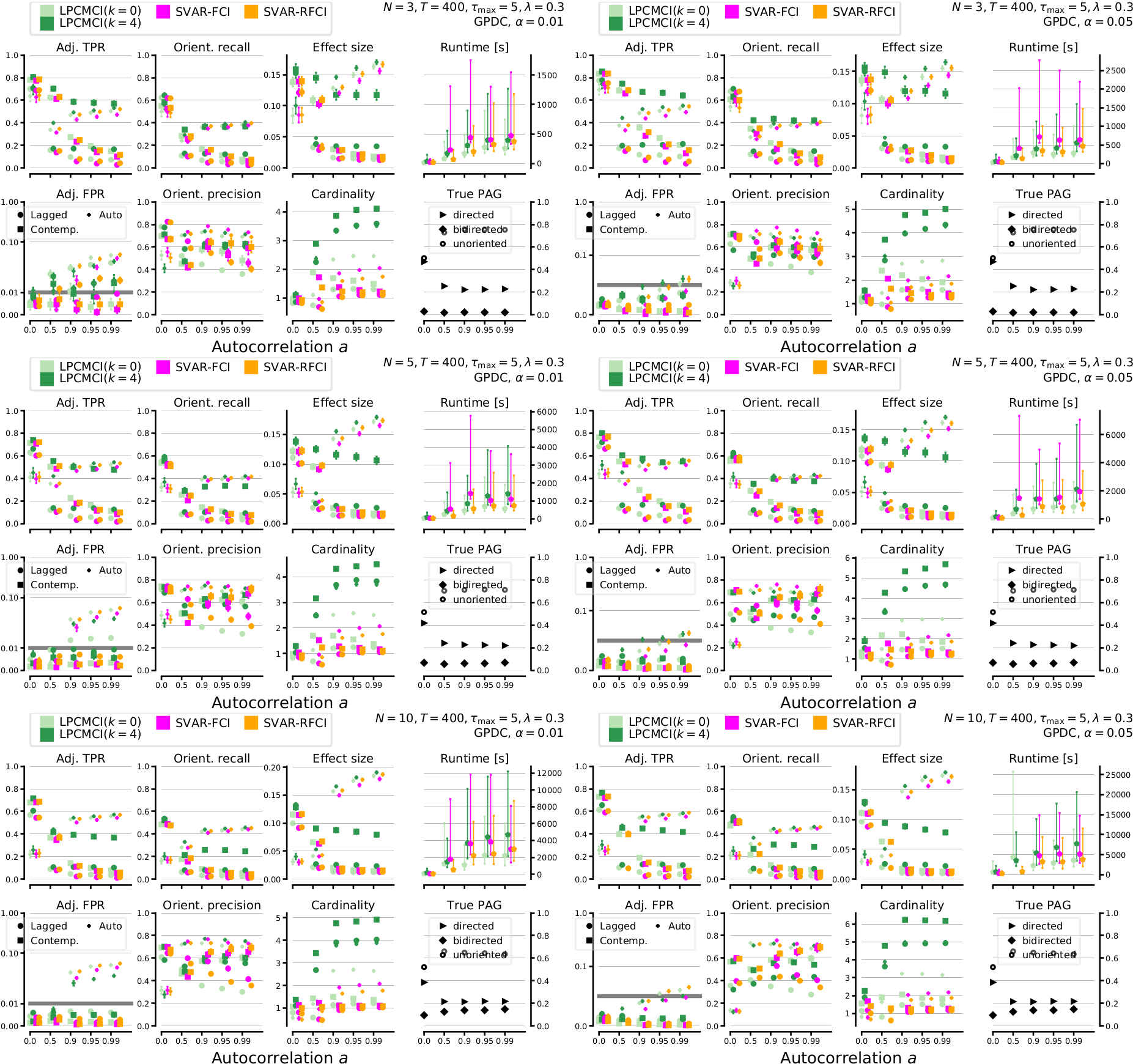}
\caption{
Results of numerical experiments for 
LPCMCI compared to SVAR-FCI and SVAR-RFCI (all with GPDC CI test \cite{Runge2018d}) for varying
 autocorrelation $a$
 for $T=400$
. The left (right) column shows results for significance level $\alpha=0.01$ ($\alpha=0.05$). 
The rows depict results for $N=3,\,5,\, 10$ (top and bottom).
All parameters are indicated in the upper right of each panel.
}
\label{fig:experiments_SM_autocorr_gpdc2}
\end{figure*}

\paragraph*{Nonlinear experiments with GPDC CI test:}
Results for the nonlinear conditional independence test GPDC \cite{Runge2018d} are shown in Figures~\ref{fig:experiments_SM_autocorr_gpdc1} ($T=200$) and \ref{fig:experiments_SM_autocorr_gpdc2} ($T=400$). Each figure depicts the results for $N=3, 5, 10$ observed variables and $\alpha=0.01, 0.05$ with varying autocorrelation on the x-axis. In these experiments we employ a variant of the model in eq.~\eqref{eqmain:numericalmodel} that features half linear and half nonlinear functions of the form $f_i(x)=(1+5 x e^{-x^2/20})x$, chosen because these tend to yield stationary dynamics. Further, a third of the noise distributions are randomly chosen to be Weibull distributions with shape parameter $2$ (instead of Normal distributions).

We find that also here LPCMCI has much higher adjacency and orientation recall than the SVAR-FCI and SVAR-RFCI baselines, especially for contemporaneous links. Precision is overall comparable, but lagged precision often higher for LPCMCI. For $N=3$ we observe partially not controlled false positives for all methods.

\begin{figure*}[t]  
\centering
\includegraphics[width=1\linewidth]{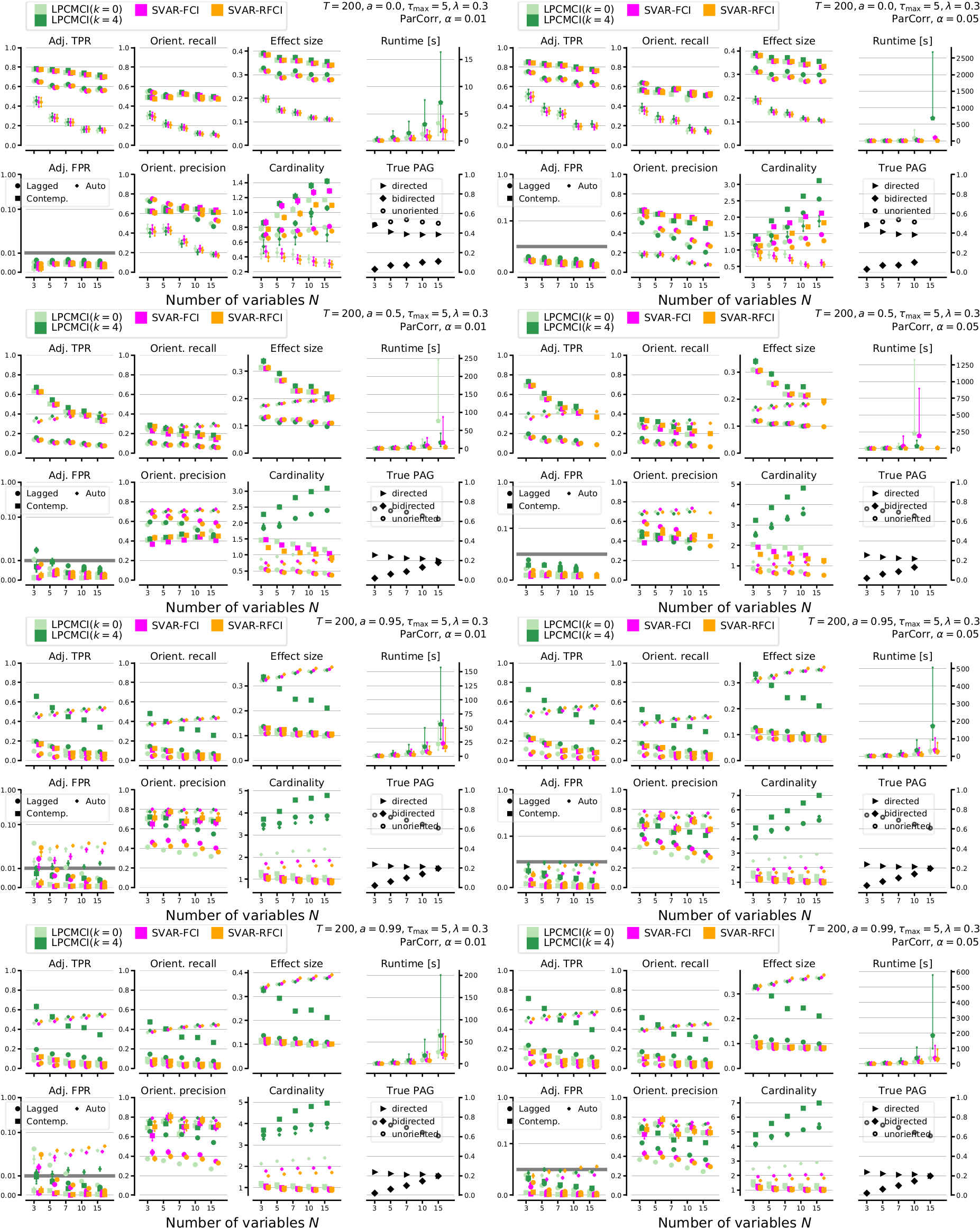}
\caption{
Results of numerical experiments for 
LPCMCI compared to SVAR-FCI and SVAR-RFCI (all with ParCorr CI test) for varying
 number of variables $N$
 for $T=200$
. The left (right) column shows results for significance level $\alpha=0.01$ ($\alpha=0.05$). 
The rows depict results for increasing autocorrelation (top to bottom).
All parameters are indicated in the upper right of each panel. Some experiments did not converge within 24hrs and are not shown.
}
\label{fig:experiments_SM_highdim1}
\end{figure*}

\begin{figure*}[t]  
\centering
\includegraphics[width=1\linewidth]{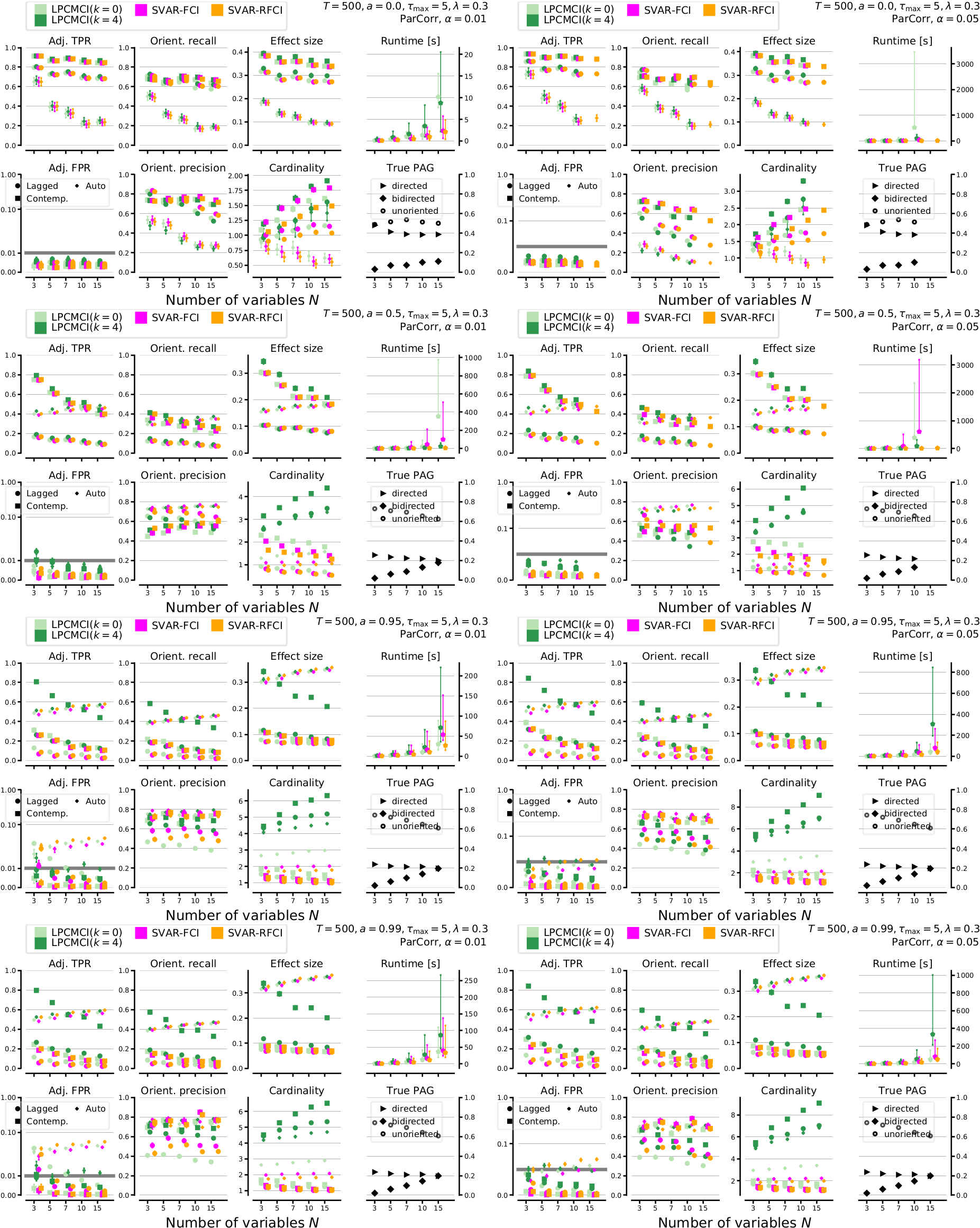}
\caption{
Results of numerical experiments for 
LPCMCI compared to SVAR-FCI and SVAR-RFCI (all with ParCorr CI test) for varying
 number of variables $N$
 for $T=500$
. The left (right) column shows results for significance level $\alpha=0.01$ ($\alpha=0.05$). 
The rows depict results for increasing autocorrelation (top to bottom).
All parameters are indicated in the upper right of each panel. Some experiments did not converge within 24hrs and are not shown.
}
\label{fig:experiments_SM_highdim2}
\end{figure*}

\begin{figure*}[t]  
\centering
\includegraphics[width=1\linewidth]{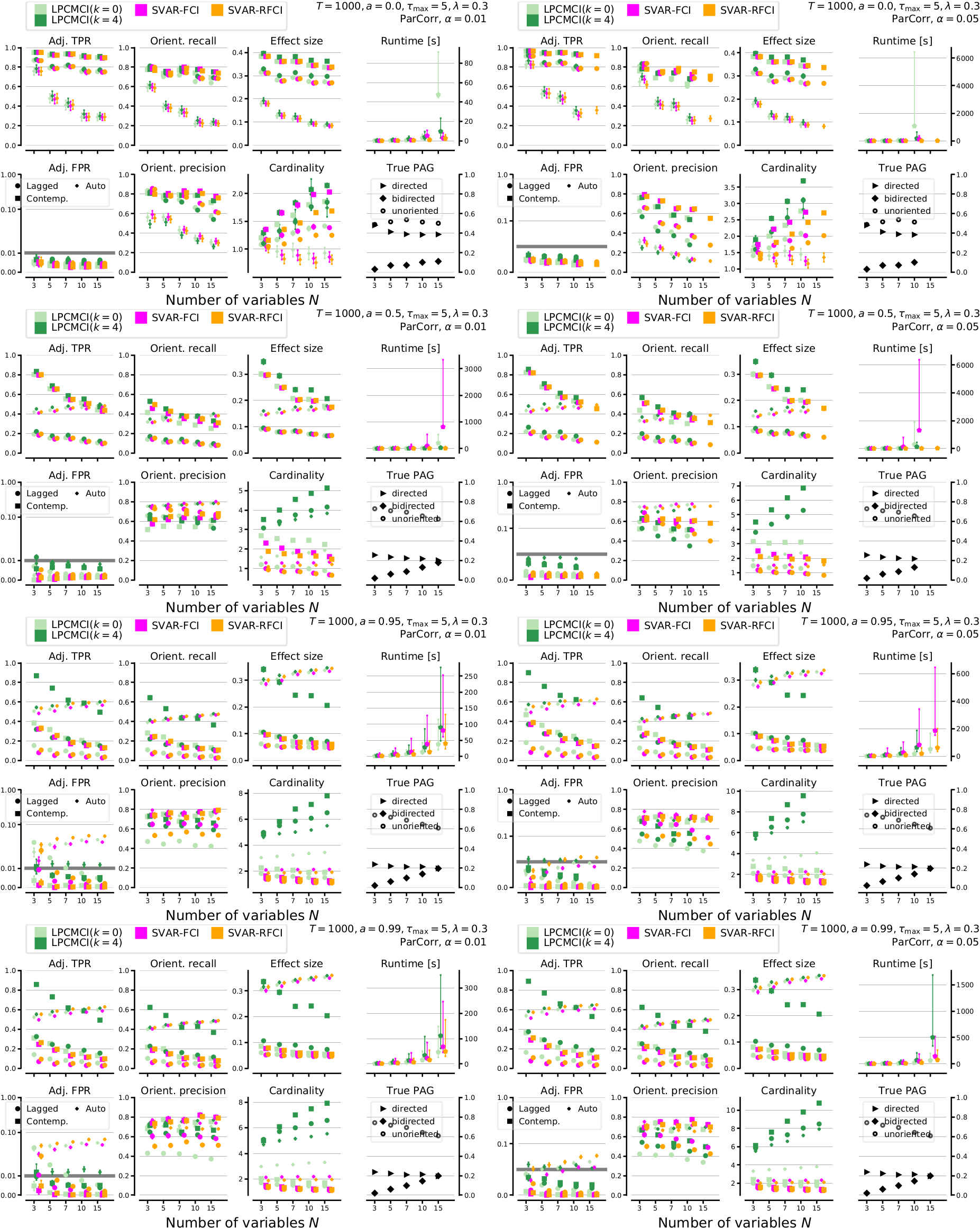}
\caption{
Results of numerical experiments for 
LPCMCI compared to SVAR-FCI and SVAR-RFCI (all with ParCorr CI test) for varying
 number of variables $N$
 for $T=1000$
. The left (right) column shows results for significance level $\alpha=0.01$ ($\alpha=0.05$). 
The rows depict results for increasing autocorrelation (top to bottom).
All parameters are indicated in the upper right of each panel. Some experiments did not converge within 24hrs and are not shown.
}
\label{fig:experiments_SM_highdim3}
\end{figure*}

\paragraph*{Linear experiments for varying number of variables $N$:}
In Figures~\ref{fig:experiments_SM_highdim1}-\ref{fig:experiments_SM_highdim3} we depict results for varying numbers of observed variables $N$ and $T=200, 500, 1000$, $a=0,0.5,0.95,0.99$, and $\alpha=0.01, 0.05$. Since the fraction of unobserved variables is kept at $\lambda = 0.3$, an increasing number of observed variables $N$ also corresponds to an increasing number of total variables $\tilde{N}$.

For the case of no autocorrelation LCPCMI has slightly higher recall and slightly lower precision at a higher runtime. For intermediate autocorrelation ($a=0.5$) the results are similar to those for $a=0$, but SVAR-FCI's runtime is higher. For $N=3,T=200,\alpha=0.01$ false positives are not controlled, but less so for LPCMCI. For higher autocorrelation LPCMCI has 0.2-0.4 higher contemporaneous recall and also substantially higher lagged recall throughout. In the highly autocorrelated regime we observe inflated false positives for SVAR-FCI and SVAR-RFCI due to ill-calibrated CI tests, similar to the PC algorithm as discussed in \cite{Runge2018d}.

\begin{figure*}[t]  
\centering
\includegraphics[width=1\linewidth]{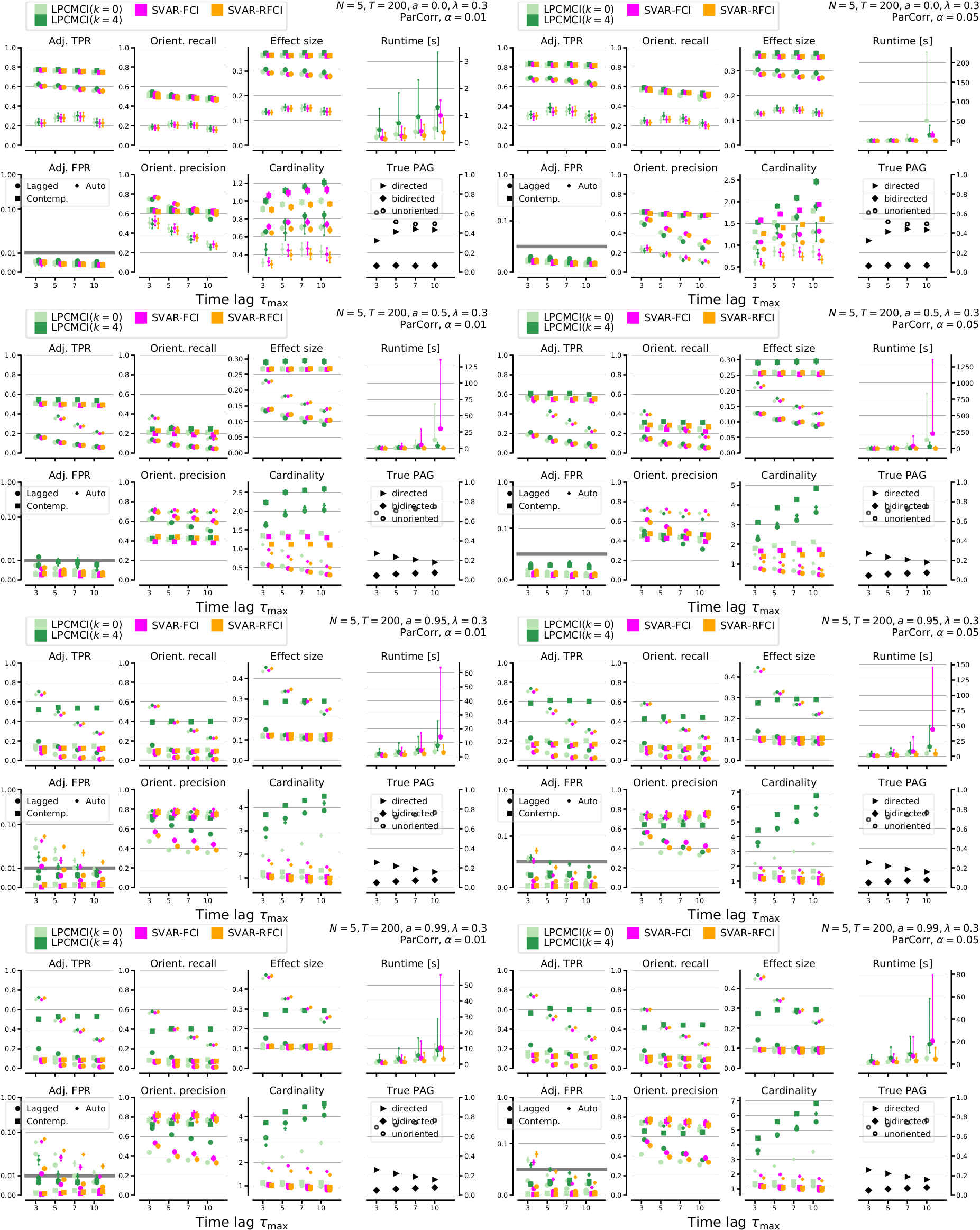}
\caption{
Results of numerical experiments for 
LPCMCI compared to SVAR-FCI and SVAR-RFCI (all with ParCorr CI test) for varying
 maximum time lag $\tau_{\max}$ 
 for $T=200$
. The left (right) column shows results for significance level $\alpha=0.01$ ($\alpha=0.05$). 
The rows depict results for increasing autocorrelation (top to bottom).
All parameters are indicated in the upper right of each panel.
}
\label{fig:experiments_SM_taumax1}
\end{figure*}

\begin{figure*}[t]  
\centering
\includegraphics[width=1\linewidth]{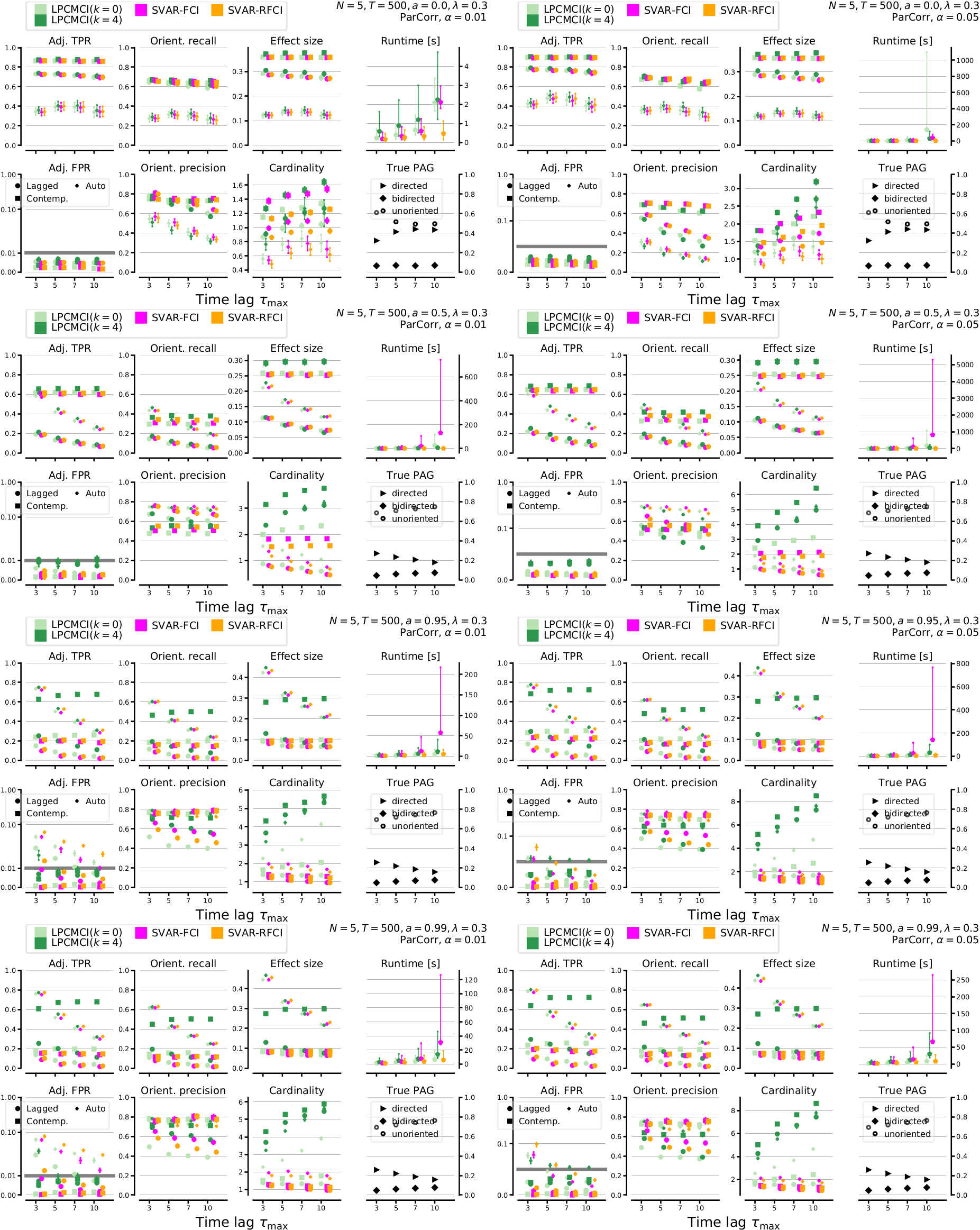}
\caption{
Results of numerical experiments for 
LPCMCI compared to SVAR-FCI and SVAR-RFCI (all with ParCorr CI test) for varying
 maximum time lag $\tau_{\max}$ 
 for $T=500$
. The left (right) column shows results for significance level $\alpha=0.01$ ($\alpha=0.05$). 
The rows depict results for increasing autocorrelation (top to bottom).
All parameters are indicated in the upper right of each panel.
}
\label{fig:experiments_SM_taumax2}
\end{figure*}

\begin{figure*}[t]  
\centering
\includegraphics[width=1\linewidth]{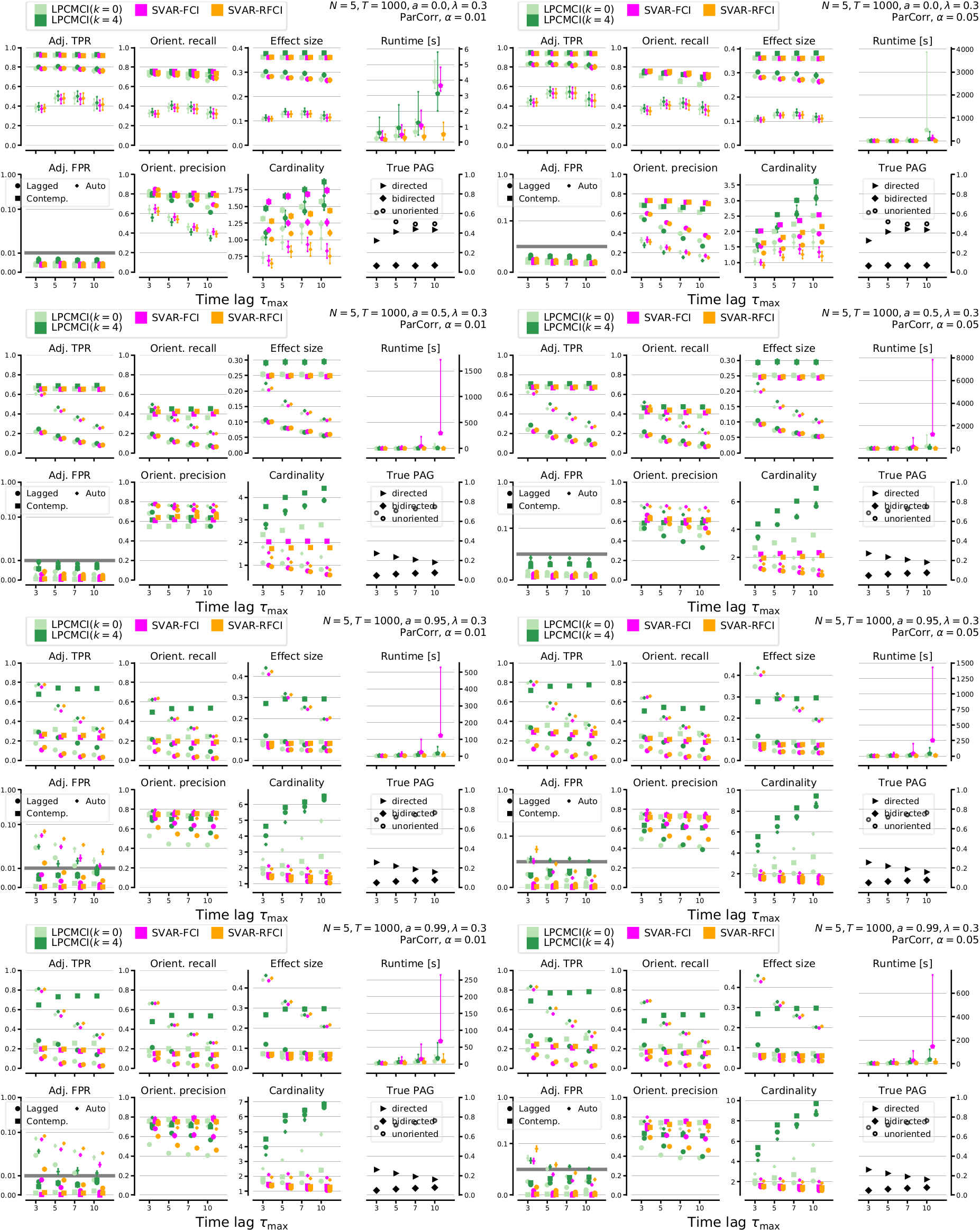}
\caption{
Results of numerical experiments for 
LPCMCI compared to SVAR-FCI and SVAR-RFCI (all with ParCorr CI test) for varying
 maximum time lag $\tau_{\max}$ 
 for $T=1000$
. The left (right) column shows results for significance level $\alpha=0.01$ ($\alpha=0.05$). 
The rows depict results for increasing autocorrelation (top to bottom).
All parameters are indicated in the upper right of each panel.
}
\label{fig:experiments_SM_taumax3}
\end{figure*}

\paragraph*{Linear experiments for varying maximum time lag $\tau_{\max}$:}
Figures~\ref{fig:experiments_SM_taumax1}-\ref{fig:experiments_SM_taumax3} show results for varying maximum time lags $\tau_{\max}$ and $T=200, 500, 1000$, $a=0,0.5,0.95,0.99$, and $\alpha=0.01, 0.05$.

For no autocorrelation all methods have almost constant contemporaneous recall, only lagged recall shows a slight decay. Note that the true PAG changes with $\tau_{\max}$. Contemporaneous precision is also largely constant, while lagged precision decreases for all methods. Runtime increases and sharply rises for LPCMCI with $k=0$, indicating that the edge removal phase of Algorithm \ref{algo:nonancestral} is faster for higher $k$, i.e., after several preliminary phases have been run. SVAR-FCI similarly features exploding runtimes for large $\tau_{\max}$, both intermediate and higher autocorrelations. Again, false positives in SVAR-FCI and SVAR-RFCI are not well controlled for small $\tau_{\max}$ and $\alpha=0.01$.

\begin{figure*}[t]  
\centering
\includegraphics[width=1\linewidth]{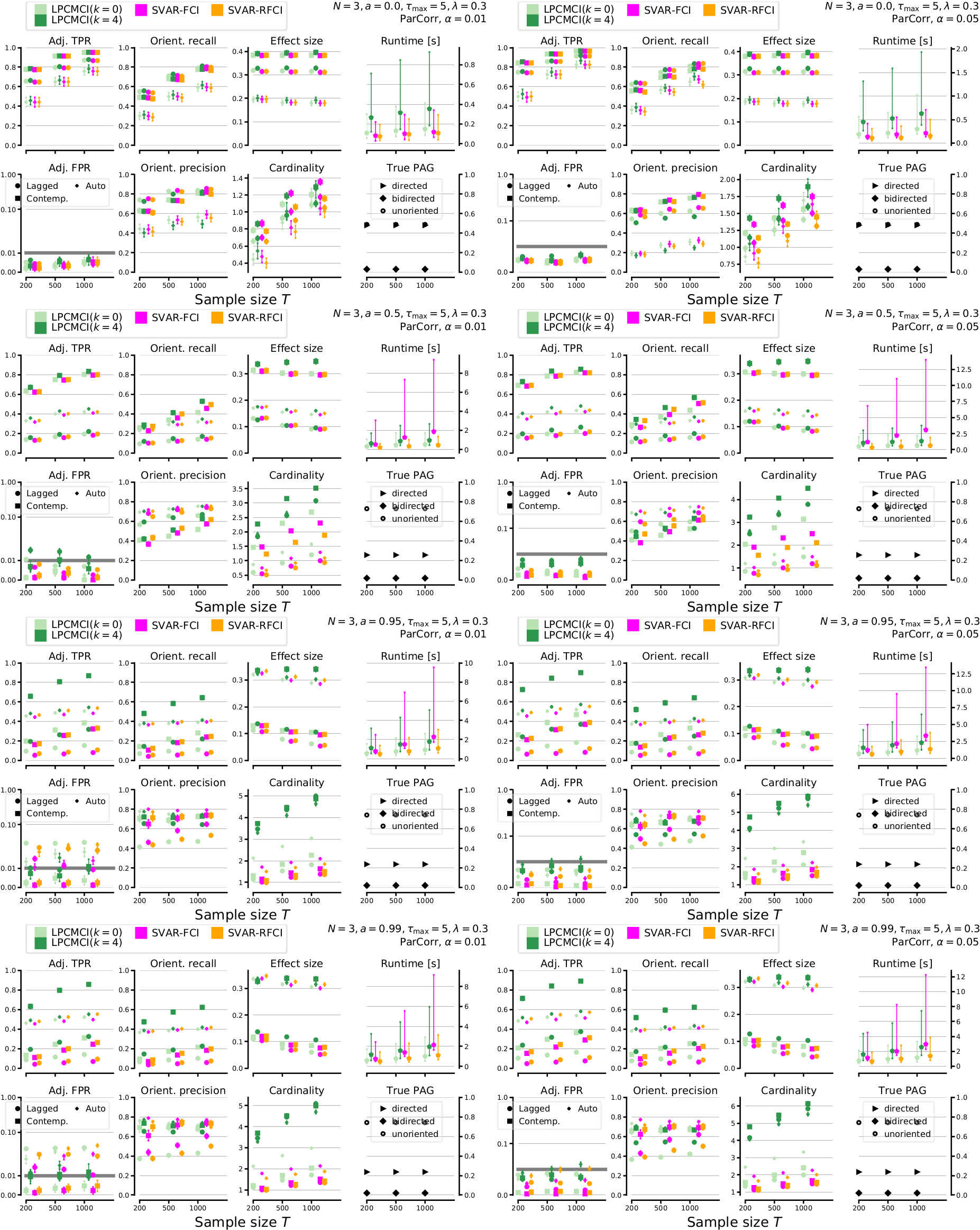}
\caption{
Results of numerical experiments for 
LPCMCI compared to SVAR-FCI and SVAR-RFCI (all with ParCorr CI test) for varying
 sample size $T$
 for $N=3$ 
. The left (right) column shows results for significance level $\alpha=0.01$ ($\alpha=0.05$). 
The rows depict results for increasing autocorrelation (top to bottom).
All parameters are indicated in the upper right of each panel.
}
\label{fig:experiments_SM_samples1}
\end{figure*}

\begin{figure*}[t]  
\centering
\includegraphics[width=1\linewidth]{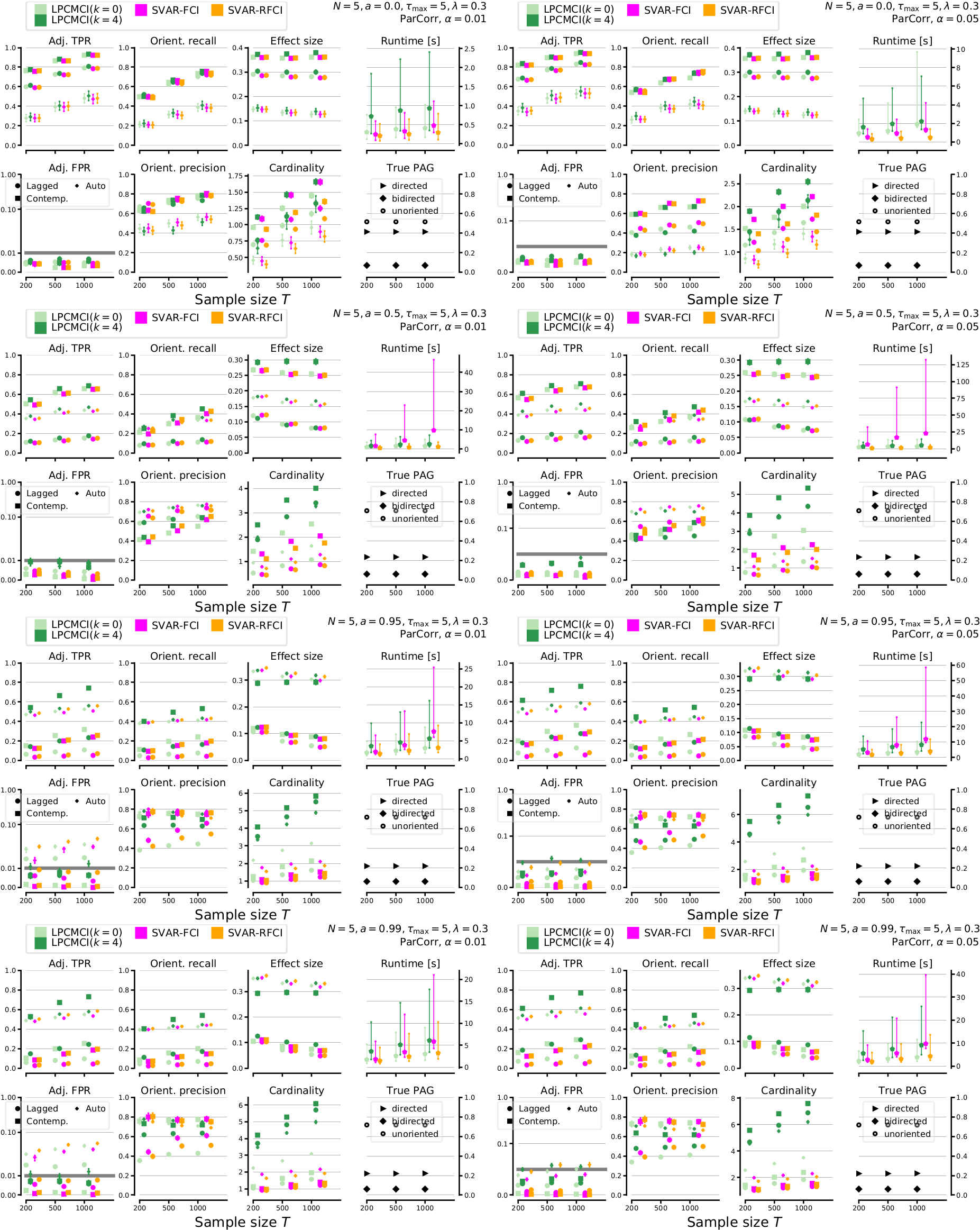}
\caption{
Results of numerical experiments for 
LPCMCI compared to SVAR-FCI and SVAR-RFCI (all with ParCorr CI test) for varying
 sample size $T$
 for $N=5$ 
. The left (right) column shows results for significance level $\alpha=0.01$ ($\alpha=0.05$). The rows depict results for increasing autocorrelation (top to bottom).
All parameters are indicated in the upper right of each panel.
}
\label{fig:experiments_SM_samples2}
\end{figure*}

\begin{figure*}[t]  
\centering
\includegraphics[width=1\linewidth]{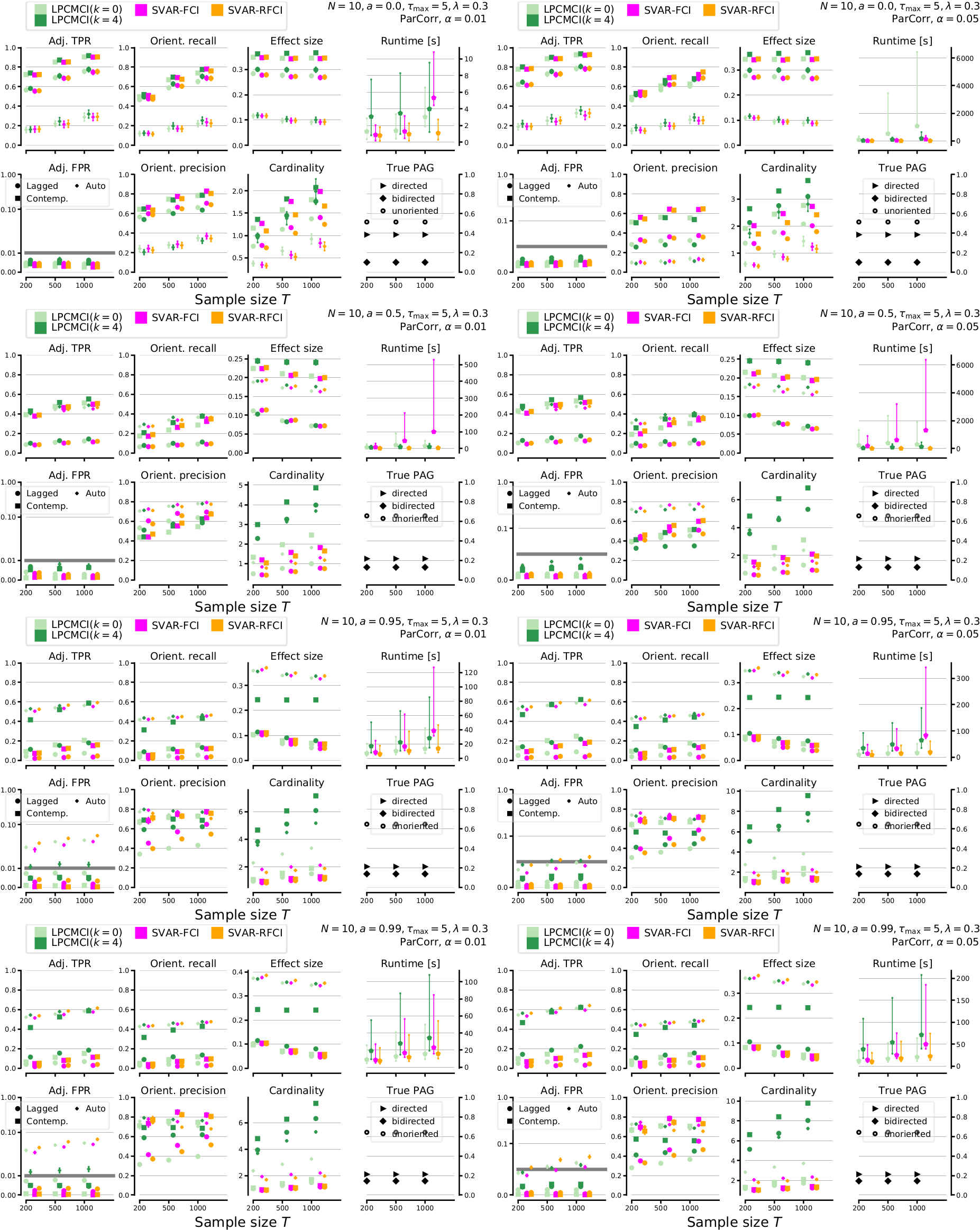}
\caption{
Results of numerical experiments for 
LPCMCI compared to SVAR-FCI and SVAR-RFCI (all with ParCorr CI test) for varying
 sample size $T$
 for $N=10$ 
. The left (right) column shows results for significance level $\alpha=0.01$ ($\alpha=0.05$). The rows depict results for increasing autocorrelation (top to bottom).
All parameters are indicated in the upper right of each panel.
}
\label{fig:experiments_SM_samples3}
\end{figure*}

\paragraph*{Linear experiments for varying sample size $T$:}
In Figures~\ref{fig:experiments_SM_samples1}-\ref{fig:experiments_SM_samples3} we depict results for varying sample sizes $T$ and $N=3, 5, 10$, $a=0,0.5,0.95,0.99$, and $\alpha=0.01, 0.05$.

As expected, both recall and precision increase with $T$. Also runtime increases, but only slowly, except for LPCMCI with $k=0$ where it explodes for $N=10$. The higher the autocorrelation, the better the increase in recall and precision for contemporaneous links. For $N = 3$ lack of false positive control (less so for LCPCMI) is visible for all sample sizes. For strong autocorrelations there is a minor decrease in the orientation precision of contemporaneous links when increasing the sample size from $T = 500$ to $T = 1000$. Since at the same time there sometimes is a slight increase in the orientation precision of lagged links, we do not see a straightforward explanation of this observation.

\begin{figure*}[t]  
\centering
\includegraphics[width=1\linewidth]{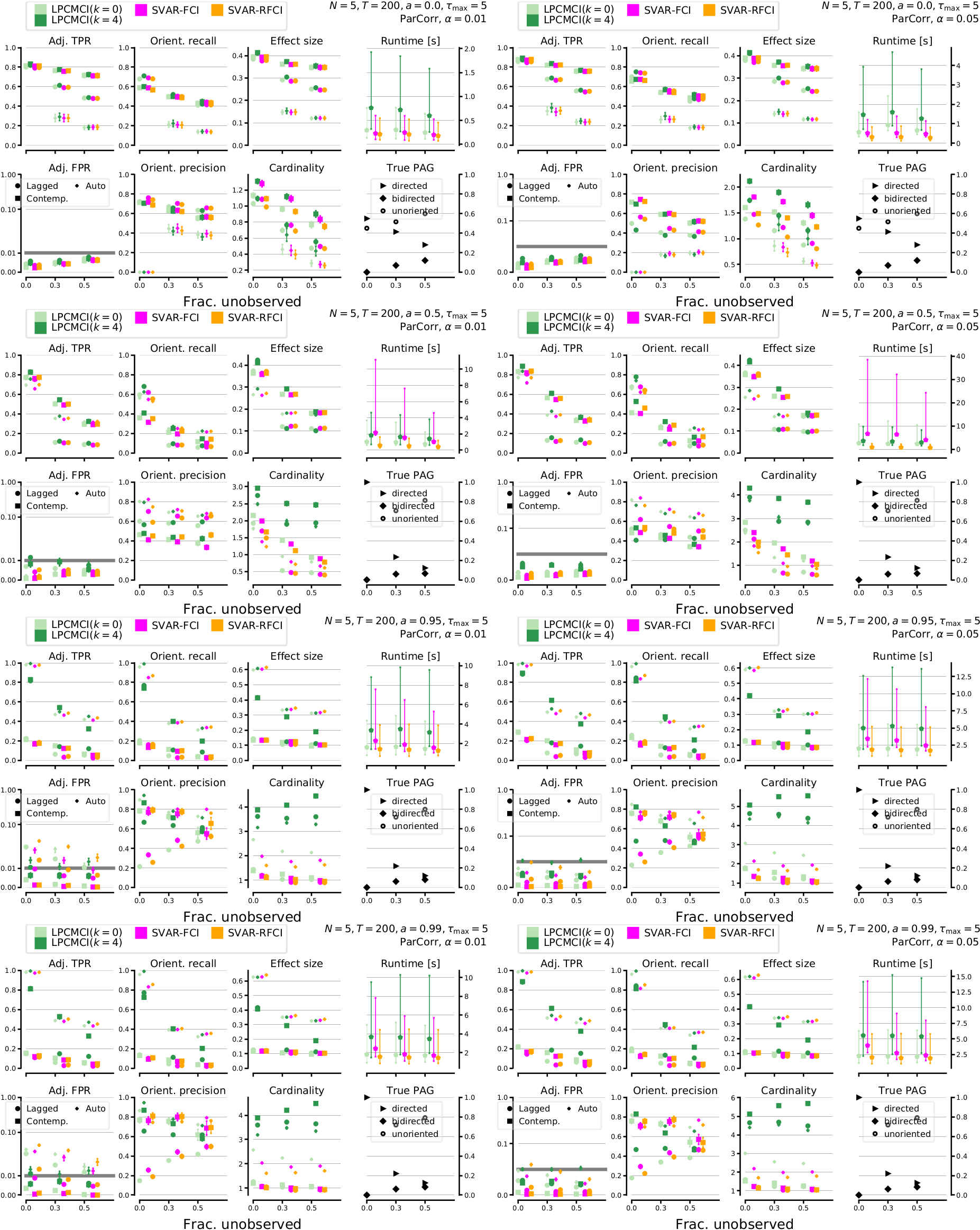}
\caption{
Results of numerical experiments for 
LPCMCI compared to SVAR-FCI and SVAR-RFCI (all with ParCorr CI test) for varying
 fraction of unobserved variables $\lambda$
 for $T=200$ and $N=5$
. The left (right) column shows results for significance level $\alpha=0.01$ ($\alpha=0.05$). 
The rows depict results for increasing autocorrelation (top to bottom).
All parameters are indicated in the upper right of each panel.
}
\label{fig:experiments_SM_unobserved1}
\end{figure*}

\begin{figure*}[t]  
\centering
\includegraphics[width=1\linewidth]{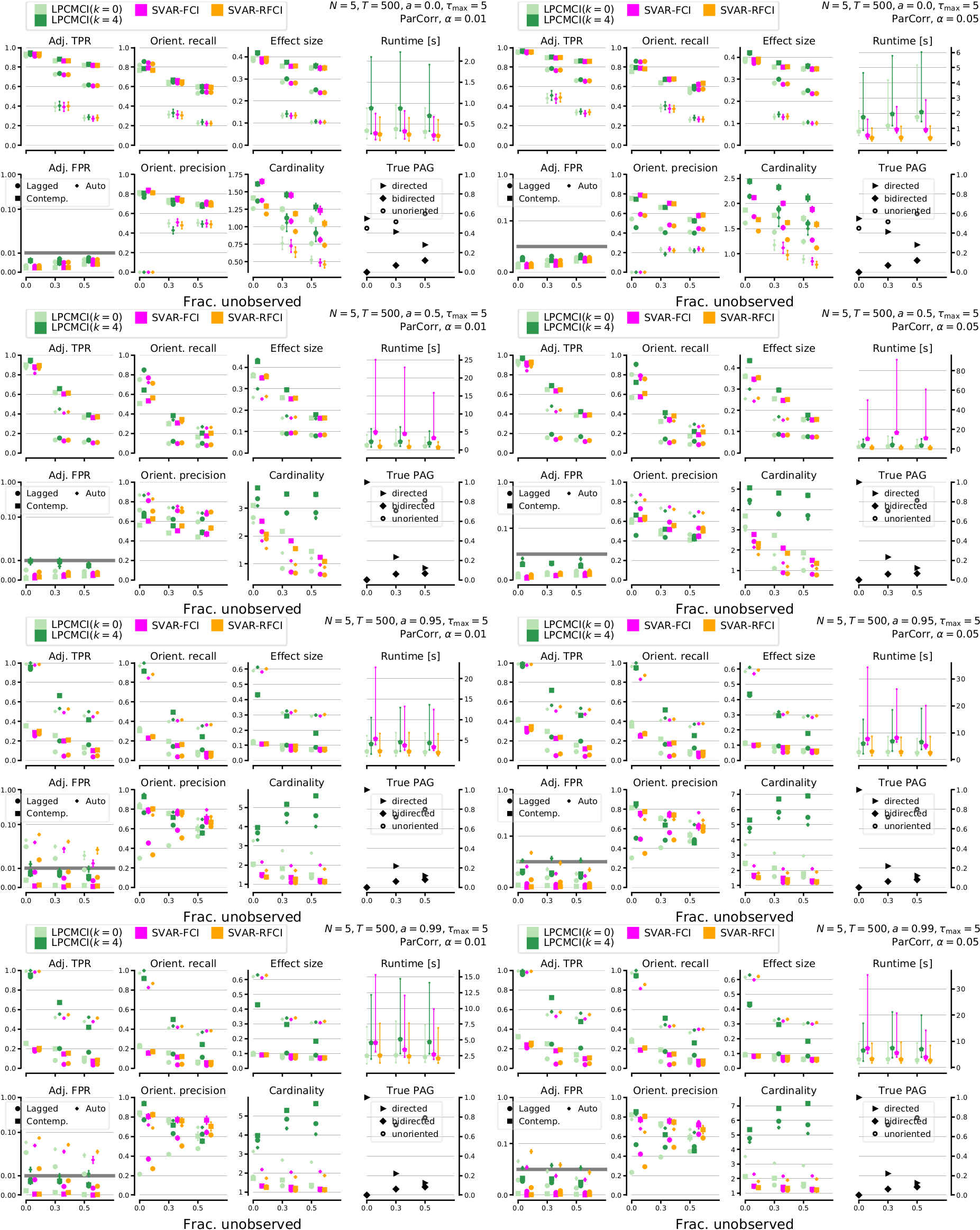}
\caption{
Results of numerical experiments for 
LPCMCI compared to SVAR-FCI and SVAR-RFCI (all with ParCorr CI test) for varying
 fraction of unobserved variables $\lambda$
 for $T=500$ and $N=5$
. The left (right) column shows results for significance level $\alpha=0.01$ ($\alpha=0.05$). 
The rows depict results for increasing autocorrelation (top to bottom).
All parameters are indicated in the upper right of each panel.
}
\label{fig:experiments_SM_unobserved2}
\end{figure*}

\begin{figure*}[t]  
\centering
\includegraphics[width=1\linewidth]{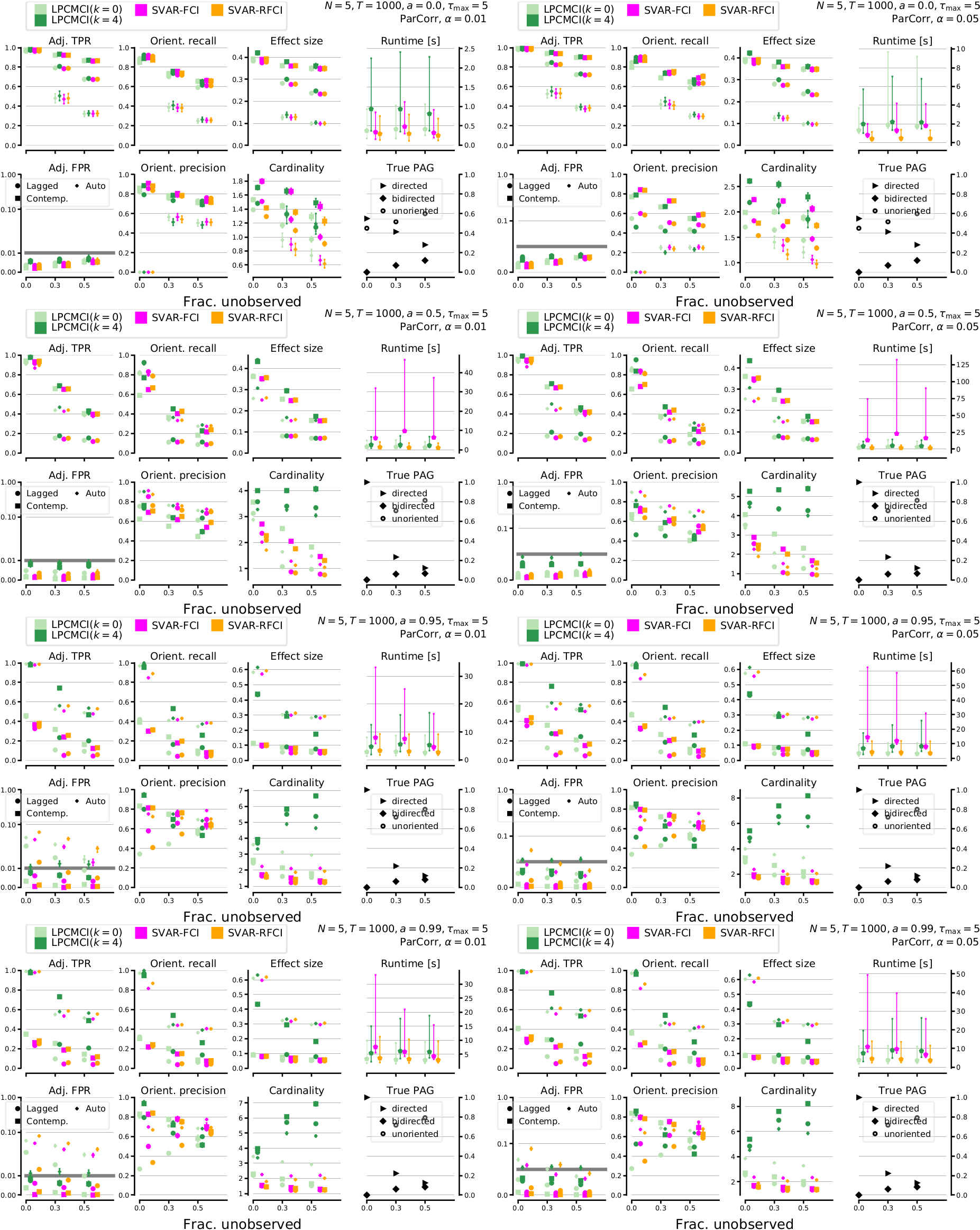}
\caption{
Results of numerical experiments for 
LPCMCI compared to SVAR-FCI and SVAR-RFCI (all with ParCorr CI test) for varying
 fraction of unobserved variables $\lambda$
 for $T=1000$ and $N=5$
. The left (right) column shows results for significance level $\alpha=0.01$ ($\alpha=0.05$). 
The rows depict results for increasing autocorrelation (top to bottom).
All parameters are indicated in the upper right of each panel.
}
\label{fig:experiments_SM_unobserved3}
\end{figure*}

\begin{figure*}[t]  
\centering
\includegraphics[width=1\linewidth]{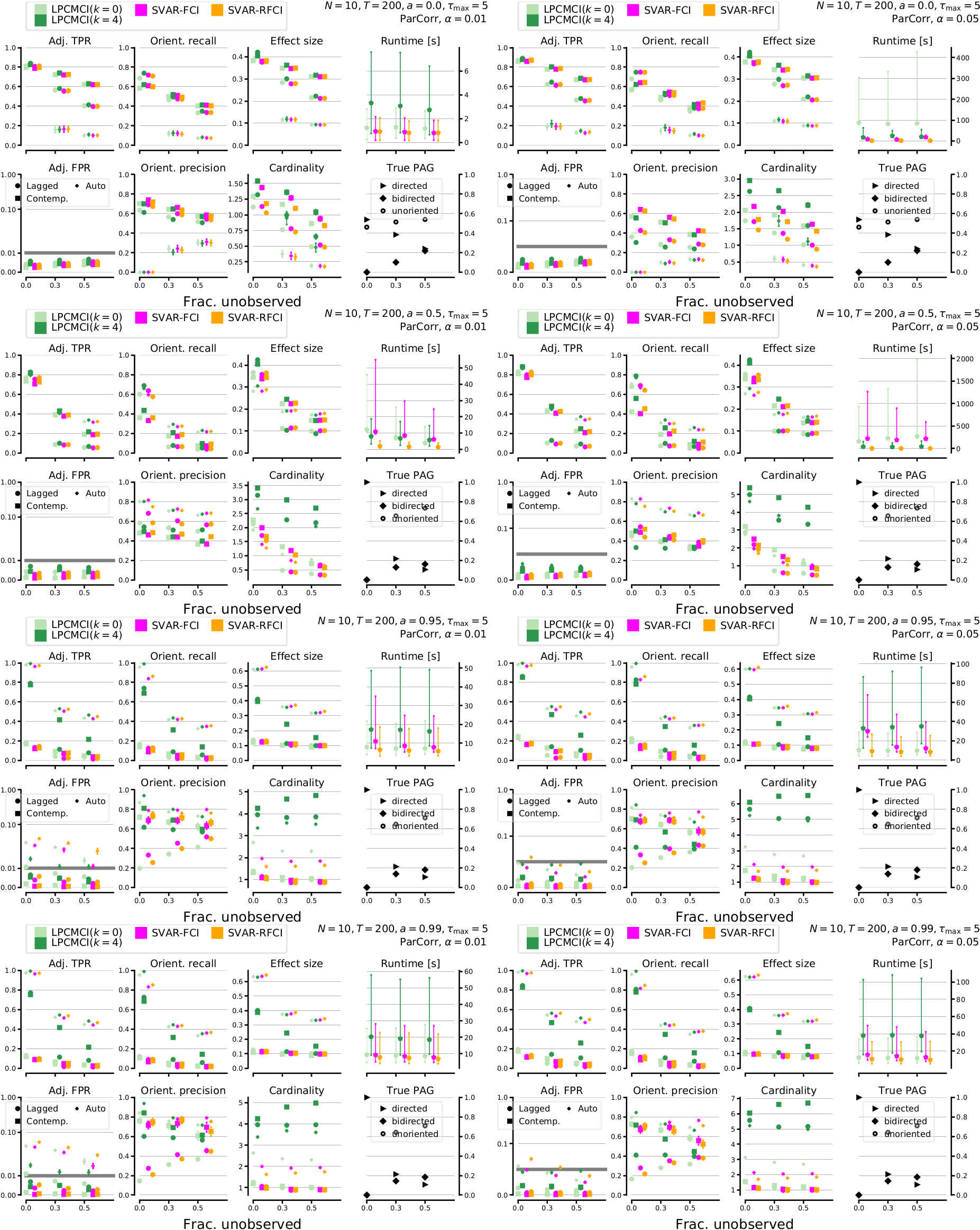}
\caption{
Results of numerical experiments for 
LPCMCI compared to SVAR-FCI and SVAR-RFCI (all with ParCorr CI test) for varying
 fraction of unobserved variables $\lambda$
 for $T=200$ and $N=10$
. The left (right) column shows results for significance level $\alpha=0.01$ ($\alpha=0.05$). 
The rows depict results for increasing autocorrelation (top to bottom).
All parameters are indicated in the upper right of each panel.
}
\label{fig:experiments_SM_unobserved4}
\end{figure*}

\begin{figure*}[t]  
\centering
\includegraphics[width=1\linewidth]{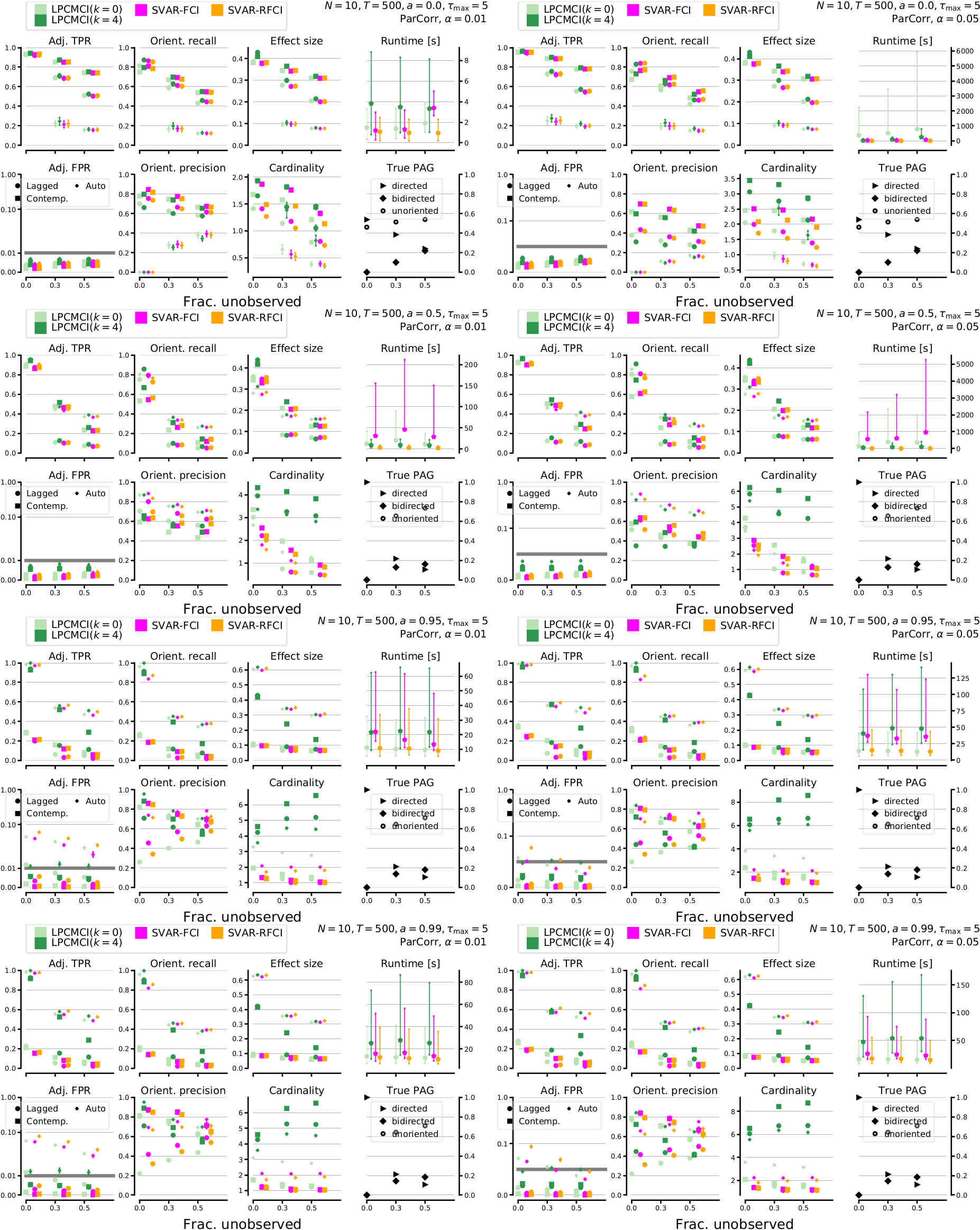}
\caption{
Results of numerical experiments for 
LPCMCI compared to SVAR-FCI and SVAR-RFCI (all with ParCorr CI test) for varying
 fraction of unobserved variables $\lambda$
 for $T=500$ and $N=10$
. The left (right) column shows results for significance level $\alpha=0.01$ ($\alpha=0.05$). 
The rows depict results for increasing autocorrelation (top to bottom).
All parameters are indicated in the upper right of each panel.
}
\label{fig:experiments_SM_unobserved5}
\end{figure*}

\begin{figure*}[t]  
\centering
\includegraphics[width=1\linewidth]{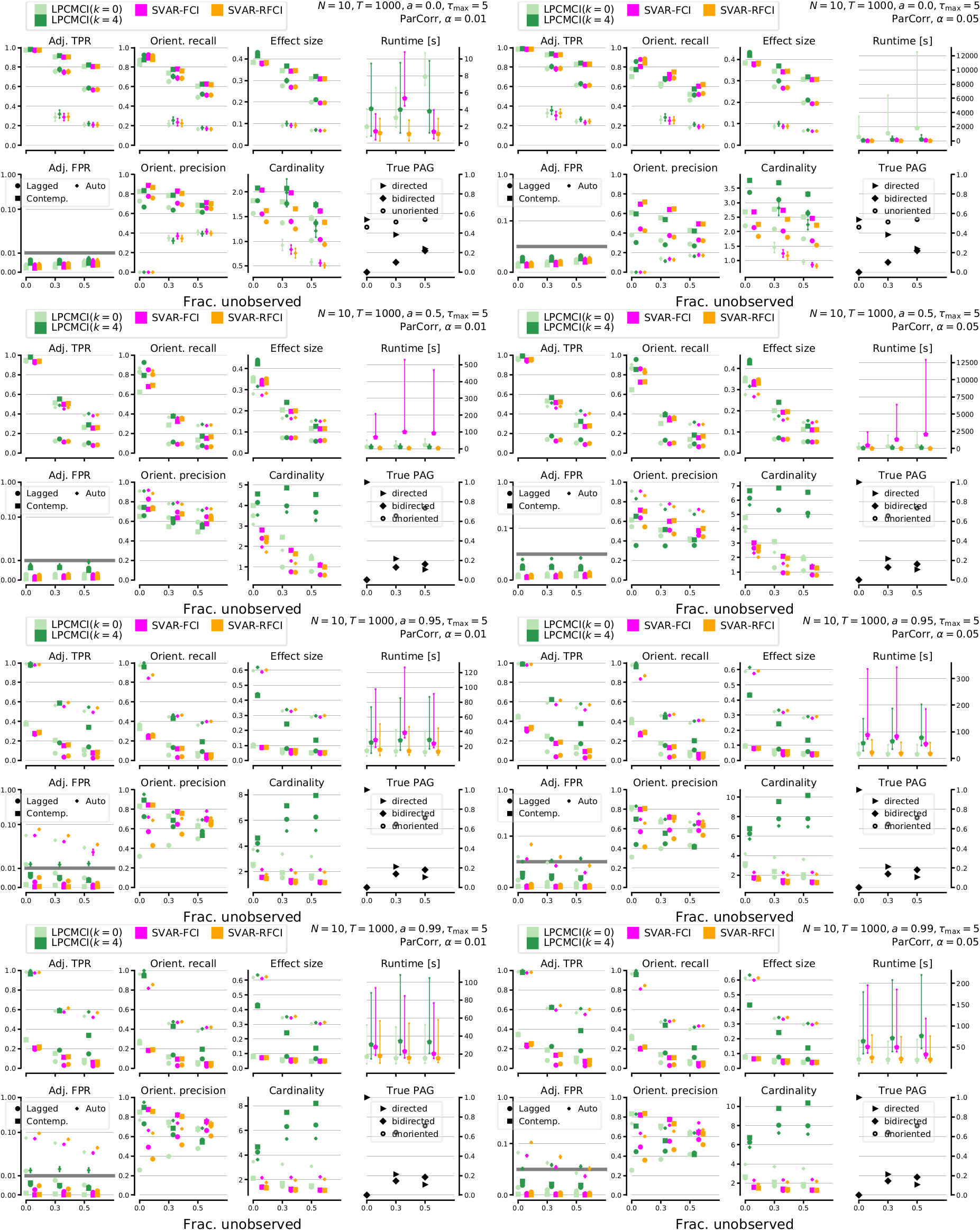}
\caption{
Results of numerical experiments for 
LPCMCI compared to SVAR-FCI and SVAR-RFCI (all with ParCorr CI test) for varying
 fraction of unobserved variables $\lambda$
 for $T=1000$ and $N=10$
. The left (right) column shows results for significance level $\alpha=0.01$ ($\alpha=0.05$). 
The rows depict results for increasing autocorrelation (top to bottom).
All parameters are indicated in the upper right of each panel.
}
\label{fig:experiments_SM_unobserved6}
\end{figure*}

\paragraph*{Linear experiments for varying the fraction of unobserved variables $\lambda$:}
In Figures~\ref{fig:experiments_SM_unobserved1}-\ref{fig:experiments_SM_unobserved6} we show results for varying fractions of unobserved variables $\lambda$ and $T=200, 500, 1000$, $N=3, 5, 10$, $a=0,0.5,0.95,0.99$, and $\alpha=0.01, 0.05$.

For no autocorrelation both recall and precision decay, while runtime is almost constant. For intermediate and strong autocorrelation we observe a strong decay in recall (even stronger for contemporaneous links), and a less stronger decay in precision. Runtime is almost constant.

\paragraph*{Linear experiments for the non-time case:}
The previous experiments already cover non-autocorrelated time series, see the various results for $a = 0$. In these cases LPCMCI shows a performance similar to the baselines. Here, we additionally analyze the truly non-time series case.

The numerical experiments are based on a purely contemporaneous variant of the model in eq.~\eqref{eqmain:numericalmodel}, namely
\begin{align} \label{eq:numericalmodel_nontimeseries}
V^j_t &= \textstyle{\sum_i} c_i V^i_t + \eta^j_t\quad \text{for}\quad j\in\{1,\ldots,\tilde{N}\} \ .
\end{align}
Note that the time index $t$ could equally well be dropped, such that the samples here conform to the \emph{i.i.d.} case. Contrary to the previous setup we randomly choose $L=\lfloor0.3\tilde{N}(\tilde{N}-1)/2\rfloor$ linear links (corresponding to non-zero coefficients $c_i$). This results in a graph with a constant link density of 30\%, which was not feasible in the time series case since it would lead to non-stationary dynamics. As before, the fraction of unobserved variables is $\lambda=0.3$.

\begin{figure*}[t]  
\centering
\includegraphics[width=1\linewidth]{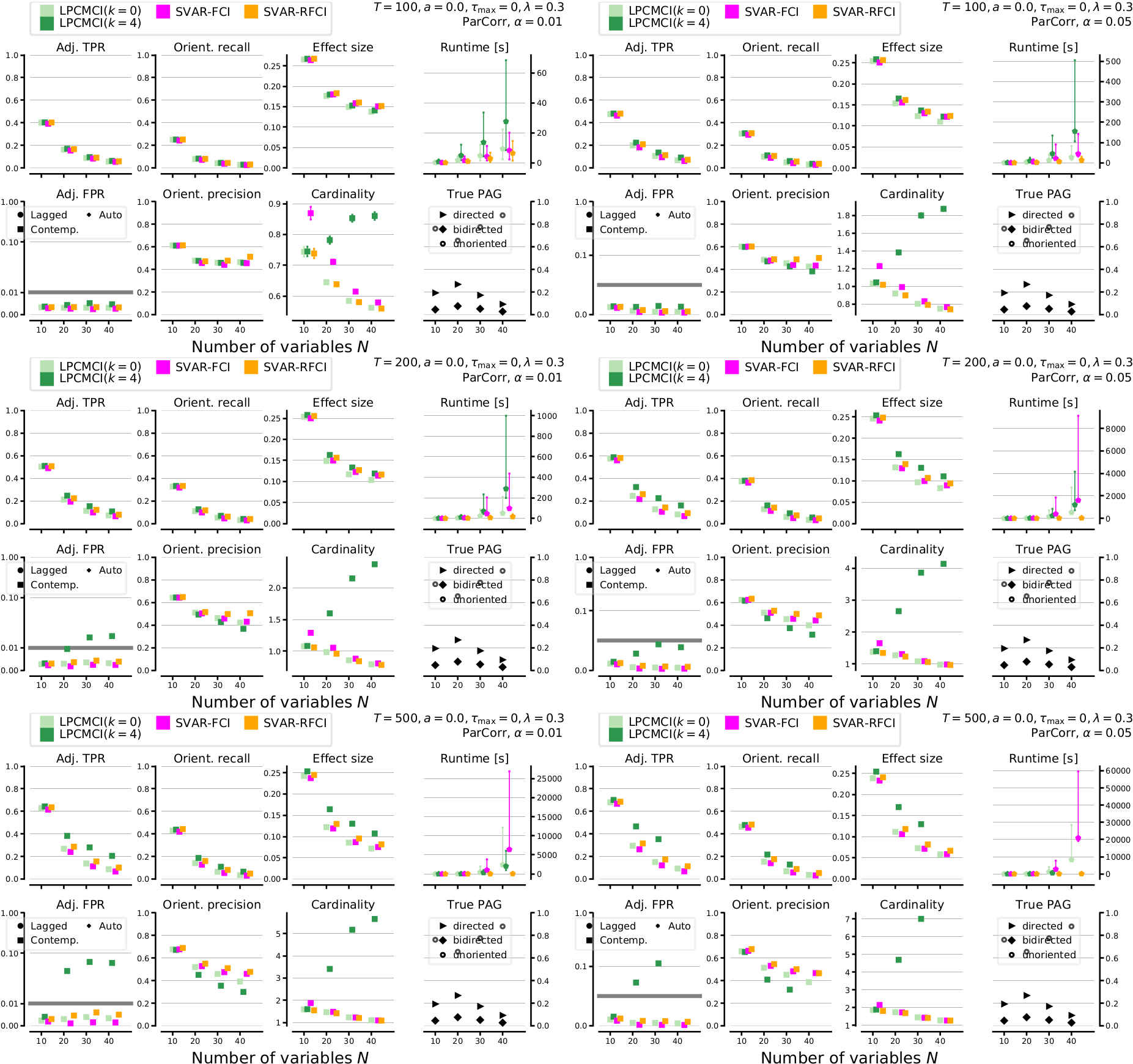}
\caption{
Results of numerical experiments in the non-time series case (see eq.~\eqref{eq:numericalmodel_nontimeseries}) for 
LPCMCI compared to SVAR-FCI and SVAR-RFCI (all with ParCorr CI test) for varying
 number of variables $N$
. The left (right) column shows results for significance level $\alpha=0.01$ ($\alpha=0.05$). 
The rows depict results for $T=100, 200, 500$ (top to bottom).
All parameters are indicated in the upper right of each panel. Some experiments did not converge within 24hrs and are not shown.
}
\label{fig:experiments_SM_nontimeseries}
\end{figure*}

In Figure~\ref{fig:experiments_SM_nontimeseries} we show results for varying numbers of observed variables $N=10, 20, 30, 40$, sample sizes $T=100, 200, 500$, and significance levels $\alpha=0.01, 0.05$. While in the non-time series case SVAR-FCI and SVAR-RFCI respectively reduce to FCI and RFCI, we keep the naming for consistency.

SVAR-FCI, SVAR-RFCI, and LPCMCI($k=0$) all perform very similar with a slight advantage for SVAR-RFCI, which also has the lowest runtimes. LPCMCI($k=4$) shows higher recall but also lower precision and partly inflated false positives. A more detailed analysis of this drop in performance of LPCMCI with higher $k$ is subject to future research. We hypothesize that with additional preliminary iterations LPCMCI determines increasingly many false ancestorships, which are then used as default conditions and thereby prevent some true conditional independencies from being found.

\paragraph*{Linear experiments for models with discrete variables:}
Recall that LPCMCI is designed to increase effect sizes (with the aim to in turn increase the detection power of true causal links) by using known causal parents as default conditions (and by avoiding to condition on known non-ancestors). The higher-dimensional conditioning sets come, however, at the price of an increased estimation dimension. This has a counteracting negative effect on detection power. As supported by the various experiments in the main text and above, for continuous variables loosing $\mathcal{O}(1)$ degrees of freedom by default conditioning is negligible to, e.g., $\mathcal{O}(100)$ sample sizes. In this case the positive effect of increased effect sizes prevails. We now study the models with discrete-valued variables, where the increased cardinality of conditioning sets is expected to have a much stronger effect on the estimation dimensions.

The numerical experiments are based on a linear and discretized variant of the model in eq.~\eqref{eqmain:numericalmodel} of the form
\begin{align} \label{eq:numericalmodel_discrete_option2}
V_t^j &= g_{n_{bin}}\left(a_j V^j_{t-1} + \textstyle{\sum_i} c_i V^i_{t-\tau_i} + \left(b^j_t - \frac{n_{bin}}{2}\right)\right)\quad \text{for}\quad j\in\{1,\ldots,\tilde{N}\} \ , 
\end{align}
where $b^j_t \sim \operatorname{Bin}(n_{bin}, 0.5)$ with $n_{bin} \in 2\mathbb{Z}$ are Binomial noises. The function $g_{n_{bin}}$ decomposes as $g_{n_{bin}} = g_{n_{bin}}^{(1)} \circ g^{(2)}$, where $g^{(2)}$ rounds to the nearest integer and $g_{n_{bin}}^{(1)}(x) = x$ for $|x| \leq n_{bin}/2$ and $g_{n_{bin}}^{(1)}(x) = \operatorname{sign}(x) \cdot n_{bin}/2$ else. This implies that each variable can take at most $n_{bin} + 1$ values. The choice of parameters $a_j$ and $c_j$ as well as the choice and fraction of unobserved variables are unaltered, i.e., as explained below eq.~\eqref{eqmain:numericalmodel} in the main text.

Statistical tests of (conditional) independencies are performed with a $G$-test, see e.g. section 10.3.1 of \cite{LearningBayesianNetworks}. The details of our implementation are as follows: When testing for CI of $X$ and $Y$ given $Z_1, \ldots, Z_n$, a separate contingency table of $X$ and $Y$ is made for each unique value $(z_1^a, \ldots, z_n^a)$ that is assumed by the conditions. Rows and columns in these contingency tables that consist of zero counts only are removed. For each such table we calculate the degrees of freedom and the test statistic as for an unconditional $G$-test and sum those values up. If the total number of degrees of freedom falls below one, the test judges independence.

\begin{figure*}[t]  
\centering
\includegraphics[width=1\linewidth]{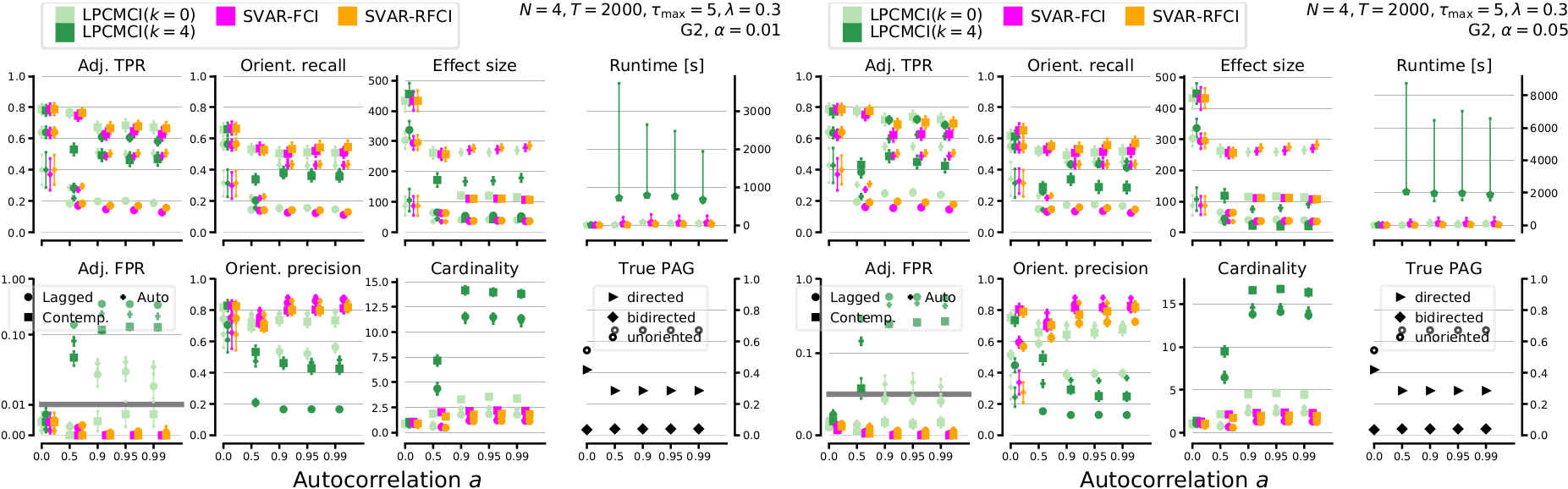}
\includegraphics[width=1\linewidth]{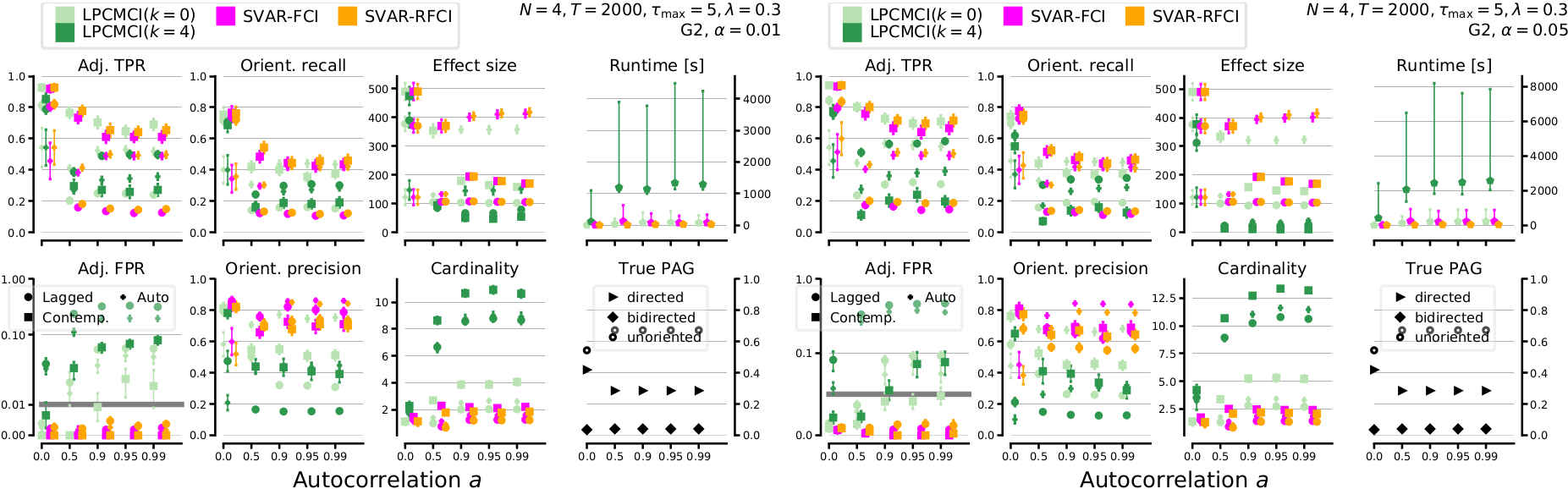}
\caption{
Results of numerical experiments with the discrete-variable model in eq.~\eqref{eq:numericalmodel_discrete_option2} for 
LPCMCI compared to SVAR-FCI and SVAR-RFCI (all with $G$-test of conditional independence) for varying
 autocorrelation $a$
 for $T=2000$ and $N=4$
. The left (right) column shows results for significance level $\alpha=0.01$ ($\alpha=0.05$). The top (bottom) row depicts the case with $n_{bin}=2$ ($n_{bin} = 4$) in eq.~\eqref{eq:numericalmodel_discrete_option2}.
}
\label{fig:experiments_SM_autocorrbinom_g2}
\end{figure*}

Figure~\ref{fig:experiments_SM_autocorrbinom_g2} shows the results for $N=4$ observed variables, sample size $T=2000$, and significance levels $\alpha=0.01, 0.05$ with varying autocorrelation on the x-axis. 

The bottom row depicts the case with $n_{bin}=4$. SVAR-FCI, SVAR-RFCI, and LPCMCI($k=0$) perform largely similar in terms of adjacency and orientation recall. LPCMCI($k=0$) has slightly higher lagged recall but also lower precision. LPCMCI($k=4$) has much lower recall (both regarding adjacencies and orientation) for contemporaneous links and autodependency links, but higher recall for lagged links. It further shows uncontrolled false positives and lower precision. One reason for this loss in performance can be found in the much higher cardinalities of CI tests in LPCMCI($k=4$). These seem to not only lead to lower power but also ill-calibrated CI tests. For $n_{bin} = 2$, depicted in the top row of Fig.~\ref{fig:experiments_SM_autocorrbinom_g2}, the overall results are similar.

These results can only be regarded as preliminary since there are different choices regarding the implementation of the $G$-test of conditional independence as well as a number of other ways to design discrete-variable models.  

\paragraph*{Comparison of LPCMCI to residualization approaches:}
The previous results demonstrate that LPCMCI shows strong gains in recall for autocorrelated continuous variables. One might wonder whether in the autocorrelated case SVAR-FCI benefits from a data preprocessing that is targeted at removing autocorrelation.

We test this idea by employing two residualization approaches: First, by fitting independent AR(1) models to each times series and then running SVAR-FCI on the residuals (SVAR-FCI-prewhite). This approach was also tested in the causally sufficient case with no contemporaneous links in \cite{Runge2018b} [Appendix B, Section 3]. Second, by using Gaussian process regression as proposed in \cite{Flaxman2015} instead of the AR(1) models (SVAR-FCI-GPwhite). More precisely, for every time series $X^j$ we fit a Gaussian process model of $X^j$ on the time index variable ($t=0,\ldots,n$ where $n$ is the sample size) and apply SVAR-FCI on the residuals. As a kernel we used the Radial Basis Function with an added unit variance White kernel. The RBF hyperparameters are optimzed as part of the default Python \texttt{scikit-klearn} implementation. These results are compared to standard LPCMCI run without prior residualization.

\begin{figure*}[t]  
\centering
\includegraphics[width=1\linewidth,page=1]{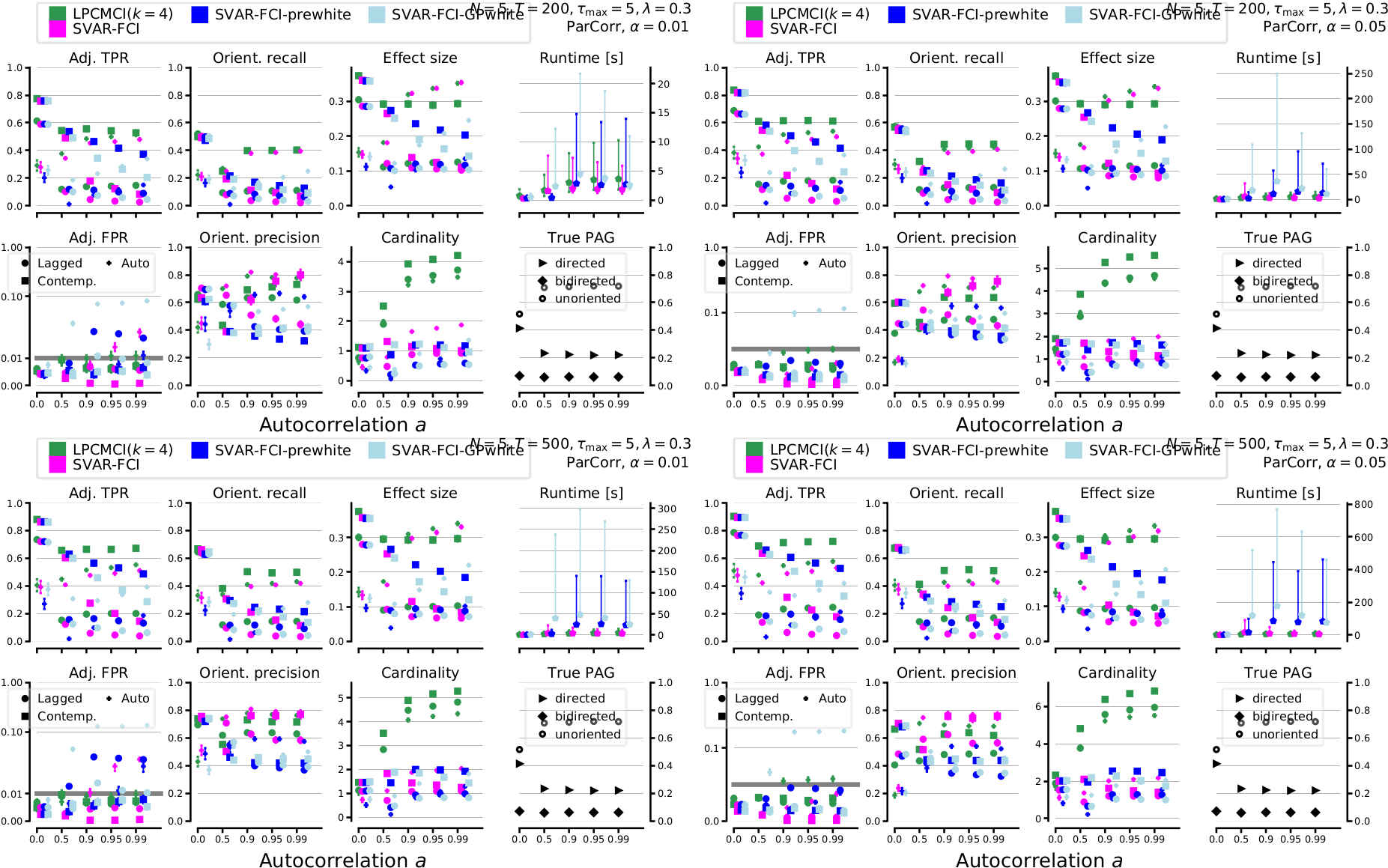}
\caption{
Results of numerical experiments for 
LPCMCI compared to SVAR-FCI, SVAR-FCI-prewhite, and SVAR-FCI-GPwhite (all with ParCorr CI test) for varying
 autocorrelation $a$
 for $N=5$. 
 The left (right) column shows results for significance level $\alpha=0.01$ ($\alpha=0.05$). 
The top (bottom) row depicts results for sample size $T=200$ ($T = 500$).
All parameters are indicated in the upper right of each panel.
}
\label{fig:experiments_SM_prewhiteautocorr}
\end{figure*}

Figure~\ref{fig:experiments_SM_prewhiteautocorr} shows the results for $N=5$ observed variables, sample sizes $T=200,\,500$, and significance levels $\alpha=0.01, 0.05$ with varying autocorrelation on the x-axis.

Both SVAR-FCI-prewhite and SVAR-FCI-GPwhite increase adjacency and orientation recall as compared to SVAR-FCI. SVAR-FCI-prewhite yields larger gains than SVAR-FCI-GPwhite, but both are still well below LPCMCI. Both SVAR-FCI-prewhite and SVAR-FCI-GPwhite have lower precision than SVAR-FCI, lead to inflated false positives, and excessively increase runtime. We further note that it is not clear what the ground truth MAG and PAG should be after residualization. This seems to require a substantially different theory.

\clearpage

\clearpage

\clearpage

\clearpage

\clearpage

\clearpage

\clearpage

\clearpage

\clearpage

\clearpage

\clearpage

\clearpage

\clearpage

\clearpage

\clearpage

\clearpage

\clearpage

\clearpage

\clearpage

\clearpage

\clearpage
\section{Figures illustrating the application to the real data example}\label{sec:real_data_supplement}
\begin{figure*}[tbhp]
\centering
\includegraphics[width=0.85\linewidth]{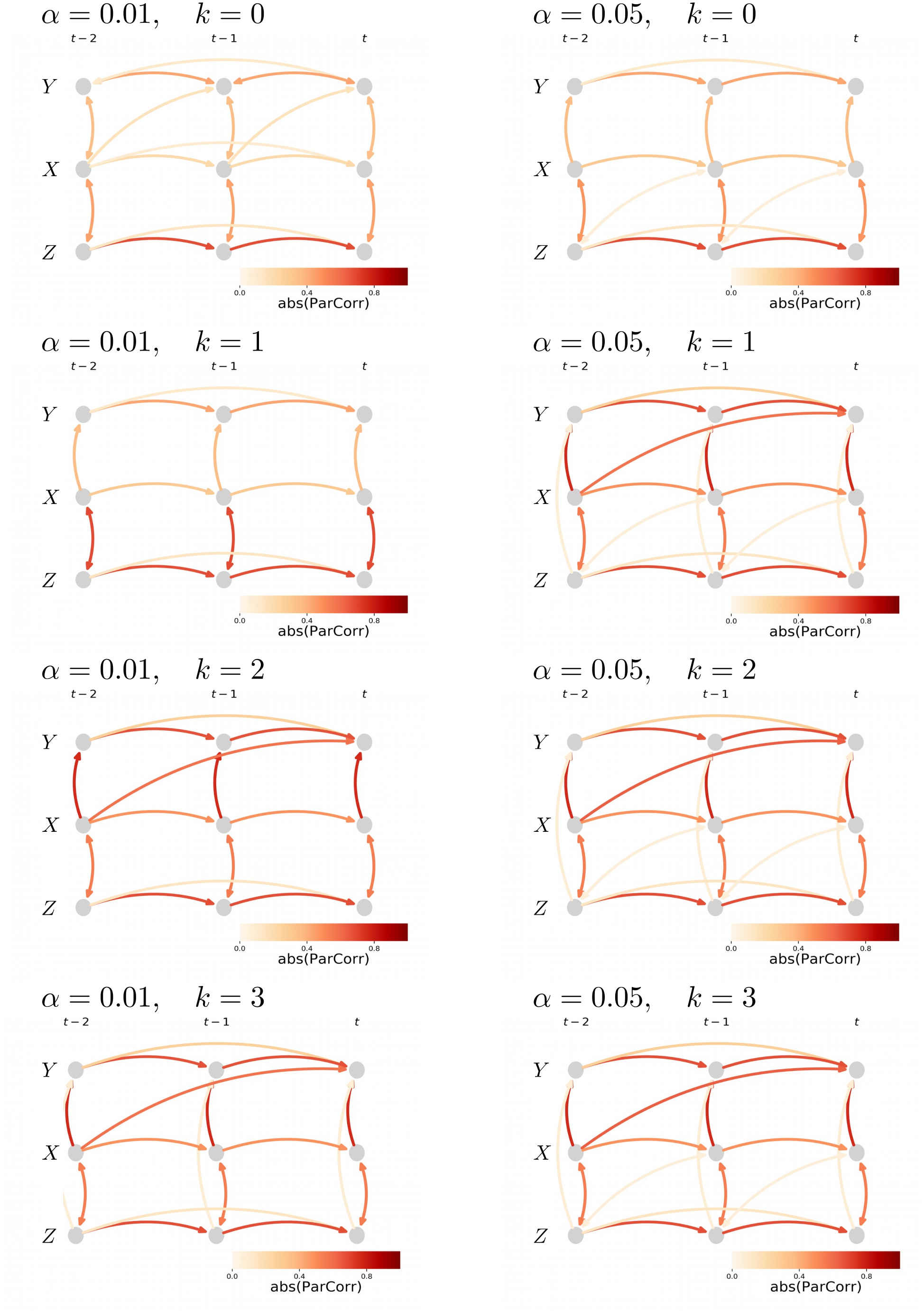}
\caption{PAGs estimated by LPCMCI($k$) on the real data example as described in Sec.~\ref{sec:real_data} of the main text. The respective values of $k$ and $\alpha$ are shown above each individual plot.
}
\label{fig:real_data_results_lpcmci}
\end{figure*}

\begin{figure*}[tbhp]
\centering
\includegraphics[width=0.9\linewidth]{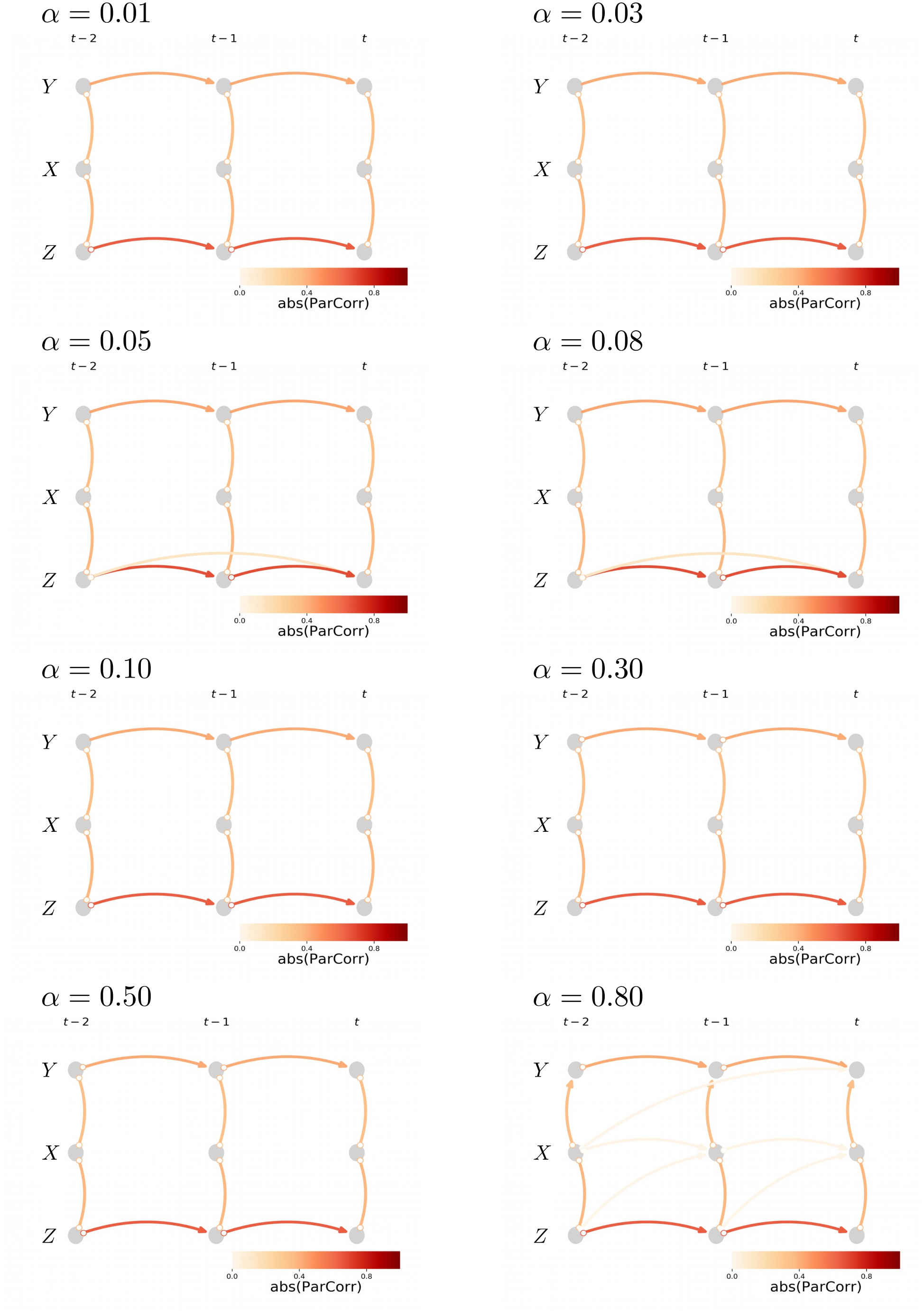}
\caption{PAGs estimated by SVAR-FCI on the real data example as described in Sec.~\ref{sec:real_data} of the main text. The respective values of $\alpha$ are shown above each individual plot.
}
\label{fig:real_data_results_fci}
\end{figure*}

This section shows the results that underlie the discussion of the application to the real data example in Sec.~\ref{sec:real_data} of the main text. Figure \ref{fig:real_data_results_lpcmci} shows the PAGs estimated by LPCMCI($k$) for $k = 0, \ldots, 3$ and $\alpha = 0.01, 0.05$. The results of LPCMCI($k = 4$), although mentioned in the main text, are not shown because they agree with those of LPCMCI($k = 3$) for both considered values of $\alpha$. Figure \ref{fig:real_data_results_fci} shows the PAGs estimated by SVAR-FCI for $\alpha = 0.01, 0.03, 0.05, 0.08, 0.1, 0.3, 0.5, 0.8$. All results are based on ParCorr CI tests and $\taumax = 2$. The link colors encode the absolute value of the minimal ParCorr test statistic of all CI tests for the respective pair of variables.

\clearpage
\section{Non-completeness of FCI with majority rule}\label{sec:noncompleteness}
\begin{figure*}[tbhp]
\centering
\includegraphics[width=0.6\linewidth]{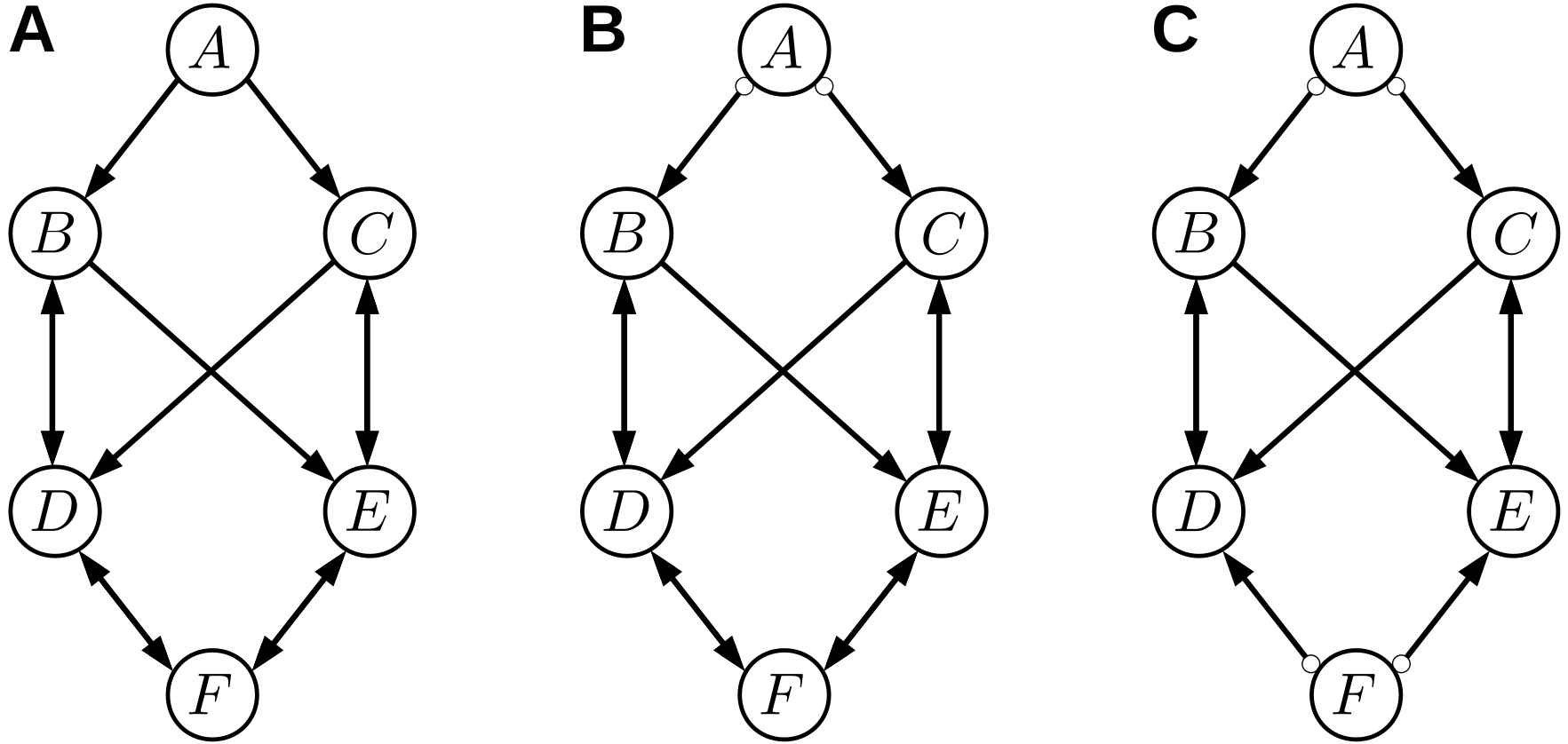}
\caption{
Example to illustrate the non-completeness of FCI with majority rule. (\textbf{A}) MAG $\mathcal{M}$. (\textbf{B}) Maximally informative PAG for $\mathcal{M}$, output of FCI \textit{without} majority rule. (\textbf{C}) Output of FCI \textit{with} majority rule.}\label{fig:noncompleteness}
\end{figure*}

In Sec.~\ref{sec:pseudocode2} it was mentioned that FCI becomes non-complete when its orientation rules in the final orientation phase are modified according to the majority rule of \cite{Colombo2014}. While this is probably known, we have not found it spelled out in the literature. Therefore, we here illustrate this point by the example given in Fig.~\ref{fig:noncompleteness}.

The left and middle part of the figure respectively show the true MAG and its fully informative PAG. As proven in \cite{Zhang2008}, the latter will be found by the standard FCI algorithm without modification according to the majority rule. Note that the two heads at node $F$ are put by the collider rule $\R0$: Since $F$ is not in the separating set $\SSet_{DE} = \{A, B, C\}$ of $D$ and $E$, the unshielded triple $D \asto F \oast E$ is oriented as collider $D \asthead F \headast E$. The output of FCI with modification according to the majority rule is shown in the right part of the figure. There, the two heads at $F$ are not found. The reason is that the majority rule instructs $\R0$ to base its decision of whether $D \asto F \oast E$ is oriented as a collider not on the separating set found during the removal phases (this is $\SSet_{DE}$) but rather on a majority vote of all separating sets of $D$ and $E$ in the adjacencies of $D$ and $E$. However, in the example there are no such separating sets since neither $D$ nor $E$ is adjacent to $A$. Therefore, $D \asto F \oast E$ is not oriented as collider by $\R0$ but rather marked as ambiguous. The heads can also not be found by $\R2$, $\R3$ and $\R4$, the other rules for putting invariant heads, because these only oriented edges that are part of a triangle. Since neither $F \ohead D$ nor $F \ohead E$ is part of a triangle, the orientations are not found. As described at the end of Sec.~\ref{sec:pseudocode2} we employ a modified majority rule in LPCMCI to guarantee both completeness and order-independence.
\end{document}